\documentclass[12pt]{article}
\usepackage{color}
\usepackage{graphicx}
\begin{document}
% ----------------------------------------------------------
% \input{title.tex}
% ===============================================================
% File title.tex .
% Last modified: 1 July 2026
% =========================================================
\Large
\noindent
Peer-Reviewed Editorial:
\begin{center}
  {\bf Axel Dieter Becke: Not just the ``B'' in ``B3LYP''}
\end{center}
\normalsize

\vspace{0.5cm}

\noindent
Mark Earl Casida\\
{\em Laboratoire de Spectrom\'etrie, Interactions et Chimie th\'eorique 
(SITh),
D\'epartement de Chimie Mol\'eculaire (DCM, UMR CNRS/UGA 5250),
Institut de Chimie Mol\'eculaire de Grenoble (ICMG, FR2607), 
Universit\'e Grenoble Alpes (UGA)
301 rue de la Chimie, BP 53, F-38041 Grenoble Cedex 9, FRANCE\\
e-mail: mark.casida@univ-grenoble-alpes.fr} 

% \vspace{0.5cm}
% 
% \noindent
% {\color{magenta}   % begin magenta
% Date of Publication: \today \hspace{0.1cm} (MS 2.15)
% }                  % end magenta

\vspace{0.5cm}

\begin{center}
{\bf Abstract}
\end{center}

There is hardly anyone doing electronic structure calculations on molecules
who has not used the famous B3LYP functional.  Many will even know that the
``B'' in B3LYP stands for Becke, but how many know the history of the B3LYP 
functional?  Behind this functional, Axel Becke has two of the most highly
cited papers in both chemistry and physics---both with several tens of 
thousands of citations---and many less cited, but arguably equally important
accomplishments, that led to B3LYP, as well as continued developments in
building ever better density-functional approximatons after the appearance
of B3LYP.  Axel has received much recognition and many honors for his
work.  In particular, in 2015, Axel was awarded Canada's highest honor for
science and engineering---namely the Gerhard Herzberg Gold Medal.  Achievements
like this require more than ``just'' brilliance; they require being the
right person in the right place at the right time.  I will argue that Axel
did indeed fulfill these requirements.  At the same time, I will try to give
some flavor of the field of electronic structure as seen through my own eyes 
and those of a few others during the period from the beginning of Axel's career
up until very recent times.

%%%%%%%
% EOF %
%%%%%%%
% \newpage
% -------------------------------------------------
% \tableofcontents
% \newpage
% =================================================
\section{Introduction}
\label{sec:intro}
% \input{intro.tex}
% ===============================================================
% File intro.tex .
% Last modified: 23 June 2026
% =========================================================

\begin{quote}
\noindent
% \fbox{
{\bf References in bold are Axel Becke's own publications
(listed in Appendix~\ref{sec:CVaxel} along with a summary
of some other highlights of his career).}
% }
\end{quote}

\vspace{0.5cm}

Like many fields, electronic structure theory has its own jargon, often
in the form of sequences of abbreviations used, among other things, as
keywords in electronic structure programs, and one of the best known of
these is ``B3LYP.''  I suspect that many more people have heard of (and may
even have used) B3LYP than have heard of Axel Becke, even though Becke
is the B in B3LYP.  But, if they have heard of Axel, do they really know
what made Axel one of the most cited authors in both chemistry and physics
and why he was awarded the Herzberg Gold Medal, Canada's highest honor for
science and engineering?  This article proposes not only to answer these
questions, but also proposes to answer the question of why Axel was the
right person in the right place at the right time to have this sort of
impact.

% ----------------------
\begin{figure}
\begin{center}
\includegraphics[width=0.4\textwidth]{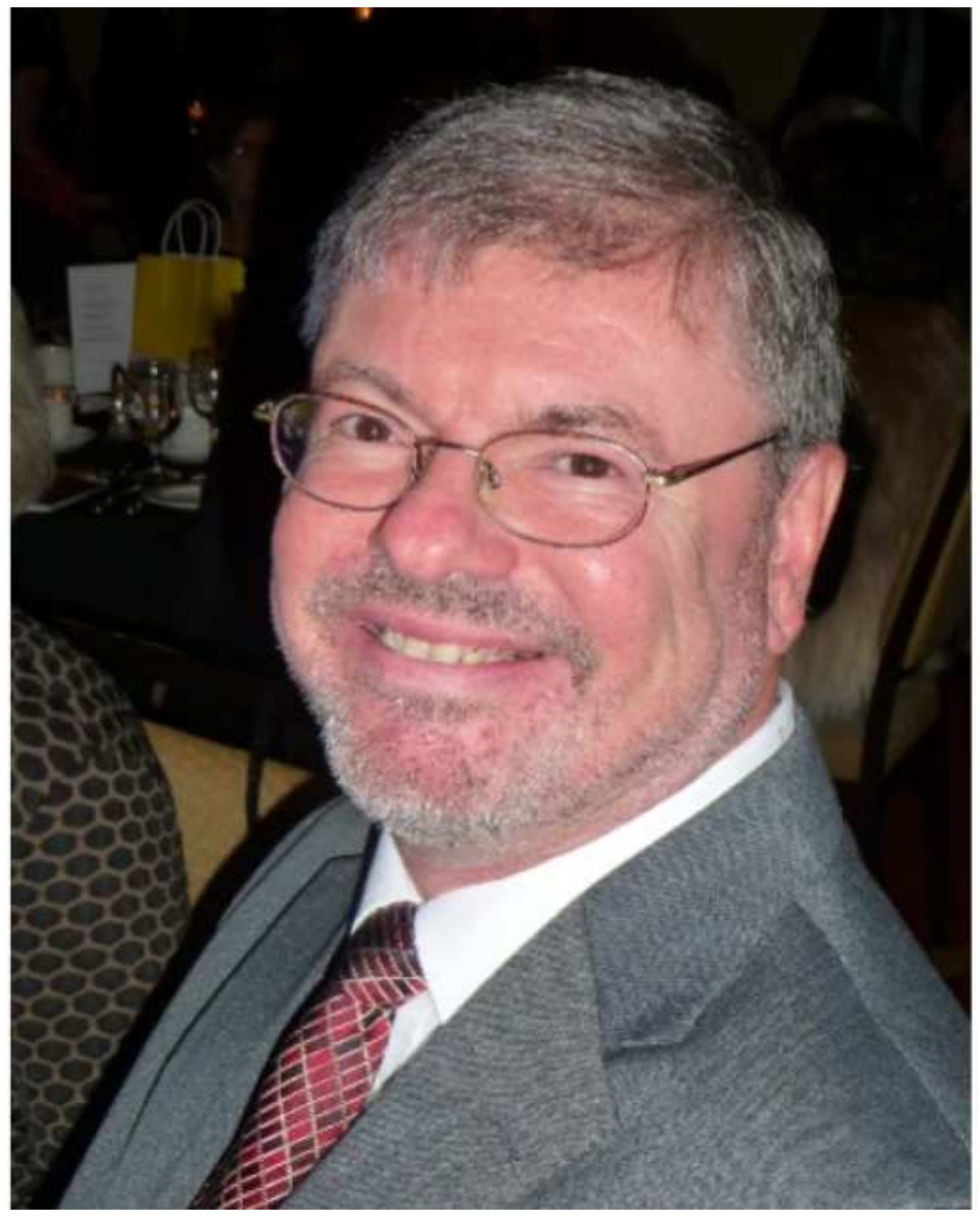}
\end{center}
\caption{
Photo of Axel Becke taken by Prof.\ Mary Anne White at the Discovery Centre Dinner in 
2015 (used with permission of Mary Anne White).
\label{fig:Axel}
}
\end{figure}
% ----------------------
Many of us were saddened to learn of the passing away of
Prof.\ Axel Becke % of Dalhousie University (whose original German 
% pronunciation must be something like ``Beckuh'', but which was pronounced 
% by most of his Canadian colleagues something like the girl's name ``Becky'') 
on 23 October 2025 ({\bf Fig.~\ref{fig:Axel}}).  This is especially true for 
those of us who had met and interacted with Axel, perhaps as part of the 
Canadian theoretical chemistry community or because of going to to 
density-functional theory (DFT\index{DFT}) conferences or for other reasons.  
Axel retired from Dalhousie University, Halifax, Nova Scotia, Canada,
and was granted emeritus status,
in 2015.  A memorial event was held for him there on 9 February 2026 in 
the form of talks by Russ Boyd, Ross Dickson, Don Weaver, Josef Zwanziger, 
Alastair Price, Erin Johnson, and Peter Becke \cite{instagram2}.  
On 25 February 2026, Lisa La Chance, Member of the Legislative Assembly,
spoke to the Nova Scotia House of Assembly in honor of Axel \cite{Instagram}.

{\em Throughout this article, I will be adding many quotes because I believe
they provide perspectives that go beyond the science while, usually, still
being about the science.  Most of these quotes come from things that Axel has
written.  Others come from people who have had a chance to get to know
Axel and can provide their own perspective.  Some are annecdotal remarks
from my own memories, not necesssarily concerning Axel directly, but which
describe what I remember about the attitudes of those around me in quantum 
chemistry as seen through my own eyes.
These latter annecdotal remarks will be marked with my initials
MEC\index{MEC} and will be {\sf in this font}.}

\begin{quote}
\noindent
{\sf
Axel was only 4 years my elder and so there are many overlaps between
Axel's career path and my own career path, as well as huge differences.
For example, Axel spent almost his entire life in Canada, whereas I grew
up in California and thereafter moved around quite a bit.  
(Appendix~\ref{sec:CVme} provides a brief summary of my own career.)  
Naturally I did not know much about Axel's early years. --- MEC
}
\end{quote}
I can only quote what is in the official obituary issued by Dalhousie:
\begin{quote}
\noindent
``Axel Dieter Becke was born in Nellingen, 
% [auf den Fildern which is now
% part of Ostfildern in the E{\ss}lingen district of the Stuttgart region], 
Germany in 1953, and came to
Canada with his parents at age 3, landing in Halifax. He was the first
of four brothers, and from the early days of his life, it would become
apparent that he would be first in whatever he decided mattered to him.

His parents, Helmut and Hannelore, created a supportive and nurturing
life and learning environment for the family. As Axel wrote for a
celebration of life for his father just a few weeks before Axel's own
death, `my parents instilled in me a love and curiosity for science as
far back as I can remember. Christmas gifts were educational, not just
for fun. Meccano sets and Lego sets.  Chemistry, electronics and physics
sets. Dad found a chemistry textbook at work once, gave it to me ---
I devoured it from cover to cover.'  He was a tinkerer in his early life,
spending much time by himself building things and playing with machines
and gadgets. But as his studies in physics progressed, he became more
interested in where formulas came from, and in the theories behind them.

While he made quantum leaps, and quantum chemistry became his life's work,
Axel had other early interests in life in which he excelled. In 1970 he
leapt almost 23 feet to set a Canadian age class record in the long jump.
He won the Canadian Accordion Competition. He had the top marks out of
high school, and went on to win the gold medal in Engineering Physics at
Queen’s University [...]'' \cite{Dal}
\end{quote}
[Some sources indicate that Axel was born in E{\ss}lingen.  There is no
contradiction:  Germany is divided into 16 partly sovereign states 
({\em L\"ander}).  Axel was born in Nellingen (formerly {\em Nellingen
auf den Fildern}) which is a suburb ({\em Stadtteil}) of the
city ({\em Stadt}) Ostfildern (established in 1975 as part of regional
reform) in the E{\ss}lingen district 
({\em Landkreis}, capital E{\ss}lingen am Neckar) of the Stuttgardt
government district ({\em Regierungsbezirk}) of the state ({\em Bundesland}) 
of Baden-W\"urttemberg.  This is part of the Swabia region of Germany and
many there speak in Swabian dialect.]

So Axel was 100\% Canadian, growing up from age 3 in Canada and establishing
his career there.  I think this deserves some elaboration because Canada
is both a large and small country.
\begin{quote}
\noindent
{\sf
One reason that I came to know Axel is that I was heavily involved with 
the {\sc deMon} suite of DFT programs, especially during my time in 
Montreal (1991-2001) but also afterwards, and so was  firmly embedded in
the small but active Canadian DFT community. Let us be sure to keep this 
in perspective:  Canada is the second largest country in the world by 
surface area (after Russia) but has a population similar to the state 
of California (Canada, 45.5 million in 2026; California, 39.4 million).  
Remarkably, Canada has had three Noble prizes in physical chemistry 
[Gerhard Herzberg, chemistry, 1971; John Polanyi, chemistry, 1986; 
Rudolph A.\ Marcus, chemistry 1992 (though Marcus became a naturalized
citizen of the U.S.A., I still think it is fair to count him because he 
grew up and was educated in Canada)].  
The Canadian theoretical chemistry community was (and still is) small 
but active and I had become a part of it until moving to France in 
2001 (and we continued to keep in touch even after that point).  This, 
and my DFT connection, was how I got to know Axel.
--- MEC
}
\end{quote}

Another way to honor Axel is with a {\em Festschrift} such as this one 
\cite{ElecStruc}.  It is within the living memory of this author of a time
when every chemist was advised to learn German.  (Sam Trickey, a physicist
from a still earlier generation, tells me that he was {\em required} to learn
German.)
This situation has 
changed as science (quite rightly!) becomes decreasingly ``Eurocentric.'' 
For those unfamiliar with the term, {\em Oxford languages \& Google} 
\cite{OLdef} defines {\em Festschrift} as ``a collection of writings published 
in honour of a scholar.''  
According to Andreas Savin, when I invited him to contribute something
to this issue, % 4 March 2026
\begin{quote}
\noindent
``I think it is good to make something for Axel. I would have preferred 
it when he was alive,'' but Axel did not want it.
% I even suggested it to Erin [Johnson], and she turned 
% me down: 
% [... I was informed that ...] Axel doesn't want it.''
\end{quote}
% But (according to Andreas) Axel did not want this type of honor while
% he was still living.
Axel Becke deserves many {\em Festschrifts} (or 
{\em Festschriften} for German-language purists) because of the huge impact 
that he has had on the development of modern DFT.  In addition to the present 
{\em Festschrift} \cite{ElecStruc}, we know of at least one other 
{\em Festschrift} in progress \cite{JCP}.  

It is appropriate to mention Ref.~\cite{NBB+2026} which will appear in the
{\em Journal of Chemical Physics} {\em Festschrift} \cite{JCP}.  This is
an article summmarizing the discussion that took place at a symposium
entitled ``DFT in Chemistry --- Dispute at Inception and Rise to Prominence''
that took place at {\em Svenska Kunliga Vetenskapsakademien} (the Swedish Royal
Academy of Science) 7-8 November 2024
with the purpose of examining past and present issues hindering DFT in
its path to its current prominence in chemistry and with the hope of 
perhaps being able to resolve any remaining issues as we look to the future.
After Axel's unexpected passing, Sture Nordholm gathered input from the various
symposim speakers to create an article in honor of Axel.  I was not at this 
symposium and the present article had already essentially reached its final 
form when I learned of Sture's article.  However, in reading the article, I
realized that some of the information communicated to me and contained in the
present article came from symposium attendees and are either, or seem to be,
verbatim identical in our two articles.  This is not plagarism, but only 
that we have both obtained our information from the same source and are
referenced appropriately.   The present article is very different from 
Sture's wonderful effort and I would recommend Sture's article to the
interested reader as a complementary view on the impact that Axel has made
in our field.

Any proper {\em Festschrift} should begin with a few words about the person
being honored so that those, perhaps less familiar with the man and his 
science, will be able to look back in future years and understand what Axel
has meant to us.  That, of course, is the purpose of this article.
In keeping with Marie Curie's famous recommendation,
\begin{quote}
\noindent
``Be less curious about people and more curious about ideas,'' \cite{C1937},
\end{quote}
I will focus mainly on ideas.  But we should also remember that the exchange of 
ideas in science is a social interaction, so this article will 
include some reminiscences of those who interacted with Axel and are still alive 
to share their memories, for key scientific elders who knew and admired
Axel's work (Nicholas Handy, Walter Kohn, John Pople, Tom Ziegler, ...) 
have already passed on. The best we can do now, besides writing down
our own memories, is to transmit what we have heard from those who have
gone before us. At other times, I will throw in some anecdotes from
my own experience in order to give a (necessarily biased) spirit of the time.

The present article is complementary to other efforts: In 2014, Axel 
published a nice perspective article of fifty years of DFT [{\bf B2014}]. 
In 2015, Canada's Natural Science
and Engineering Research Council (NSERC\index{NSERC}) made a 3 1/2 minute 
video of an interview with Axel \cite{NSERCvideo} after he won the Herzberg 
Medal, Canada's highest honor for scientific achievement.  There is also
a recent video \cite{FreasierVideo} presenting a popular review of Axel's 
science.  The Tuesday session of the 2026 session of the Canadian Symposium 
on Theoretical and Computational Chemistry (CSTCC\index{CSTCC}) will
be dedicated to a symposium in honor of Axel \cite{CATC}.  It would be
surprising if there are not also other sessions in his honor!

The organization of the rest of this article is as follows: The next
section describes the place of DFT in electronic structure theory up
until the point when Axel began his Ph.D.\ thesis work in this area.
Section~\ref{sec:golden} describes the exciting period where Axel played 
a key role in the coming of age and the ultimate acceptance of DFT in
the greater quantum chemistry community.  It should be clear by the
end of that section why Axel was the right person, in the right place,
at the right time to contribute the way he did.  But it was certainly not
the end of Axel's contributions!  Section~\ref{sec:ladder} continues with
more recent efforts made by Axel to create, and his choice of strategy
for making, increasingly accurate functionals.  Section~\ref{sec:misc} 
covers a few choice topics that did not fit neatly into the previous sections.
Section~\ref{sec:conclude} concludes. 

% =================================================
\section{Electronic Structure Theory Before Axel}
\label{sec:before}
% \input{before.tex}
% ===============================================================
% File before.tex .
% Last modified: 18 July 2026
% =========================================================

This section reviews the state of electronic structure theory up to about the
time when Axel began his Ph.D.\ thesis.  Central to this theory
is solving the nonrelativistic Schr\"odinger
equation for $N$-electrons in the Born-Oppenheimer (frozen nuclei)
approximation,
\begin{equation}
   {\hat H} \Psi_I(1,2,\cdots,N) = E_I \Psi_I(1,2,\cdots,N) \, ,
   \label{eq:intro.1}
\end{equation}
where $i=(x_i,y_i,z_i,\sigma_i)$ are the space ${\vec r}_i = (x_i,y_i,z_i)$
and spin $\sigma_i=\uparrow,\downarrow$ coordinates of
electron $i$.  It is well known that analytic solutions of
Eq.~(\ref{eq:intro.1}) do not exist beyond $N=1$ (unless we count the
analytic solution of ``harmonium,'' which consists of two electrons in
a spherically-symmetric harmonic oscillator potential \cite{YYL2024}).  However very accurate numerical solutions obtained by the Hylleraas approach 
exist up to $N\approx 4$ or so.  Thereafter approximate solutions are necessary. 

% \input{./tables/flavors.tex}
% ========================================
% File: flavors.tex
% Last updated: 13 July 2026
% ========================================

% -----------------------------------------------
\begin{table}
\begin{center}
\begin{tabular}{cc}
\hline \hline
Flavor & Number of Parameters \\
\hline
{\em ab initio} & 3 ($m_e$, $e$, and $\hbar$) for the whole periodic table \\
DFT & the same 3 plus several parameters for the whole periodic table \\
semi-empirical & several parameters per element in the table \\
\hline \hline
\end{tabular}\\
\caption{
\label{tab:flavors}
The three major flavors of electronic structure methods.
Many will argue that there are also ``hidden parameters'' to be found in
the numerical methods used --- that is, in the construction of basis sets,
in the choice of order of perturbation, in number of excitations to be 
included etc.  Obviously these have not been included in this table.
}
\end{center}
\end{table}
% -----------------------------------------------
One way to distinguish the different approaches to developing approximate
solutions is by counting how many fitting parameters are involved.  This
leads to the three flavors of electronic structure theory shown in 
{\bf Table~\ref{tab:flavors}}---namely {\em ab initio}, DFT, and semi-empirical. Axel's work focused on DFT.  The quality of density-functional approximations
(DFAs\index{DFA}) is often judged by comparing calculated observables against
either the experimentally-measured values or against high-quality {\em ab
initio} calculations.  Semi-empirical calculations are sometimes ``improved''
by adding semi-empirical parameters (DFT+Hubbard) or DFT is ``extended''
by deliberately designing a DFT-like semi-empirical theory and then fitting
parameters so as to retain results as close as possible to the corresponding
DFT results [density-functional tight binding (DFTB\index{DFTB})].
We will first review these two flavors of electronic structure theory and then,
in a separate subsection, have a deeper look into DFT.

% ----------------------------------------------------------
\subsection{The other flavors of electronic structure theory}
% ----------------------------------------------------------

\begin{quote}
\noindent
{\sf
My first exposure to quantum chemistry was doing undergraduate research
with Henry F.\ Schaefer III in the late 1970s.  This consisted of
{\em ab initio} calculations on small exotic gas phase molecules.
Our goal was to get ``the right answer for the right reason''
\cite{D2019} as opposed to just getting the right answer without understanding
what the method was telling us about the physics, which often seemed to be
the case in semi-empirical theories.  (Note, however, that this did not
prevent me from an appropriate use of DFTB semi-empirical theory in later
years!)  I was unaware of DFT in any of its forms in the 1970s.
--- MEC
}
\end{quote}

% \input{tables/unified.tex}
% ========================================
% File: unified.tex
% Last updated: 13 July 2026
% ========================================

\begin{table}
\begin{center}
\begin{tabular}{cccccc}
\hline \hline
Year & \multicolumn{3}{c}{Flavor$^a$} & Milestone & References \\
\cline{2-4}
     & AI & DFT & SE &        &    \\
\hline
1987 & X  &   &     & Quadratic Configuration Interaction \cite{PHR1987} \\
1986 &    &   &  X  & Density-Functional Tight Binding (DFTB\index{DFTB}) &
\cite{SEB1986,EPJ+1998,KSF2005,YYY+2007} \\
1980 &    & X &     &Vosko-Wilk-Nusair LDA parameterization & \cite{VWN1980} \\
1980 &    & X &     & Ceperley-Alder quantum Monte Carlo & \cite{CA1980} \\
1975 & X  &   &     & Davidson diagonalization for CI & \cite{D1975,D1989,MRD1992} \\
1973 &    &   &  X & Zerner's Intermediate Neglect of Differential Overlap (ZINDO\index{ZINDO}) & \cite{RZ1973,Z1991,DZ2001} \\
1971 &    & X &    & X$\alpha$ & \cite{S1951,G1954,HVO1969,SW1971,S1972,S1972b} \\
1966 & X  &   &    & Coupled-Cluster & \cite{C1966} \\
1965 &    & X &    &  Kohn-Sham method & \cite{KS1965}\\
1965 &    &   &  X & Complete Neglect of Differential Overlap (CNDO\index{CNDO}) &
   \cite{PS1965,PSS1965,PS1966,SS1967,PB1970} \\
1964 &    & X &    & Hohenberg-Kohn theorems & \cite{HK1964} \\
1963 &    &   &  X & Extended H\"uckel Theory & \cite{H1963a,H1964b,H1964c,H1964d} \\
1963 &    &   &  X & Hubbard Model & \cite{H1963b} \\
1953 &    &   &  X  & Pariser-Parr-Pople (PPP\index{PPP}) & \cite{PP1953a,PP1953b,P1953} \\
1951 & X  &   &     & Roothaan-Hall SCF equations & \cite{H1951,R1951,R1960} \\
1951 &    & X &     & Slater potential & \cite{S1951} \\ 
1949 & X  &   &     &   Coulson-Fischer VB-MO equivalence & \cite{CF1949} \\
1934 & X  &   &     &   M{\o}ller-Plesset perturbation theory & \cite{MP1934} \\
1931 &    &   &  X  & Simple H\"uckel Theory & \cite{H1931,H1932,W1941} \\
1929 & X  &   &     &   Hylleraas solution for He & \cite{H1963} \\
1929 & X  &   &     &   Lennard-Jones paper on diatomic molecules (MO theory) & \cite{L1929} \\
1927 & X  &   &     &   Heitler-London wave function for H$_2$ (VB theory) & \cite{HL1927} \\
1927 &   &  X  &  & Thomas-Fermi-Dirac theory & \cite{F1927,T1927,F1928,D1929} \\
1926 & X  &   &     &   Schr\"odinger equation & \cite{S1926a,S1926b} \\
\hline \hline
\end{tabular}\\
$^a$AI = {\em ab initio}, DFT = density-functional theory,
SE = semi-empirical
\caption{
\label{tab:unified}
A brief (and no doubt incomplete and highly arbitrary) history of electronic 
structure theory, classified according to the three flavors of 
Table~\ref{tab:flavors}.
}
\end{center}
\end{table}
{\bf Table~\ref{tab:unified}} gives a brief (and necessarily biased) overview
of the history of the development of the three major flavors of electronic
structure theory.

% --------------------------------
\subsubsection{\em Ab Initio}
% --------------------------------
The oldest approach is also the one which is most familiar to most
people.  Hence this section is mainly review and 
the establishment of the notation that will be used elsewhere in this
article.
{\em Ab initio} electronic structure theory has only three parameters---namely 
the mass of the electron $m_e$, the (absolute value of) the charge of 
the electron $e$, and $\hbar =h /( 2\pi) $ (Planck's constant divided 
by $2 \pi$) in the gaussian system ($4\pi\epsilon_0=1$) 
of electomagnetic units.  In the corresponding set of Hartree atomic 
units, $m_e = e = \hbar = 1$.  

A brief history of this {\em ab initio} (Latin for ``from the beginning'')
or first-principles approach to the electronic structure problem is summarized
in Table~\ref{tab:unified}.  Schr\"odinger solved his equation
analytically for the harmonic oscillator, rigid rotor, and for the hydrogen
atom \cite{S1926a,S1926b}.  Already, the following year, Heitler and London
found an approximate solution for the H$_2$ molecule which provided a
qualitatively correct potential energy curve \cite{HL1927}.  
By 1929, Hylleraas had found an essentially exact numerical solution of the 
Schr\"odinger equation for the He atom \cite{H1963}.  In fact, Hylleraas 
invented many of the numerical methods which would be later become part of the
toolbox of the most highly accurate {\em ab initio} calculations.  However
the Heitler-London paper would become the seminal paper for the classic 
valence-bond (VB\index{VB}) method of electronic structure theory 
whose identification of spin-coupling diagrams with Lewis dot structures 
gave birth to the concept of resonance structures \cite{P1939,W1944}.  
These and the localized 
orbitals naturally produced in VB theory are far from obsolete as they may be 
found in any University-level chemistry textbook.  Molecular orbital 
(MO\index{MO}) theory rapidly established itself as a competing theory for
{\em ab initio} calculations \cite{L1929} and it is the one currently
favored by most electronic structure theorists.  All University-level 
chemistry textbooks also contain an explanation of MO diagrams for diatomic
molecules.  

One way to explain MO theory
is based upon the use of the variational principle to find the lowest energy
Slater determinant of orthonormal orbitals,
\begin{equation}
   \Phi(1,2,\cdots,N) = \frac{1}{N!} \left| \begin{array}{ccccc}
   \psi_{i_1}(1) & \psi_{i_1}(2) & \cdots & \psi_{i_1}(N) \\
   \psi_{i_2}(1) & \psi_{i_2}(2) & \cdots & \psi_{i_2}(N) \\
   \vdots        & \vdots        & \ddots & \vdots \\
   \psi_{i_N}(1) & \psi_{i_N}(2) & \cdots & \psi_{i_N}(N) 
   \end{array} \right| \, ,
   \label{eq:intro.2}
\end{equation}
which is sometimes abbreviated as,
\begin{equation}
  \Phi = \vert i_1, i_2, ..., i_N \vert \, .
  \label{eq:intro.2.5}
\end{equation}
% Here $i = ({\vec r}_i,\sigma_i)$ is shorthand for the space and spin 
% coordinates of electron $i$.
The corresponding Hartree-Fock (HF\index{HF}) energy expression is,
\begin{equation}
  E_{HF} = E_H + E_x \, ,
  \label{eq:intro.3}
\end{equation}
where the Hartree energy is given by,
\begin{equation}
 E_H = \sum_i n_i \langle \psi_i \vert \hat{t} + v \vert \psi_i \rangle
 + \frac{1}{2} \int \int \frac{\rho(1) \rho(2)}{r_{1,2}} \, d1 d2 \, .
 \label{eq:intro.4}
\end{equation}
Here $\hat{t}$ is the kinetic energy operator, $v$ is the external 
potential (i.e., everything external to the system of $N$ electrons
and, in particular for molecules, the attraction to the nuclei),
and $n_i$ is the occupation number of orbital $\psi_i$.  The exchange
energy,
\begin{equation}
  E_x = -\frac{1}{2} \int \int \frac{\vert \gamma(1,2) \vert^2}{r_{1,2}} \, 
  d1 d2 \, ,
  \label{eq:intro.5}
\end{equation}
where the mathematical quantity,
\begin{equation}
  \gamma(1,2) = \sum_i n_i \psi_i(1) \psi_i^*(2) \, ,
  \label{eq:intro.6}
\end{equation}
is the one-electron reduced density matrix (1-RDM\index{1-RDM}) and its
diagonal element is the density,
\begin{equation}
  \rho(1) = \gamma(1,1) \, .
  \label{eq:intro.7}
\end{equation}

Up to this point, I am really using spin-orbitals, but most of DFT is done
explicitly in terms of the separate space and spin parts.  This means
that the spin-orbital index $i$ is a composite object which we can
rewrite as $i\sigma$ where $i$ now labels just the spatial part and $\sigma$
labels the spin part.  This latter formalism is the most frequently 
used internally in electronic structure theory codes.  Then,
\begin{equation}
  \psi_{i\sigma}(1) = \psi_i^{\sigma}({\vec r}_1) \sigma(1) \, ,
  \label{eq:intro.7.1}
\end{equation}
and,
\begin{eqnarray}
  \gamma(1,2) & = & \sum_{i\sigma} n_i^\sigma \psi_i^\sigma({\vec r}_1)
  \psi_i^{\sigma *}({\vec r}_2) \sigma(1) \sigma^*(2) \, .
  \label{eq:intro.7.2}
\end{eqnarray}
DFT will frequently ``integrate'' (really sum) over the spin variable
to get just the spatial part,
\begin{equation}
  \gamma({\vec r}_1,{\vec r}_2)  =  \sum_\sigma \left( \sum_i n_i^\sigma 
  \psi_i^\sigma({\vec r}_1) \psi_i^{\sigma \, *}({\vec r}_2) \right) 
   =  \sum_\sigma \gamma_\sigma({\vec r}_1,{\vec r}_2) \, .
  \label{eq:intro.7.3}
\end{equation}
A consequence of the separation of the 1-RDM into a linear combination
of a spin-up  and a spin-down 1-RDM, is that the exchange energy
also separates into the sum of two parts,
\begin{equation}
  E_x = E_x^\uparrow + E_x^\downarrow \, .
  \label{eq:intro.7.4}
\end{equation}
Of course, the density also separates into two parts,
\begin{equation}
  \rho({\vec r}) = \rho_\uparrow({\vec r}) + \rho_\downarrow({\vec r}) \, .
  \label{eq:intro.7.5}
\end{equation}
We will often use the spin-orbital notation when we want to be brief
and the spatial-orbital notation when we want to be more explicit.

\begin{quote}
\noindent
{\sf
My next exposure to quantum chemistry was my Ph.D.\ work with Prof.\ John
E.\ Harriman, during the first half of the 1980s.  I chose to work on the
mathematical properties of reduced density matrices.  John was very
much of a mathematician, but (like Henry F.\ Schaefer before him) he saw
me as even more of a mathematician (perhaps because I chose to minor in
mathematics).   

During my time in Montreal or shortly thereafter, Axel and I were talking 
over lunch at some conference and Axel asked me, ``Where did you come from?''  
He thought he knew all the major players in DFT and I suppose that he was 
beginning to see me as important in DFT.  The answer, of course, is that 
I had come up from the very different background of reduced density matrix 
and Green's function theory (which are sometimes viewed as a sort of
time-dependent reduced density matrix), something which left me well 
prepared to follow the intricacies behind the working of DFT.
--- MEC
}
\end{quote}

\begin{figure}
\begin{center}
\includegraphics[width=0.4\textwidth]{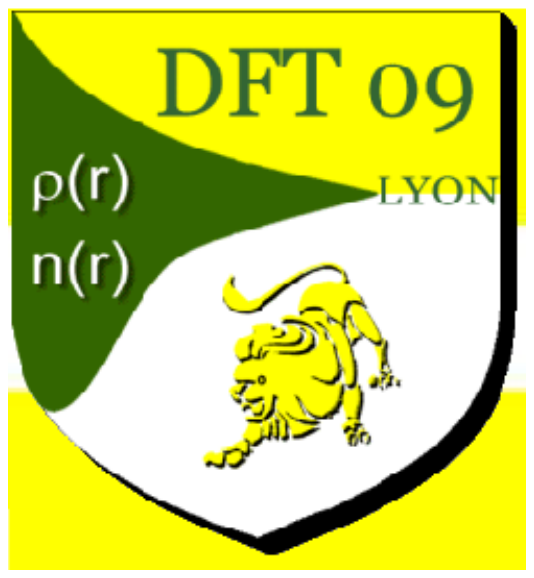}
\end{center}
\caption{
The logo from the book of abstracts of the 13th edition of the International
Conference on the Applications of Density Functional Theory in Chemistry and
Physics (DFT'09), 31 August-4 September 2009, in Lyon, France.
\label{fig:DFT09}
}
\end{figure}

Minimizing $E_{HF}$ subject to the constraint of orthonormal orbitals
leads to the HF orbital equation,
\begin{equation}
  \left( \hat{h}_H + \hat{\Sigma}_x \right) \psi_i(1) 
  = \epsilon_i \psi_i(1) \, ,
  \label{eq:intro.8}
\end{equation}
where 
\begin{equation}
  \hat{h}_H = \hat{h}_C + v_H \, ,
  \label{eq:intro.9}
\end{equation}
is the Hartree hamiltonian .  Here 
$\hat{h}_C = \hat{t}+v$ is the so-called core hamiltonian which 
contains everything except the electron-electron interactions (i.e., 
the core hamiltonian contains
the kinetic energy $\hat{t}$ and attraction to the nuclei $v$), 
$v_H = J$ is the Hartree
(in physics) or Coulomb (in chemistry) potential defined by,
\begin{equation}
  v_H \psi(1) = \int \frac{\rho(2)}{r_{1,2}} \, d2 \, \psi(1) \, ,
  \label{eq:intro.10}
\end{equation}
and $\hat{\Sigma}_x = -\hat{K}$ is the exchange self-energy (in physics) or 
the exchange operator (in chemistry) and is defined in spin-orbital notation
as,
\begin{equation}
  \hat{\Sigma}_x \psi(1)  =  -\int \frac{\gamma(1,2)}{r_{1,2}} \psi(2) \, d2 
  \, ,
  \label{eq:intro.11.3}
\end{equation}
or in spatial orbital notation as,
\begin{equation}
  \hat{\Sigma}_x^\sigma \psi^\tau(\vec{r}_1)  =  -\delta_{\sigma,\tau} 
  \int \frac{\gamma_{\sigma}
  (\vec{r}_1,\vec{r}_2)}{r_{1,2}} \psi^\sigma({\vec r}_2) \, d\vec{r}_2 \, .
  \label{eq:intro.11}
\end{equation}
Another difference between chemistry and physics is that the density is
usually indicated by $\rho$ in the chemistry literature and by $n$ in the 
physics literature.   Advantage was taken of this in composing
the logo ({\bf Fig.~\ref{fig:DFT09}}) for the DFT09 meeting in Lyon 
in order to emphasize that the meeting spanned both the chemistry and
the physics communities.
One of the organizers of this meeting was Henry Chermette who had known
Axel since their first meeting at the North Atlantic Treaty Organization
(NATO\index{NATO}) Advanced Study Institute 
Density Functional Methods in Physics, in 1983 at Alcabideche, Portugal.
Henry wrote to me:
\begin{quote}
\noindent
``I invited him to give the first plenary lecture in the DFT09 (2009) in 
Lyon (450 participants), and he [Axel] told me that it was the first time 
he was invited to give the first talk of a congress. His modesty was to be 
underlined.

I remember his talk, very pedagogical, with use of handmade transparencies 
written with very clear color felt tips, whereas that was (already) the 
beginning of power-point presentations.'' 
\end{quote}
% I think that this might come from the idea that $n$
% is a number density and $\rho$ is a charge density, but this is pure
% speculation.
% ----------------------
% ----------------------

Roothaan's introduction of a finite orbital basis set [frequently called
the linear combination of atomic orbitals (LCAO\index{LCAO}) approximation,
especially in the older chemistry literature]
has made the (approximate) solution of the Hartree-Fock problem tractable
for molecules \cite{R1960}.  This is often known as a self-consistent
field (SCF\index{SCF}) calculation to distinguish it from the an exact
solution of the Hartree-Fock MO problem.  Very importantly, the introduction 
of a basis set replaces solving differential equations with matrix equations.
This is also commonly done in DFT.

It is now widely recognized that neither a single VB wave function nor a single
Slater determinant of MOs is adequate for chemically accurate calculations.
This is solved in VB theory by including linear combinations of wave functions
corresponding to different resonance structures, thereby creating the VB
$N!$ problem.  The solution in MO theory is to include a linear combination
of Slater determinants in a configuration interaction (CI\index{CI}) expansion,
thereby creating the CI $N!$ problem.  Nevertheless Coulson and Fischer
showed that the two approaches become identical when 
higher order terms are included \cite{CF1949}.  One key point is that 
special Krylov subspace methods, such as the block Davidson 
method \cite{D1975,D1989,MRD1992},
allow the solution of large CI matrix eigenvalue problems.   It was later
recognized that CI suffers from problems with both size-consistency and
size-extensivity which can be solved by many-body perturbation theory
(MBPT\index{MBPT}) either in its primitive M{\o}ller-Plesset form \cite{MP1934}
or (better yet) in its coupled-cluster (CC\index{CC}) form \cite{C1966}.
This review of {\em ab initio} electronic structure theory is necessarily
only brief.  
It should be pointed out that the different approximations (HF, VB, MBPT,
CI, CC, etc.) as well as the choice of basis set are sometimes viewed as
a {\em de facto} additional set of fitting parameters.  However {\em ab initio}
ideally provides a systematic way to converge to the exact solution, even 
if this is not always possible to put into practice.
A more detailed review of some of these now classic concepts 
may be found elsewhere, such as in the book of Szabo and Ostlund \cite{SO1996}.

Interestingly solid-state quantum physicists often refer to DFT as 
{\em ab initio}. Axel did the same thing:
\begin{quote}
\noindent
``I will refer to HF and post-HF methods as {\em wave-function} methods, and the
theory as {\em wave-function theory} (WFT).  The more common terminology,
{\em ab initio} theory, is regrettable as it denigrates density-functional
theory. [...] DFT is as `{\em ab initio}' as WFT.'' [{\bf B2014}]
\end{quote}
What I believe Axel meant by this is that the systemmatic construction of DFAs 
is (usually) based upon physical and mathematical arguments that place constraints on the
exchange-correlation (xc\index{xc}) hole and therefore DFT is as rigorous
as is WFT.  
Although I appreciate Axel's point, we will follow the usual practice in
quantum chemistry which distinguishes DFT from {\em ab initio}.

% --------------------------------
\subsubsection{Semi-Empirical}
% --------------------------------

Another flavor of electronic structure theory is the semi-empirical 
approach.  In a sense, this approach began with the very beginning of the
application of quantum mechanics to trying to obtain a better understanding
of the electronic structure of molecules.  Some integrals were too hard to
calculate and so were estimated from experiment.  In other cases, the matrix
reformulation of the electronic structure problem lent itself to correcting
oversimplified assumptions by either fitting integral values to experiment or
to more sophisticated electronic structure calculations.  Ref.~\cite{BJ2005}
provides a very nice review of semiempirical theory up to the year 2005 and I
have used it to help me to try to capture some key dates in 
Table~\ref{tab:unified}.  By
its very nature, semiempirical theory aims at treating systems which are too
large to handle by {\em ab initio} or DFT methodology.  So why do it?
Let me paraphrase an answer given to me by my friend and colleague (and 
sometimes science guru) because I find she gives particularly clear 
explanations:
\begin{quote}
\noindent
``The best use of semiempirical theories is either (i) to show that a simple 
theory suffices to explain some phenomenon or (ii) to carry out calculations 
which could not be done any other way.'' --- Lucia Reining 
(personal communicaton, 2019, Paris, France)
\end{quote}

While the place in this article of a rapid review of {\em ab initio} calculations 
should be clear because of how often DFT is compared against {\em ab initio}
benchmarks, there are two reasons for mentioning semi-empirical methods here.
One is the DFT+Hubbard (i.e., DFT+U and DFT+U+V) method \cite{AZA1991,LAZ1995,DBS+1998,CC2010} 
which combines DFT with the Hubbard semi-empirical model in order
to correct bandgap calculations in solids containing 
transition metals.  The other reason is the existence of the 
DFTB semi-empirical model which is specifically designed 
to behave as much as possible like DFT and is usually fit to DFT calculations 
but uses additional semi-empirical approximations to extend DFT-like 
calculations to much larger molecules than would otherwise be possible 
\cite{PFK+1995,KM2009,GGE2013}. Axel Becke did not dabble in these things
(though he once mentioned to me that he had been invited to talk about
DFT+U and was confused about how to respond!) but Axel's work has had
a great impact on DFTB which is designed to resemble DFT as much as possible
and whose parameterizations are based upon DFT calculations.

% ------------------------------------------------------------
\subsection{Density-Functional Theory}
% -----------------------------------------------------------

In this section, we go back to a time before Axel's involvement in DFT
and have a look at the development
of DFT up to the moment when Axel would begin his Ph.D.\ where he 
specialized in, and began to make tremendous contributions to, DFT.

% -----------------------------------------------------------------
% \subsubsection{Thomas-Fermi-Dirac and the Hohenberg-Kohn Theorems}
% -----------------------------------------------------------------
\subsubsection{History and Formalism}

\begin{quote}
\noindent
{\sf
During our weekly research discussions when I was a doctoral student
with John Harriman in Wisconsin in the 1980s, we often discussed
DFT as well as reduced density matrix theory.   Questions of $N$- and
$v$-representability in DFT interested us, but I was skeptical that DFT
could ever replace {\em ab initio} methods.  John wrote
in the acknowledgement of an article on densities, operators,
and basis sets \cite{H1986},
\begin{quote}
\noindent
``I would like to thank Mark E.\ Casida for the proof of
Theorem 4 and for a number of helpful discussions.''
\end{quote}
Was this my first contribution to DFT?
--- MEC
}
\end{quote}

Although modern DFT is often defined by the key 
sequential papers of Hohenberg and 
Kohn (HK\index{HK}) \cite{HK1964} and Kohn and Sham (KS\index{KS}) 
\cite{KS1965}, DFT existed {\em de facto} before these two seminal papers.
Orbital-free DFT had already been formulated in the early work of Thomas, 
Fermi, and Dirac (TFD\index{TFD}) \cite{F1927,T1927,F1928,D1929} as an
approximation based, in part, on phase-space arguments. 
Hohenberg and Kohn \cite{HK1964} set TFD theory on a rigorous basis by their 
proof of two theorems. 
\begin{quote}
\noindent
{\em Theorem (HK1)}. The external potential of the nondegenerate ground state
of $N$ electrons is determined by the density up to an arbitrary additive 
constant.
\end{quote}
E.\ Bright Wilson famously noted during a conference discussion 
that this is a trivial 
observation for a molecule because 
the integral of the density gives the number of electrons, the cusps in 
the density gives the positions of the nuclei, and the slope at each cusp 
gives the charge of the nucleus.  
(There is no article where Wilson makes this argument, but the 
remark was probably made during the 1963 Sanibel conference in honor 
of Egil Hylleraas: P.-O.\ L\"owdin
attributes the argument to Wilson on page 30 of Ref.~\cite{L1986}; 
I learned it by word of mouth from my thesis director, J.E.\ Harriman)
However HK1 is not limited to a 
molecular external potential (i.e., all that is not composed of electrons).
Interestingly HK1 does not tell us how to match up the arbitrary additive
constants for a system composed of separate non-overlapping densities.
The second theorem is 
\begin{quote}
\noindent
{\em Theorem (HK2)}.  Given the external potential $v$, the ground state
density may be found by minimizing the functional,
\begin{equation}
  E_v[\rho] = F[\rho] + \int v(1) \rho(1) \, d1 \, ,
  \label{eq:intro.12}
\end{equation}
where the functional $F$ is universal in the sense that it is independent
of $v$.
\end{quote}
The notation $F[f]$ indicates that $F$ is a function{\em al}
of the function $f$---that is, a function of a function (i.e., a rule 
that assigns a number to each function in a set of functions). In order to 
understand this more clearly, compare
$F(\vec{q})=F(q_1,q_2,q_3, \cdots)$ which is a function of a vector, that is
of a discrete set of variables, with $F[f]$ which is a simultaneous function
of all of $f(x)$ over the continuous set of values of $x$.  [The function $f$ (or $[f]$ to emphasize its function nature) is the ``vector'' whose components 
are labeled by $x$. From this point
of view, Eq.~(\ref{eq:intro.17}) might be better written as 
$v_{xc}^\sigma([\rho_\uparrow],[\rho_\downarrow],{\vec r}) = \delta
E_{xc}^\sigma([\rho_\uparrow],[\rho_\downarrow])/\delta \rho_\sigma({\vec r})$,
but writing $E_{xc}([\rho])$ looks odd compared with $E_{xc}[\rho]$.]  
Functionals and their derivatives are the subject of the calculus of 
variations \cite{GF1963}.

While it is often said that the functional $F$ is unknown, the Levy-Lieb
\cite{L1979,L1983}
constrained variational principle gives the explicit formula,
\begin{equation}
  F[\rho] = \min_{\Psi \rightarrow \rho} \langle \Psi \vert {\hat T} + V_{e,e}
  \vert \Psi \rangle \, ,
  \label{eq:intro.13}
\end{equation}
where ${\hat T}$ is the $N$-electron kinetic energy operator and $V_{e,e}$
is the electron repulsion operator.  The minimum (really the infimum)
is over all normalized N-electron wave functions 
($\langle \Psi \vert \Psi \rangle = 1$) which produce the charge density $\rho$.  Thus $F$ is {\em known} but just too
complicated to be useful in practical calculations.  But it does tell us
what we want to approximate when making DFAs.
Note the distinction between DFT and DFAs:  DFT is
formally exact but useless as it stands.  But it tells us what we are trying
to approximate with our DFAs.   
\begin{quote}
\noindent
``The target is not always placed to be hit, but to serve as a place to 
aim or as a direction.'' --- Joseph Joubert (p.~221 of Ref.~\cite{C1838},
translation my own)
\end{quote}
TFD and HK are the foundation for orbital-free DFT.  But it turns out to be
very hard to find an accurate kinetic energy DFA.  However see Ref.~\cite{MLTP2023} for an update on modern approaches to orbital-free DFT.

% -----------------------------------------------------------------
% \subsubsection{Kohn-Sham Theory}
% -----------------------------------------------------------------
This is why most DFT calculations are done within the KS formulation which
is superficially similar to HF theory (and even more so to pure Hartree
theory).  A fictitious system of $N$ noninteracting electrons (obeying the
Pauli exclusion principle!) is imagined whose ground state has the same
density as the real system of $N$ interacting electrons.  The corresponding
energy expression is then,
\begin{equation}
  E_{KS} = E_H + E_{xc} \, .
  \label{eq:intro.14}
\end{equation}
The xc-energy includes not only the
usual correlation energy but also the difference of the kinetic energies 
of the interacting $T[\rho]$ and noninteracting electrons $T_s$,
\begin{equation}
  E_{xc}[\rho] = F[\rho] - \frac{1}{2} \int \int \frac{\rho(1)\rho(2)}{r_{1,2}}
  \, d1 d2 + T[\rho] - T_s \, ,
  \label{eq:intro.15}
\end{equation}
where $T_s$ is to be calculated from the orthonormal KS orbitals which 
minimize the KS energy.  Minimizing subject to the orthonormal orbital
constraint gives the KS orbital equation,
\begin{equation}
  \left( {\hat h}_H + v_{xc}[\rho] \right) \psi_i(1) = \epsilon_i \psi_i(1) \, ,
  \label{eq:intro.16}
\end{equation}
where the xc-potential,
\begin{equation}
  v_{xc}^\sigma[\rho_\uparrow,\rho_\downarrow]({\vec r})  =  
  \frac{\delta E_{xc}[\rho_\uparrow,\rho_\downarrow]}{\delta \rho_\sigma({\vec r})} \, ,
  \label{eq:intro.17}
\end{equation}
is the functional derivative of the xc-energy with respect to the density.
% (Functional derivatives are discussed in Ref.~\cite{GF1963}.)
This is a relatively small contribution to the total electronic energy,
so that a large percentage error in $E_{xc}$ is still a small error in the
total energy $E_{KS}$.  

% ----------------------
\begin{figure}
\begin{center}
\includegraphics[width=0.6\textwidth]{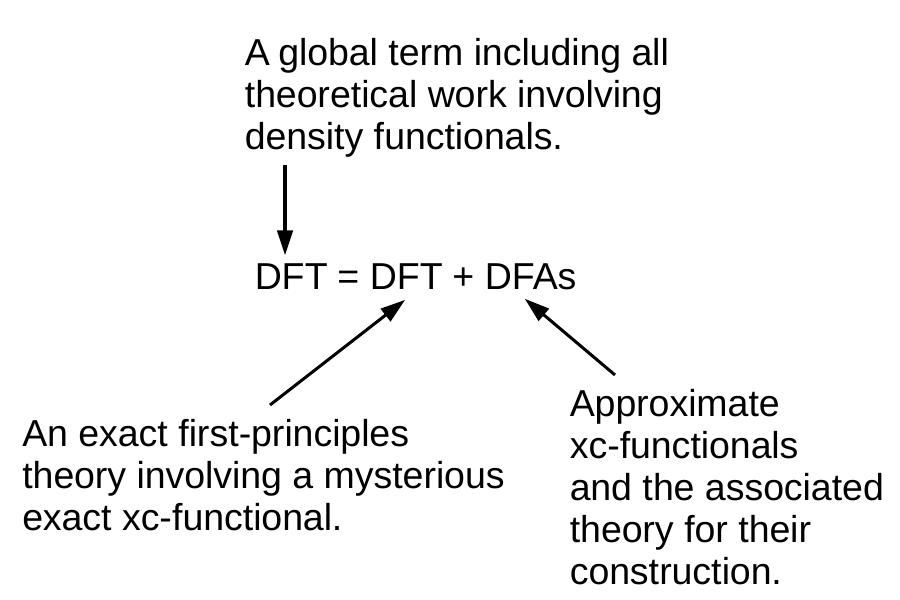}
\end{center}
\caption{
The most confusing equation in DFT.
\label{fig:DFTA}
}
\end{figure}
% ----------------------
{\bf Figure~\ref{fig:DFTA}} shows ``the most confusing equation in DFT.''
If the DFT on the left-hand and right-hand sides were really the same, then
we could just cancel them and find that DFAs = 0.  But they are {\em not} the
same.  The DFT on the right-hand side is formal DFT (and should perhaps be
more correctly called FDFT\index{FDFT}?)  It consists of the
exact first-principles theory which does fit into the definition of {\em 
ab initio}.  It also supposes that we know the exact xc-functional and
can use it in practical calculations.  But we do not have this, so another
part of practical DFT (PDFT\index{PDFT}?) --- the DFT on the left-hand
side --- is the theoretical work needed to make approximations.  As it is
common practice to use DFT for both FDFT and for PDFT, the reader is cautioned
to keep in mind which meaning is intended at any given time.

% -----------------------------------------------------------------
\subsubsection{Hartree-Fock-Slater or X$\alpha$}
% -----------------------------------------------------------------
We still need an expression for $E_{xc}[\rho]$.  The oldest DFA is the local
(spin) density approximation (LDA\index{LDA}).  This can be done for any
type of energy, but we will just focus here on the exchange energy.  The
LDA assumes that the total exchange energy $E_x[\rho]$ of a molecule is
the sum of the number of electrons $\rho(1)\, d1$ in a small volume times
the exchange energy per electron $\epsilon_x^{\mbox{HEG}}(\rho(1))$ of a 
homogeneous electron gas [HEG\index{HEG}, also known as the uniform electron
gas (UEG\index{UEG}) or (informally) as ``jellium''] of the same density,
\begin{equation}
  E_x[\rho] = \int \rho(1) \epsilon_x^{\mbox{HEG}} (\rho(1)) \, d1 \, .
  \label{eq:intro.18}
\end{equation}
The HEG is an infinite system consisting of a fixed electron density $\rho=N/V$
whose charge is neutralized by a uniform positive background.  Dirac had shown that 
\cite{D1929},
\begin{equation}
  \epsilon_x^{\mbox{HEG}} (\rho) = -\frac{3}{4} \left(\frac{3}{\pi}\right)^{1/3}
  \rho^{1/3} \, ,
  \label{eq:intro.19}
\end{equation}
not including spin. Including spin, the exchange LDA is,
\begin{equation}
  E_x[\rho] % = -\frac{3}{4} \left(\frac{6}{\pi}\right)^{1/3} 
            %  \int \rho^{4/3}(1) \, d1
            = -\frac{3}{2} \left(\frac{6}{\pi}\right)^{1/3} \sum_\sigma 
              \int \rho^{4/3}_\sigma({\vec r}) \, d{\vec r} \, .
  \label{eq:intro.20}
\end{equation}
% where the two spin components have been separated for clarity: 
% $\rho(1) = \left( \rho_\uparrow({\vec r}_1) , 
% \rho_\downarrow({\vec r}_2) \right)$.
Then,
\begin{eqnarray}
  E_x & = & E_x^\uparrow + E_x^\downarrow \nonumber \\
  E_x^\sigma & = & -\frac{1}{2} \int \frac{\vert \gamma_\sigma(1,2)\vert^2}{r_{1,2}} \, d1 d2 
  \approx -\frac{3}{2} \left(\frac{6}{\pi}\right)^{1/3} 
              \int \rho^{4/3}_\sigma({\vec r}) \, d{\vec r}
  \label{eq:intro.21}
\end{eqnarray}
As,
\begin{equation}
  E_H = \frac{1}{2} \int v_H(\vec r) \, d{\vec r} \, ,
  \label{eq:intro.21a}
\end{equation}
and,
\begin{equation}
  E_x^\sigma = \frac{1}{2} \int \left( -\int 
  \frac{\vert \gamma({\vec r}_1, {\vec r}_2) \vert^2 }{r_{1,2} \rho({\vec r}_1)}
  \, d{\vec r}_2 \right) \rho_\sigma({\vec r}_1) \, d{\vec r}_1
  = \frac{1}{2} \int v_x^\sigma({\vec r}) \rho_\sigma({\vec r}) \, d{\vec r} \, ,
  \label{eq:intro.21b}
\end{equation}
Slater made the educated guess \cite{S1951,S1972} that the ``Slater potential'',
\begin{equation}
  v_x^\sigma({\vec r}_1) = -\int 
  \frac{\vert \gamma({\vec r}_1, {\vec r}_2) \vert^2 }{r_{1,2} \rho({\vec r}_1)}
  \, d{\vec r}_2  \approx -\frac{3}{2} \left(\frac{6}{\pi}\right)^{1/3} \rho_\sigma^{1/3}(\vec{r}_1) \, .
  \label{eq:intro.21c}
\end{equation}
However this differs from the KS (G\'asp\'ar \cite{G1954}) potential
\begin{equation}
  v_x^\sigma[\rho_\sigma]({\vec r}) = \frac{\delta E_x^\sigma[\rho_\sigma]}{\delta
  \rho_\sigma({\vec r})} = -\left( \frac{6}{\pi} \right)^{1/3} \rho_\sigma^{1/3}({\vec r})
  \, ,
  \label{eq:intro.25}
\end{equation}
by a factor of 2/3.
Finally Slater proposed to compromise \cite{SW1971} and use,
\begin{equation}
 v_x^\sigma[\rho_\sigma]({\vec r}) = -\alpha \frac{3}{2} \left( \frac{6}{\pi} \right)^{1/3} \rho_\sigma^{1/3}({\vec r}) \, .
 \label{eq:intro.26}
\end{equation}
Rather than taking $\alpha = 2/3 = 0.66666$, it was proposed to fit $\alpha$
to give energies as close as HF energies for atoms.  This 
resulted in the method variously known as Hartree-Fock-Slater (HFS\index{HFS})
or X$\alpha$\index{X$\alpha$} with $0.70 < \alpha < 0.77$ \cite{S1972b}.
A more recent proposition is to take $\alpha =  0.7091$ \cite{ZD2004}.
Henry Chermette reminded me of a more important observation that, whereas
$\alpha=0.77$ for the hydrogen atom, $\alpha$ increasingly approaches the ideal
2/3 value as the atom becomes heavier, with $\alpha=0.68$ for bismuth.
Despite obvious similarities between Slater's approach and DFT, ``Prof.\ Slater
did not, I think, view X$\alpha$ as a form of DFT.  He always presented
the X$\alpha$ total energy in the intellectual context of approximate
Hartree-Fock theory'' \cite{T2010}.

Let us now focus more closely on quantum chemists' reactions to DFT in its 
HFS (X$\alpha$) form.  As already mentioned above, science is a society in 
which ideas are constantly being discussed, criticized, defended, etc.  
At the same time, it is also subject to fads.  What I am about to describe 
is as best I can remember what has been passed down to me from my elders 
(mainly Tom Ziegler and John Harriman) regarding the rise and fall of the 
popularity of X$\alpha$ in the quantum chemistry community!
See however Ref.~\cite{C1992} for a technical review of X$\alpha$.

The late 1960s and early 1970s saw the advent of photoelectron spectroscopy
as a tool for the experimental investigation of molecular
orbitals within molecules \cite{TBBB1970}.  Not only did photoelectron
spectroscopy give information
about different electron binding energies (BEs\index{BE}, roughly the negative 
of MO energies according to Koopmans' theorem \cite{K1934}) but it also gave 
some information about the momentum distribution and hence about the character
of the different MOs.  According to Tom Ziegler, one of the first popular
applications of X$\alpha$ was as an efficient and surprisingly accurate way 
to calculate photoelectron spectra.  Interestingly, as such, X$\alpha$ was
being used to extract information about the cation excited states.

\begin{quote}
\noindent
{\sf
During the second half of the 1980s, I was first a postdoctoral fellow
and then a research associate with Delano P.\ Chong at the University of
British Columbia (UBC\index{UBC}).  My initial work involved collaborating
with Chris Brion's experimental group doing electron momentum spectroscopy
(EMS\index{EMS}).  This is conceptually similar to photoelectron spectroscopy 
but photons are replaced with high energy electrons in a binary (e,2e) 
collision experiment.  The detailed theory behind EMS is clearer than 
that behind photoelectron spectroscopy which allows it to provide more 
detailed information regarding ionization cross sections.
My main tool was the theory of one-electron Green's functions which may
be thought of as 1-RDMs but with additional time-dependence.  However,
towards the end of my time at UBC, we proposed the target-KS approximation
as an alternative to the target-HF approximation then often used to analyze
EMS cross sections, thus attributing an approximate physical meaning to KS
orbitals.
--- MEC
}
\end{quote}

Some of the comments in Sture's article \cite{NBB+2026} indicate that 
X$\alpha$ gained popularity in the 1970s quantum chemistry community
because it was faster than Hartree-Fock calculations of the time, could
be applied to larger molecules, and seemed to give excellent results.
Quantum chemists then wanted to see if X$\alpha$ would also be useful for 
calculating other properties and 
this is where they ran into difficulties with the method.  Some properties
were incorrectly predicted by X$\alpha$.  This was also a time of major
advances in {\em ab initio} theory.  Many quantum chemists decided that 
they had ``burned their fingers'' trying out X$\alpha$ theory and did not
ever want to touch it again.  Instead they would focus on the {\em ab initio}
approach.  DFT, at least in its HFS (X$\alpha$) form had become {\em 
persona non grata} for a large part of the quantum chemistry community.

I believe that there were some other things going on that made many
quantum chemists uncomfortable with X$\alpha$.  Besides uncertainties 
about the best value of $\alpha$ to use, there were also questions about 
the numerical quality of X$\alpha$ programs which, in retrospect, may not 
have been the greatest problem with the method \cite{B2026}.  
Certainly a very important problem with the ``muffin-tinning'' of multiple
scattering X$\alpha$ was that the potential $v_s$ was discontinuous.  It
was composed of usually spherically symmetric atomic spheres which could
be easily handled by numerical methods developed for atoms that separated
angular and radial degrees of freedom and interatomic
regions where the wave functions could be handled via plane wave 
expansions \cite{C2009}.  This discontinuous potential, besides being 
physically crude, did not lend itself to the development of analytical 
derivatives \cite{YOGS1994} which were becoming essential in the late 1970s 
as automatic geometry optimizations became routine in quantum chemistry codes.
Furthermore, Sam Trickey mentions in his article \cite{T2010} that 
muffin-tinning actually leads to linear water, instead of the 
familiar correct bent structure.

The muffin-tinning approach was also in stark contrast with {\em ab initio} 
programs using gaussian-type orbital (GTO\index{GTO}) basis sets where 16 
significant 
figure agreement could be expected between programs
written by different authors.  This had turned out to be highly useful in the
early days for identifying and fixing programming bugs.  At the same time,
there were LCAO X$\alpha$
programs that retained some of the characteristics of {\em ab initio}
quantum chemistry codes but which complicated the question of testing
of DFAs because of worries about reaching the basis set limit.

X$\alpha$ also seemed incompatible with the philosophy of the time where
{\em ab initio} quantum chemists, such as Henry F.\ Schaeffer~III, were pushing
for calculations that were as accurate as possible for small molecules where
comparison with experiment could test the limits of accuracy of the best
quantum mechanical models.

\begin{quote}
\noindent
{\sf
Shortly after my arrival in France in 2001, I learned much to my surprise 
that the theoretical solid state physics community had also rejected DFT 
as ``good science.''  Their reason was very different from that of 
{\em ab initio} quantum chemists.  Driven, in part, by the discovery
of high temperature superconductors and the quantum fractional Hall
effect, the solid state physics community sought to understand strong 
correlation by seeking near exact solutions of highly-correlated systems 
described by
semi-empirical hamiltonians.  DFT simply did not fit into this framework.
However by the time of my arrival in Grenoble, the younger generations
of chemists and physicists were both ready to welcome DFT as a new and
quite respectable tool in their fields.
}
\end{quote}

% ----------------------------------------------------------
\subsection{Two Solitudes}
% ----------------------------------------------------------

At the risk of repetition, I think it is important to be clear about
what was happening with DFT during the period between roughly 1970 and 1991.  
While many
quantum chemists tried and then turned away from DFT as if it were some
sort of foul creature, a strong interdisciplinary DFT community, 
composed of solid state physicists (``physicists'') and theoretical
physical chemists/chemical physicists (``chemists''), continued on.  This is a 
bit like the situation described in Hugh MacLennan's 1945 novel {\em 
Two Solitudes} where French- and English-speaking Canadians lived in close
proximity but failed to communicate adequately \cite{M1945}.  

Here, instead of French and English,
we had a strongly united DFT community with its own meetings and computer
programs and what I will call the quantum chemistry community with its 
programs.  Note that quantum chemistry programs often included both {\em
ab initio} and semi-empirical methods, but not the ``horrible'' DFT methods.

Meanwhile the DFT community was developing better numerical methods for
solving the KS equations that did not involve muffin-tinning with some
of the earliest such numerical methods coming from Gunnarsson and co-workers
\cite{GJ1976,GHJ1977a,GHJ1977b}.  Others were developing programs using
Slater-type orbitals (STOs\index{STO}) or GTOs
with or without auxiliary-function techniques 
\cite{EP1970,BER1973,SF1975,DCS1979,DCS1979b}.  The local density
approximation (LDA\index{LDA}) correlation
functional had also been introduced early on, for example in work 
involving the ``jellium 
edge'' (divides space in two with the HEG on one side and empty space on the
other) \cite{LP1977,LP1980}.  Often this LDA correlation functional
was from a random-phase approximation (RPA\index{RPA}) estimate or some other
method developed to explore the many-body physics of the HEG.  After the
Ceperley-Alder quantum Monte Carlo calculations \cite{CA1980}, chemists and
physicists often used different parameterizations of the same Ceperley-Alder
results.  There was also work both to firm up the rigorous basis of DFT 
and continued work developing new and improved DFAs.  

Axel Becke did not 
arrive on the scene as an isolated DFT researcher but as part of a larger
community.  However Axel managed to do something that no one else before
him had managed to do: He broke down the barrier between the DFT and
quantum chemistry communities.  This is part of the story to be told in the
next section.  Also, since the focus of this article is on Axel and not on
DFT in general, I will necessarily omit a great deal of the good things
done by others in the DFT community.
A few books on DFT are Refs.~\cite{PY1989,DG1990,KH2000}.
These books contain information about some of Axel's work up to around the
turn of the century, but are too old to contain information about Axel's most
recent work.
Some more recent review articles on DFT are Refs.~\cite{J2015,THS+2022}.

% =================================================
\section{From Resistance to Acceptance by {\em Ab Initio} Quantum Chemists}
\label{sec:golden}
% \input{golden.tex}
% ===============================================================
% File golden.tex .
% Last modified: 14 July 2026
% ===============================================================

\begin{quote}
\noindent
{\sf
As a doctoral student, I was already a member both of the American Chemical
Society (ACS\index{ACS}) and of the American Physical Society (APS\index{APS}),
memberships that I maintain to this day.  Through their newsletters, I was
becoming aware that something was happening in the quantum chemistry community.
A young man was beginning to make quite an impact.  I did not yet know
that this young man was Axel Becke.  He had done two remarkable things.
The first was to write a basis-set free DFT program for diatomics
with more reliable numerical methods than had previously been the case,
thereby dissipating some concerns that earlier conclusions about X$\alpha$
calculations were due either to problems involving finite basis sets
or due to insufficiently accurate numerical calculations on
grids (i.e., ``muffin tinning'').  The second thing that he was doing
was showing how to improve density functionals to make them more useful.
--- MEC
}
\end{quote}

Many of the most famous contributions of Axel Becke began with his
doctoral studies and continued on through his probable impact on 
the Nobel prize of 1998.

% -------------------------------------------------------
\subsection{The Road to {\sc NUMOL}}
% -------------------------------------------------------

\begin{quote}
\noindent
``The theory of quantum mechanics also explained all kinds of details, 
such as why an oxygen atom combines with two hydrogen atoms to make water, 
and so on. Quantum mechanics thus supplied the theory behind chemistry. 
So, fundamental theoretical chemistry is really physics.'' --- 
Richard P.\ Feynman \cite{F1985}
\end{quote}
Many of the contemporaries of Axel who ended up in theoretical chemistry
had a strong interest in physics.
As an undergraduate student doing research in theoretical chemistry, I
often wondered what was the difference between Physical Chemistry and
Chemical Physics and asked my mentor, Henry F.\ Schaefer III,
if I were a Physical Chemist or a Chemical Physicist?  His answer was
that I was definitely a Chemical Physicist, which puzzled me for a long
time.  I found the solution to this puzzle many years later in a beautiful 
historical article by Rowlinson who explained that the mathematics of 
quantum mechanics was long considered to be too hard for physical chemists 
and so was done in physics departments \cite{R2009}.  This pretty much ended 
during the 1970s when more and more chemistry departments hired quantum 
chemists so that now days there is no real distinction between Physical 
Chemistry and Chemical Physics.  This also means that the continued
healthy development of Physical Chemistry/Chemical Physics will likely
require maintaining a healthy grounding in physics, mathematics, and
computer science for chemistry students for some time to come.

Axel Becke certainly started his University studies in physics, not chemistry.
Axel obtained his B.Sc.\ in Engineering Physics from Queen's University,
Kingston, Ontario, Canada, in 1975.  He wrote that,
\begin{quote}
\noindent
``My association with Queen's began in the years 1971-1975, as an 
Engineering Physics student in the Faculty of Applied Science.  [...]
to be honest, chemistry was not among my favorite courses.  I was determined
to be an engineer some day.'' \cite{Queens}
\end{quote}
Sports and {\sc FORTRAN} coding seem to have been highlights of this
period for Axel:
\begin{quote}
\noindent
``I was the studious type, but managed to find time to train and travel
with the track team. [...] The rest of my spare time was spent in the basement
of Dupuis, writing {\sc FORTRAN} code, punching card decks, ripping printouts
off the Dupuis printer.  I was a computer nerd, writing programs for anything
that seemed amusing, from solving linear systems to integrating the trajectories
of artillery shells.'' \cite{Queens}
\end{quote}
He published an application to solid state semiconductor 
physics [{\bf SKB1975}]. 

Axel sought advice regarding his Masters studies:
\begin{quote}
\noindent
``Nuclear physics was an exciting frontier at the time, and Boris
Castel suggested that I contact Donald Sprung at McMaster as a 
potential supervisor.  His advice was taken, and with an [National 
Research Council of Canada] NRC\index{NRC}  1967 
Science Scholarship in hand, I headed for McMaster in the fall of 1975.''
\cite{Queens}
\end{quote}
Axel obtained his M.Sc.\ degree from McMaster University, Hamilton, Ontario,
Canada, in Theoretical Physics in 1977.  His Masters thesis was done under
D.W.L.\ Sprung and was entitled, ``Eikonal Distorted Wave Approximation for 
High Energy Electron Scattering from Spherical Nuclei''.  His two publications
from this period [{\bf BS1977}, {\bf KBBS1979}] concern the least squares fitting
problem in the context of nuclear physics.

Axel's defended his Ph.D.\ thesis in Theoretical Physics in July of
1981 at McMaster University, Hamilton, Ontario, Canada, under the
direction of D.W.L.\ Sprung.  The title was ``Numerical Hartree-Fock-Slater
Calculations on Diatomic Molecules''.  The acknowledgement of Axel's
thesis thanks,
\begin{quote}
\noindent
``[...] my superviser, Dr.\ D.W.L.\ Sprung, for his assistance and support 
of this research, and also members of my supervisory committee, Drs.\ R.K.\
Bhaduri, R.F.W.\ Bader, and D.P.\ Santry.  Thanks also to Dr.\ Tom Ziegler
for particularly helpful discussions.'' \cite{B1981}
\end{quote}
Doctoral students may find it interesting that Axel's thesis was sent out to 
an external examiner and he was asked to rework the thesis before defending:
\begin{quote}
``My Ph.D.\ defense was suspended by the external examiner for many months,
due to insufficient referencing in the thesis and too few applications of
my new benchmarking tool.  Though the setback was a disappointment, the 
external examiner was right.

The thesis was improved significantly and, for the first time, accurate 
and reliable Xalpha spectroscopic properties of diatomic molecules were available
and published [{\bf B1982}, {\bf 1983b}].'' \cite{Queens}
\end{quote}
It is important to keep in mind that this was a period were most quantum
chemists had abandonned HFS (X$\alpha$) but where Axel, with his background
in numerical work in nuclear physics thought he could make a contribution.
The methods that Axel used were described in the abstract of the thesis:
\begin{quote}
\noindent
``A completely numerical method has been developed for the calculation
of Hartree-Fock-Slater wave functions in diatomic systems.  The method is 
numerical in the sense that no 
LCAO basis sets are employed.  All molecular functions are
represented by cubic spline interpolants on a two-dimensional discrete mesh
in prolate spheroidal coordinates.  The method is mathematically simple,
and numerical accuracy is very easily controlled by adjusting the number
of mesh points.  Futhermore, it is easly applied to other local 
exchange-correlation theories beyond the Hartree-Fock-Slater approximation.''
\cite{Queens}
\end{quote}
Many theoreticians will recognize this as the usual finite element method
(FEM\index{FEM}) used by engineers to study bridge mechanics or in by 
fluid dynamicists to study flow patterns. 
The last sentence reminds us that the Vosko-Wilke-Nusair (VWN\index{VWN})
parameterization of the Ceperley-Alder quantum Monte Carlo results for the
correlation energy of the electron gas had only just appeared in 1980
\cite{CA1980,VWN1980}.  But I think it best to let Axel express the situation
himself:
\begin{quote}
\noindent
``As an NSERC
1967 Science Scholar, my supervisor granted me considerable leeway to set
my own course and explore my own options.  Slater's book on electronic
structure held ever increasing appeal.  In the 1970s, he and K.H.\ Johnson
enthusiastically promoted a new computational approach to molecular
structure called the `Xalpha' method.  An antecedent of what is today
known as Density-Functional Theory (DFT), Xalpha offered tremendous 
economization over the Hartree-Fock LCAO technology which was the mainstay
of computational chemistry.  Evolved from the so-called `muffin-tin'
approximation of solid-state physics, the approach of Slater and
Johnson was elegant and numerically beautiful to my mind.  I was hooked
on the problem of molecular structure methodology.

As the literature on the Xalpha method grew, it became apparent that the
muffin-tin aspects of the Slater and Johnson approach were far too crude
for molecules.  The numerical methods of the quantum chemistry on spectroscopic
properties such as bond lengths, bond energies, and 
vibrational frequencies, simply could not be met by `muffin-tinning' the
molecular potential.  The underlying theory, however, which had been proposed
by Slater in the early 1950s, remained of interest.  Improved implementations
of the X$\alpha$ theory by others, using LCAO instead of muffin-tin 
strategies, began to appear in the late 1970s.

They delivered tantalizing molecular properties, superior to those of 
Hartree-Fock theory and at a much reduced cost.'' \cite{Queens}
\end{quote}
Examples of the LCAO X$\alpha$ schemes mentioned by Axel are given in 
Refs.~\cite{BER1973,SF1975,DCS1979b}.  As discussed in the 
Sec.~\ref{sec:before}, there were multiple concerns about X$\alpha$.
Some came from the ``muffin-tinning'' --- a sort of crude numerical integration
scheme.  Some came from the use of LCAO basis sets and auxiliary basis sets
used to describe the xc-potential via a
least squares fitting procedure over a grid.  And, I suspect, some came from
a sort of desire to push LCAO calculations in {\em ab initio} chemistry to
16 significant figure accuracy that could be reproduced anywhere on any
machine allowing double precision calculations.

A Ph.D.\ can also be a time to build research connections.  In this context,
it is interesting that Axel wrote:
\begin{quote}
\noindent
Tom Ziegler, a computational inorganic chemist, paused at McMaster as an
NSERC Postdoctoral Fellow on his way to the University of Calgary (where
he is today).  I happened to meet Tom Ziegler at an informal seminar in
the Senior Sciences building given by his good friend Evert Jan Baerends,
from Amsterdam.  Both were prominent researchers in the field of Xalpha
computational chemistry.  Baerends had created the first of the LCAO Xalpha
programs, now known as the Amsterdam Density-Functional program 
({\sc ADF}\index{ADF}), and Ziegler was a major user and collaborator.
I chatted with them about my diatomic Xalpha project and had periodic 
conversations with Tom Ziegler thereafter.'' \cite{Queens}
\end{quote}
Evert Jan Baerends still recalls this meeting with Axel:
\begin{quote}
\noindent
``In 1981 I gave a talk at McMaster university on the
subject of X$\alpha$ (also called Hartree-Fock-Slater, HFS) calculations. 
I vividly remember the piercing dark eyes of the graduate student who came 
to me after my talk telling me he felt it was worthwhile to concentrate on 
numerically solving the Hartree- Fock-Slater equations. This was at a time 
when X$\alpha$ (or LDA) was thoroughly impopular in the quantum chemistry 
community. The student was Axel Becke.''
\cite{NBB+2026}
% \cite{B2026}
\end{quote}
Note that Tom Ziegler passed away in 2015.

At this point, Axel had already accomplished something very important: 
He had established the true limits of X$\alpha$, in (``the infinite basis 
set limit'' of) a completely numerical program. 
% [At this point, I should
% mention that the division between a numerical solution over a grid and
% a basis set solution is neither simple nor really clean. For example, 
% it is normal in the finite element method (FEM\index{FEM}) to introduce 
% interpolation functions between grid points and these interpolation points 
% may be seen as basis functions. Conversely some algorithms for the rapid 
% evaluation of integrals involve evaluating the functions to be integrated 
% over just a few key grid points.  However, it is 
% often easier to obtain convergence numerically over a grid than by doing 
% all integrals analytically and expanding the underlying basis set.  The 
% catch, though, is that numerical methods on a grid almost invariably 
% involve a certain amount of numerical noise which requires careful analysis 
% to reduce and this analysis is usually done in the context of some 
% assumptions about what types of (basis) functions are to be integrated. 
% Notwithstanding these remarks, we will continue to call numerical calculations
% over a grid ``basis-set free'' or ``fully numerical.'']
I will skip what was learned here
so that we may continue concentrating on Axel's algorithic contributions.
Axel had moved to the faculty of Queen's University in Kingston, Ontario,
Canada where he had a doctoral student, Ross Dickson, during the period
1986-1992.  Ross's thesis title was ``Tests of a Basis-Set-Free
Approach to Local Density Functional Calculations of the Structures of 
Polyatomic Molecules'' \cite{D1992}.  Ross also continued on for 
another year as a postdoctoral fellow/research associate with Axel. 
Later Ross would return as a postdoctoral fellow/research associate
with Axel % in Dalhousie University, Nova Scotia, Canada 
during the period 2003-2006.  However the next part concerns only the 
earlier period
which helped transform Axel's fully numerical program for diatomics
into a fully numerical program for polyatomics.  Ross recalls about
Axel that,
\begin{quote}
\noindent
``He was very patient and generous with me, his first grad 
student.  We had a relaxed and good-humored relationship for the rest 
of his life, but never what one would call `close'--- He was (in my 
experience) quite a private man who liked working alone.  One can see 
traces of this in his publication history where his most prominent 
papers are all single-authored.'' (personal communication)
\end{quote}
Let me back that up with my own experience and what I have heard from
other friends and colleagues:  Axel was in no way stand-offish nor
put-offish.   He was friendly and always happy to engage in conversation,
at least with people he knew.  But he also had a very private side
and this extended to his work on DFT where he liked to work alone on
many things and did not like to share ideas. 
Remarkably, of his
95 publications, fully 46\% (44 publications) are single author papers.
Another indication of this aspect of Axel's personality to me from 
Andreas Savin who, together with Merkle and Preuss, had obtained a copy
of {\sc NUMOL} done some work.  Axel was invited to be a co-author as
he had contributed the {\sc NUMOL} program, but Axel did not feel that
this contribution was adequate justification for a co-authorship.  In the
end, the article \cite{MSP1992} was published without Axel as co-author.

One of Axel's most important numerical contributions was the Becke grid
[{\bf 1988b}] which allows accurate numerical integration over an entire 
polyatomic molecule without ``muffin-tinning''.  The idea is simple and
elegant.  Rather than dividing molecules up in to spherical atoms connected
by flat areas (rather like a muffin tin), divide the molecule up into
polyhedral volumes $V_i$ separated by planes.  But let the volumes 
interpenetrate by defining a function $f_i$ which is unity inside the 
polyhedron $\vec{r} \in V_i$, zero well outside the polyhedron $\vec{r} \notin V_i$, but between zero and one on the borders between
polyhedra $\vec{r} \in V_i \cap V_j$ so as to create ``fuzzy'' slightly 
overlapping polyhedra satisfying,
\begin{equation}
  \sum_i f_i({\vec r}) = 1 \, ,
  \label{eq:axel.1}
\end{equation}
at every point in space.
It follows that any integral over all space can be re-expressed as an 
integral over fuzzy atoms,
\begin{equation}
  \int g(\vec{r}) \, d\vec{r}  =  \sum_i \int f_i(\vec{r}) g(\vec{r}) \, 
  d\vec{r} 
 = \sum_i \int_{\vec{r} \in V_i} f_i(\vec{r}) g(\vec{r}) \, d\vec{r} \, .
  \label{eq:axel.2}
\end{equation}
Note that the final integrals 
are formally over all space, but 
in practice are limited to where the function $f_i(\vec{r})$ is nonzero
which is only in and around the atomic volume $V_i$.  Becke's numerical 
integration is universally used in nearly all quantum chemistry programs.  

This brings me to a point which
may be of some interest.  The DFT community was a fairly small and united
community, separate from the {\em ab initio} (and semi-empirical) communities
of the time.  Science was sometimes propagated not just by publications but
literally by exchanging {\sc FORTRAN} code.  For example, it is my understanding
that the Lebedev grid used in the {\sc deMon} codes for the Becke integration
came directly from Axel himself.  I can think of other examples, but they have
no place here as they did not involve Axel.  Further improvement over a
polyatomic grid was reported in [{\bf PBS1994}].

This and other improvements [{\bf BD1988}] led to the completion of a fully 
numerical program for polyatomics called {\sc NUMOL} for NUmerical MOLecules 
[{\bf B1989}, {\bf BD1990}].  It seems appropriate here to interject a personal
memory where Axel was very helpful.  During my Vancouver years, we had done
some very careful work using two different LCAO DFT programs ({\sc DMol} and 
{\sc deMon}) and found that DFT gave remarkably good hyperpolarizabilities
for the molecules that we tested \cite{GDC+1993}.  In a way, this is suspicious 
because vibrational effects are now known to be important for calculating
hyperpolarizabilities, so it is not so strange that our work was questioned
\cite{CMHA1993} but the claim that we had not pushed the basis set limit
hard enough was disturbing as we had been very careful in our work.  Three
years later Dickson and Becke came to the rescue with {\sc NUMOL} [{\bf DB1996}].
Their work vindicated the correctness of our earlier calculations:
\begin{quote}
\noindent
``Higher polarizabilities $\beta$, and $\gamma$, though, clearly require larger
basis sets. We have therefore calculated reference values for
these properties in the local spin-density approximation. We
observe surprising agreement with experiment for $\beta$, using only
the simple LDA, which must be considered fortuitous until
further investigations into vibrational effects are carried out.'' 
[{\bf DB1996}]
\end{quote}
Once again, this proved the value of basis set free calculations.
The closeness to the experimental results is most likely due to error 
cancelling effects in these molecules.  However our calculations did give 
close to the basis set limit for the molecules studied.

It seems appropriate to add something here about Axel's attitude towards
programming and program distribution.  As we have seen, Axel started with
punched cards.  But I remember him best from the era when he did most (if not
all) of his work on a single workstation on his desk.  I would say that he
liked programming and algorithms but like the traditional physicist who sees
the computer as a tool for testing out ideas.  This may be contrasted with
the present era of big programming projects and questions of open-source
software.  The question of open-source software is quite controversial, at
least from the point of view of major European grants.  On the one hand,
the granting agency wishes to see the production of a product which is
distributed --- hence open source publishing and open source code is preferred.
On the other hand, the granting agency wishes to see collaboration with 
industry who, as a rule, wish to protect their interests by {\em not} sharing
key programs that help them win against the competition.
In reaction to an article called ``Open Source and Open Data Should Be Standard Practice'' \cite{G2015} (see also \cite{KHF+2015}), Axel wrote an answer 
called ``Open Source? Or Open Season'' \cite{openseason}.  I received a 
private copy from Filipp Furche with the understanding that the document was 
intended by Axel to be private and not for publication, which I respect.  
Suffice it to say that Axel was not a fan of {\em mandatory}
open-source programs and felt that time spent maintaining code for 
public distribution was not always in the interests of doing science.  
This might also help to explain why {\sc NUMOL} is not better known.
Filipp also wrote to me about Axel:
\begin{quote}
\noindent
``I also think he was concerned about `citizen scientists', self-appointed 
experts, and administrators encroaching onto scientific freedom of expression 
and our ability to decide what happens to our work.''
\end{quote}

% -------------------------------------------
\subsection{Local Density Approximation}
% -------------------------------------------

In the mid-1980s, Axel used his basis-set free numerical program to evaluate the quality of
state-of-the-art DFT [{\bf B1986c}], namely the local (spin) 
density approximation (LDA\index{LDA}) using the VWN \cite{VWN1980} 
parameterization of
the Ceperley-Alder quantum Monte Carlo results for the HEG.  Kohn and Sham 
had been pessimistic regarding the value of their LDA for chemistry:
\begin{quote}
\noindent
``We do not expect an accurate description of chemical bonding.'' \cite{KS1965}
\end{quote}
Axel's results provided important confirmation that the LDA worked 
for molecular geometries and vibrational frequencies:
\begin{quote}
\noindent
``In general, the LDA bond lengths and vibrational frequencies are 
remarkably good. The rms [i.e., root mean square] deviation from
experiment for the 13 molecules in Table I is only 0.05
bohr for the bond lengths and 80 cm$^{-1}$ for the vibrational
frequencies. On the other hand, the LDA dissociation energies 
tend to be much too large, with an rms deviation
from experiment of 1.2 eV.'' [{\bf 1986c}]
\end{quote}
1 eV is 96.5 kJ/mol.  Given that the C-C bond energy
in ethane (H$_3$C-CH$_3$) is about 368 kJ/mol, the LDA overbinding error
is clearly too large to study chemical reactivity, even though geometries and
vibrational frequencies are quite reasonable.
\begin{quote}
\noindent
``Having established numerically reliable LDA spectroscopic 
properties for first- and second-row molecules, our challenge now 
is to improve the LDA results by considering nonlocal corrections such 
as that of Langreth and Mehl.'' [{\bf B1986c}]
\end{quote}
Axel further reported on the performance of basis-free LDA calculations for
molecules in [{\bf DB1993}] and [{\bf B1989b}].

% -------------------------------------------
\subsection{Generalized Gradient Approximations}
% -------------------------------------------

The xc-energy is traditionally written as,
\begin{equation}
  E_{xc}[\rho] = \int \epsilon_{xc}[\rho](1) \rho(1) \, d1 \, ,
  \label{eq:GGA.1}
\end{equation}
where $\epsilon_{xc}[\rho](1)$ is the xc energy per electron.  [Note that 
$\epsilon_{xc}[\rho](1)$ is not unique as several different choices can
give the same $E_{xc}[\rho]$ (i.e., there is a gauge problem).]  Kohn and Sham had considered not just the LDA
but also a semi-local functional that would also depend upon the derivatives
of the density at any given point.  In one dimension, this means that the
functional derivative of 
\begin{equation}
  f[\rho](x) = f(\rho(x),\rho^{\prime}(x),\rho^{\prime \prime}(x), \cdots) \, ,
  \label{eq:GGA.2}
\end{equation}
is, after integration by parts,
\begin{equation}
   \frac{\delta f(x)}{\delta \rho(y)} 
    = \frac{\partial f(x)}{\partial \rho(y)}
    - \frac{\partial}{\partial y} \frac{\partial f}{\partial \rho^\prime(y)}
    + \frac{\partial^2}{\partial y^2} \frac{\partial f}{\partial \rho^{\prime \prime}(y)}
    + \mbox{HOT} \, ,
    \label{eq:GGA.3}
\end{equation}
where ``HOT'' stands for ``higher-order terms'' \cite{GF1963}.  
[The negative sign in Eq.~(\ref{eq:GGA.3}) is due to the integration by
parts.] Kohn and Sham concluded
that the next gradient correction should be \cite{KS1965}
\begin{equation}
  \epsilon_{xc}[\rho](\vec r) = \epsilon^{(0)}_{xc}(\rho(\vec r)) 
  + \epsilon_{xc}^{(2)}(\rho({\vec r})) \vert {\vec \nabla} \rho(\vec r) \vert^2
  + \mbox{HOT} \, .
  \label{eq:GGA.4}
\end{equation}
Here $\epsilon^{(0)}_{xc}$ is the LDA while the rest is the gradient
expansion approximation (GEA\index{GEA}).  In the exchange-only case,
Eq.~(\ref{eq:GGA.4}) may be rewritten to lowest order in the 
X$\alpha\beta$\index{X$\alpha\beta$} form,
\begin{equation}
  \epsilon_x^\sigma[\rho_\sigma](\vec r) = 
  -\frac{3}{4} \alpha \left( \frac{6}{\pi} \right)^{1/3} 
  \rho_\sigma^{1/3}(\vec r)
  - \beta \left( x_\sigma(\vec r) \right)^2 \rho_\sigma^{1/3}(\vec r) \, ,
  \label{eq:GGA.5}
\end{equation}
where,
\begin{equation}
   x_\sigma(\vec r) = \frac{\vert {\vec \nabla} \rho_\sigma (\vec r) \vert}{\rho_\sigma^{4/3} }
  \, ,
  \label{eq:GGA.6}
\end{equation}
is the reduced gradient.  Up to a constant, the reduced gradient
\begin{equation}
   x_\sigma \propto \frac{\vert {\vec \nabla} \rho_\sigma \vert/\rho_\sigma}{k_F^\sigma} =\frac{s_\sigma}{2} \, ,
  \label{eq:GGA.7}
\end{equation}
% is unitless.  Here 
where the local Fermi wave number is,
\begin{equation}
  k_F^\sigma = \left( 3\pi^2 \rho_\sigma \right)^{1/3} \, ,
  \label{eq:GGA.8}
\end{equation}
and $s_\sigma$ is different definition of the reduced gradient which is
also used in the DFT literature.
This means that one way to interpret the X$\alpha\beta$ form is that 
the GEA will converge as long as $\vert {\vec \nabla} \rho_\sigma \vert/\rho_\sigma$
varies slowly in comparison to the Fermi wave number of the HEG of density $\rho_\sigma$.
Interestingly this works reasonably well inside an atom but not on the ``surface''
of the atom \cite{PG1985}.

Axel began by exploring the X$\alpha\beta$ approximation \cite{HVO1969}.  
In [{\bf B1983}], he derived 
a theoretical value of $\beta = 0.00293$ which agreed well with the values of 0.0022-0.0034
calculated for atoms.  In [{\bf B1985}] the X$\alpha\beta$ exchange functional is combined
with the LDA correlation functional to obtain improved molecular dissociation energies.
% Of particular interest is appearance of two important concepts: (1) the two-body  probability
% density which is related to the xc-hole and (2) the kinetic energy density
% \begin{equation}
%   \tau_\sigma({\vec r}) = \sum_i n_{i\sigma} \vert \psi_{i\sigma} ({\vec r}) \vert^2 \, .
%   \label{eq:GGA.9}
% \end{equation}

It was beginning to be understood that the GEA diverged, meaning that some
type of resummation would be needed.  The resultant functionals are called
generalized gradient approximations (GGAs\index{GGA}) and date back to the
early 1980s \cite{LP1980,LM1981,PY1986}.  In [{\bf B1986}]
proposes a particular GGA functional:
\begin{equation}
  E_x = E_x^{\mbox{LDA}} - \beta \sum_\sigma \int 
   \frac{\vert {\vec \nabla} \rho_\sigma \vert^2}{\rho_\sigma^{4/3}}
  \left[ 1 + \gamma \frac{\vert {\vec \nabla} \rho_\sigma \vert^2}
  {\rho_\sigma^{8/3}} \right]^{-4/5} \, d{\vec r} \, .
  \label{eq:GGA.10}
\end{equation}
Two points stand out for me in this article.  

The first is the explicit discussion of the use of the 
xc-hole \cite{LP1975,GL1976,LP1977} in the derivation of a functional.  
The xc-hole is most properly defined in terms of the adiabatic connection 
formalism (ACF\index{ACF}).  A parameter $0 \leq \lambda \leq 1$ is 
introduced in the hamiltonian,
\begin{equation}
  \hat{H} = {\hat H}_s + \lambda V_{ee} + W_\lambda[\rho] \, ,
  \label{eq:GGA.11}
\end{equation}
where ${\hat H}_s = \sum_{i=1,N} {\hat h}_s$ is the KS hamiltonian [``s''
stands for ``single particle'' (Walter Kohn, personal communcation)], 
$V_{ee} = \sum_{i,j=1,N}^{i<j} 1/r_{i,j}$
is the usual electron repulsion term, and $W_\lambda[\rho]$ is a restoring
potential designed to keep the ground state density equal to $\rho$ for
all values of $\lambda$.  Note that the density was not kept fixed
in the earliest form of the ACF \cite{HJ1974,JS1974} but that the ACF was 
soon modified \cite{LP1975,GL1976,LP1977} to include the restoring potential.
Varying $\lambda$ takes the hamiltonian from that
of the fictious system of noninteracting electrons ($\lambda=0$) 
to the hamiltonian of the real system of interacting electrons ($\lambda=1$).
The xc-energy,
\begin{equation}
  E_{xc}[\rho] = \frac{1}{2} \int \int \frac{\rho(1) \rho_{xc}(1,2)}{r_{1,2}} \,
  d1 d2 \, ,
  \label{eq:GGA.12}
\end{equation}
where the xc-hole,
\begin{equation}
  \rho_{xc}(1,2) = \frac{\int_0^1 \langle \Psi_\lambda[\rho] \vert 
  {\tilde \rho}(1) {\tilde \rho}(2) \vert \Psi_\lambda[\rho] \rangle 
  \, d\lambda}{\rho(1)}
  - \delta(1-2) \, ,
  \label{eq:GGA.13}
\end{equation}
and,
\begin{equation}
  {\tilde \rho}(1) = {\hat \rho}(1) - \rho(1) \, ,
  \label{eq:GGA.14}
\end{equation}
where ${\hat \rho}(1)$ is the second-quantized density-operator and
$\rho(1) = \langle \Psi \vert {\hat \rho}(1) \vert \Psi \rangle$ is the
density.
Ignoring the $\lambda$-dependence of $\Psi$ gives,
\begin{equation}
  \rho_{xc}(1,2) = \frac{\Gamma(1,2;1,2)}{\rho(1)} - \rho(2) \, ,
  \label{eq:GGA.15}
\end{equation}
where $\Gamma(1,2;1',2')$ is the 2-electron reduced density matrix
[normalized to $N(N-1)$].
In the HF limit,
\begin{equation}
  \rho_{xc}(1,2) = -\frac{\gamma(1,2)\gamma(2,1)}{\rho(1)} \, .
  \label{eq:GGA.16}
\end{equation}
Note that only the spherical average of the xc-hole,
\begin{equation}
   \rho^{xc}_{0,0}({\vec r}, \Delta r)  =  
   \frac{1}{4\pi} \int \rho_{xc}({\vec r},{\vec r}^{\, \prime}) \, d\Omega
   \nonumber \\
   \label{eq:GGA.17a}
\end{equation}
enters into the energy calculation,
\begin{equation}
   \epsilon_{xc}({\vec r})  =  \sqrt{4}{\pi} \int_0^\infty 
   \rho^{xc}_{0,0}({\vec r}, y) y \, dy \, .
   \label{eq:GGA.17b}
\end{equation}
It ``contains'' (i.e., excludes) exactly one electron,
\begin{equation}
  \sqrt{4\pi} \int_0^\infty \rho^{xc}_{0,0}({\vec r}, y) y^2
  \, dy = -1 \, ,
  \label{eq:GGA.18}
\end{equation}
and often is of similar appearance in molecules as in the HEG.
This has much to do with the success of DFT.  Moreover it is probably not 
too much of an overstatement to say that what differentiates DFT from 
wave function theory (WFT\index{WFT}) is that much of the work done
on creating DFAs focuses on understanding the xc-hole and engineering 
improved approximations for this object.

A second point that I find remarkable in this article is the development
of a series approximation for the local kinetic energy,
\begin{equation}
  \tau_\sigma = \frac{3}{5} \left( 6 \pi^2 \right)^{2/3} \rho_\sigma^{5/3}
  +\frac{1}{3} \nabla^2 \rho_\sigma + \frac{1}{36} \frac{\vert {\vec \nabla} \rho_\sigma \vert^2}{\rho_\sigma} + \mbox{HOT} \, .
  \label{eq:GGA.19}
\end{equation}
Axel claims no priority for this well known result.  But DFAs for, or directly
involving, the kinetic energy density come up so frequently in Axel's work 
that I felt it deserved some mention here.

Axel's work on gradient corrected functionals attracted the attention of 
Tom Ziegler.  But let us read Axel's own words:
\begin{quote}
\noindent
``Shortly after my arrival at Queen's, the investigations of gradient
corrections to Xalpha theory began to pay off.  Tom Ziegler quickly
implemented my (yet unpublished) gradient-corrected theory into the
Amsterdam Density-Functional ({\sc ADF}\index{ADF}) Program and commenced
wide testing on molecules not amenable to my diatomic program.  The results
were very encouraging.  His primary research interest was organometallic
chemistry, for which traditional Hartree-Fock-based computational
methods were out of reach.  He applied the new density functionals to
challenging and practical problems in organometallic chemistry, and scored
impressive successes.

[...]

I derived tremendous satisfaction and encouragement from Ziegler's work,
and I resolved to refine and improve gradient corrections as far as 
possible.'' \cite{Queens}
\end{quote}
I note the following publications with Tom Ziegler: [{\bf TTB1987}, 
{\bf ZTB1989}, {\bf ZTFB1989}].
  
% Axel along with Roussel went on to study gradient corrected
% functionals for correlation {\bf B1988c} as well as for exchange
% {\bf BR1989}.  We note that the latter case is an example of meta GGA
% (mGGA\index{mGGA}) --- that is, a GGA with an additional explicit dependence
% on the local kinetic energy, $\tau$.  In {\bf B2009} shows how the use
% of a mGGA can allow inclusion of relativistic effects in a gauge-invariant 
% manner.

We now arrive at the very famous B88 functional published in 
\framebox{\bf B1988}.  I have placed a box around the reference because 
this article, together with one other of Axel's articles (boxed below), 
has been cited an extraordinary number of times (51~499 times as of 
14 May 2026 according to the {\em Web of Science}).  I should make it clear 
that Axel was not the only one developing DFAs at this
time.  Two other people who I admire immensely from these early
years are Mel Levy and John Perdew (to give but two names among many).  
Levy and Perdew's strategy is to try to find as
many {\em exact conditions} that must be satisfied by the exact density
functional as possible, and then to build an {\em ab initio} functional 
which would satisfy as many of these conditions as they could 
in the hope that DFAs built
for solids would work for atoms and molecules and vice versa.  Axel's 
B88 functional did not fall into this category as it was only based
upon reasonable physical arguments and any needed parameters were just
fit.  This x-only functional,
\begin{equation}
  E_x = E_x^{\mbox{LDA}} - \beta \sum_\sigma \rho_\sigma^{4/3}
  \frac{x_\sigma^2}{1 + 6 \beta x_\sigma \mbox{sinh}^{-1} x_\sigma }
  \, d{\vec r}
  \, ,
  \label{eq:GGA.20}
\end{equation}
involves the previously mentioned reduced gradient [Eq.~(\ref{eq:GGA.7})].
This functional has 
the property that it becomes the X$\alpha\beta$ functional when
$x_\sigma \ll 1$ but that $\epsilon_{xc}$ approaches the expected limit 
of $-1/(2r)$ at large distances ($r$) from a molecule for densities that fall 
off exponentially (as they do in molecules).
The article contains a disclaimer regarding their value for $\beta$ which,
arguably, is inappropriate for periodic systems:
\begin{quote}
\noindent
``As our interests are restricted to atomic
and molecular applications only, we make no attempt to
incorporate the Sham value of $\beta$ into our functionals as
other authors have done.'' [{\bf B1988}]
\end{quote}
% Axel's work on GGAs continued on with several other articles 
% {\bf B1992b} and {\bf B1992}.

\begin{quote}
\noindent
{\sf
My final time at the University of British Columbia (UBC) around 1990 involved 
working with two DFT programs ({\sc DMol} and
{\sc deMon}) to evaluate the ability of DFT to calculate hyperpolarizabilities
and as a tool for calculating EMS cross sections.  There is a now famous quote:
\begin{quote}
\noindent
``All models are wrong, but some are useful.'' ---
George E.P.\ Box \cite{N1996}
\end{quote}
Finally I saw the usefulness of DFT for practical calculations!
But there were still two separate communities in quantum chemistry:
the {\em ab initio} community refused to trust and hence to touch DFT
while a parallel DFT community developed its own quantum chemistry programs.
--- MEC
}
\end{quote}

% -------------------------------------------
\subsection{Hybrid Functionals}
% -------------------------------------------

1993 was a remarkable year for DFT.  Axel seemed to have concluded that
GGAs had hit a brick wall, at least as far as thermochemistry was concerned.
He returned to the adiabatic connection formalism which he chose to write as,
\begin{equation}
  E_{xc} = \int_0^1 U^\lambda_{xc} \, d\lambda \, .
  \label{eq:hybrid.1}
\end{equation}
Comparing with Eq.~(\ref{eq:GGA.13}), we see that
\begin{equation}
  U^\lambda_{xc} = \frac{1}{2} \left[ 
  \int \int \langle \Psi_\lambda[\rho] \vert 
  {\tilde \rho}(1) {\tilde \rho}(2) \vert \Psi_\lambda[\rho] \rangle
  \, d1 d2 - N \right] \, .
  \label{eq:hybrid.2}
\end{equation}
At $\lambda=0$, this is just the HF exchange energy evaluated with the 
Slater determinant of Kohn-Sham orbitals.  Axel initially proposed his
half-and-half functional [{\bf B1993b}],
\begin{equation}
  E_{xc} = \frac{1}{2} \left( E_x^{\mbox{HF}} + E_{xc}^{\mbox{GGA}} \right) \, ,
  \label{eq:hybrid.3}
\end{equation}
in a short talk given in 1992 at a Canadian theoretical chemistry conference
in Montreal (Emil Proynov, personal communication).
Of course, this is just a guess as there is no particular reason to chose
a 50/50 mixing and this mixing percentage was nothing
more than a first crude guess.  (A more detailed analysis of hybrid
methods is given in Ref.~\cite{GL1997}.)
More importantly, Eq.~(\ref{eq:hybrid.3}) implies an orbital-dependence 
beyond just a density-dependence, so it is
no longer pure DFT.  However, Axel did specify KS orbitals and KS orbitals
are normally associated with a multiplicative single particle potential.
In fact, a private conversation that I had with Axel 
confirmed that what he really had in mind was,
\begin{equation}
  E_{xc} = \frac{1}{2} \left( E_x^{\mbox{OEP}} + E_{xc}^{\mbox{GGA}} \right) \, .
  \label{eq:hybrid.4}
\end{equation}
where OEP\index{OEP} is the optimized effective potential---that is, the 
local (multiplicative) potential whose orbitals minimize the HF (or, in this
case, the KS) energy
expression.  Using the OEP definition would not change the energy much, but
it would change the way the xc-potential is calculated as it would no longer
be a straightforward functional derivative but rather would be subject to 
the OEP procedure for finding the appropriate local xc-potential.  Since
this is rarely necessary, we have now left the realm of KS DFT and entered
into a new {\em generalized} KS (GKS\index{GKS}) DFT \cite{SGV+1996} where 
the orbital energies no longer have the same interpretation as in conventional 
KS DFT.  This disturbing observation caused Peter Gill to write an obituary 
for DFT, born as TFD theory and killed by Axel's hybrid theory \cite{G2001}.
Interestingly, later, Axel and his doctoral student Erin Johnson found a simple
but accurate approximation to the OEP [{\bf BJ2006}].

% ----------------------
\begin{figure}
\begin{center}
\includegraphics[width=0.6\textwidth]{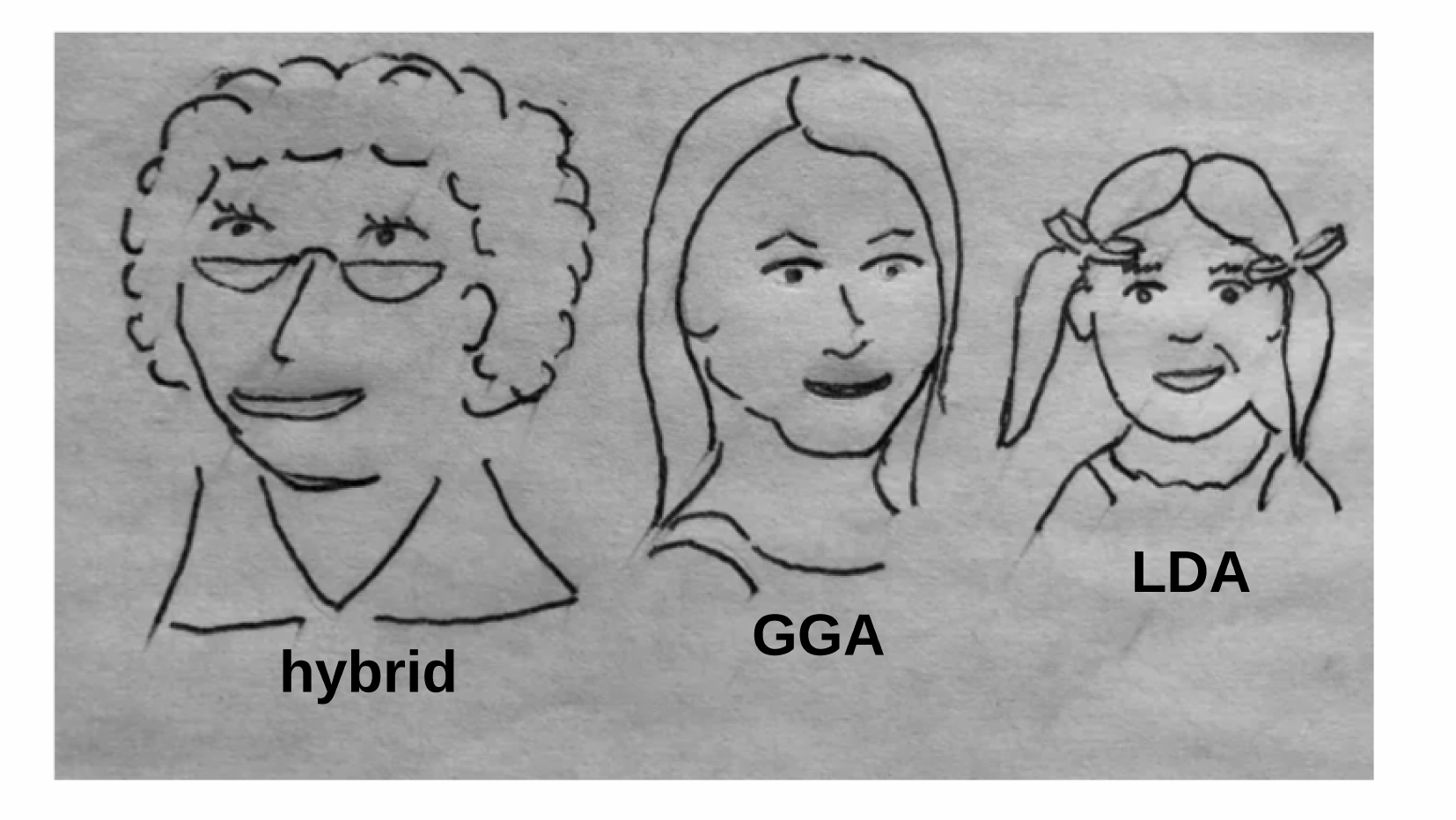}
\end{center}
\caption{
Three generations of density-functional approximations.
(Any resemblance to real people is purely co\"{\i}ncidental.)
\label{fig:3gen}
}
\end{figure}
% ----------------------
Axel further improved upon his half-and-half functional with his B3PW91
functional \framebox{\bf B1993} (cited 84~895 as of 14 May 2026 according 
to the {\em Web of Science} \cite{web})
\begin{equation}
  E_{xc} = E_{xc}^{\mbox{LDA}} + a_0 \left( E_x^{\mbox{HF}} 
  - E_x^{\mbox{LDA}} \right) + a_x \Delta E_x^{\mbox{B88}}
  + a_c \Delta E_c^{PW91} \, ,
  \label{eq:hybrid.5}
\end{equation}
where $\Delta E_x^{\mbox{B88}}$ is the [{\bf B1988}] x-GGA correction
and $\Delta E_c^{\mbox{PW91}}$ is the Perdew-Wang 1991 functional
\cite{PCV+1992} and the three parameters $a_0$, $a_x$, and $a_c$
are fit.  Also $E_{xc}^{\mbox{LDA}}$ uses the Perdew-Wang
parameterization of the LDA \cite{PW1992}.  This B3PW91 hybrid
functional was shown to give near {\em ab initio} thermochemical accuracy 
(i.e., 5 kcal/mol for atomization energies).
Axel was justifiably proud of what he had accomplished and began to talk
of three generations of functionals {\bf Fig.~\ref{fig:3gen}} which reminds
me of some of our own family photos showing mother, daughter, and grandmother,
all in one photo.

What happened next is one of the stranger stories in the history of DFAs
when Gaussian Inc.\ rushed to put Axel's functional into their {\sc Gaussian}
program!  Their introduction of DFT into {\sc Gaussian} had not always gone
smoothly.  For example, quantum chemistry DFT programs had long used the 
VWN parameterization \cite{VWN1980} of the 
Ceperley-Alder quantum Monte Carlo results \cite{CA1980} for the HEG.  
This had always
been called VWN.  But the VWN article actually contained two parameterizations.
The less good one (never used by DFT experts) was a parameterization of the
results of the RPA for the HEG.
The correct parameters for the VWN fitting of the Ceperley-Alder results are
found in the caption of Fig.~5 of Ref.~\cite{VWN1980}.  Confusingly 
{\sc Gaussian} programmed the RPA parameterization and called it VWN.  
Once they had realized their error, then they also introduced the parameters
from the caption of Ref.~\cite{VWN1980} Fig.~5 and called that VWN5.  So {\sc Gaussian}'s VWN5
is everyone else's VWN.

They also hurried to introduce Axel's hybrid functional, but they did not
have all of the pieces that they needed, so they replaced the Perdew-Wang
parameterization of $E_{xc}^{\mbox{LDA}}$ by their VWN (i.e., RPA) and
they also replaced $\Delta E_c^{\mbox{PW91}}$ with $\Delta E_c^{\mbox{LYP}}$,
the Lee-Yang-Parr \cite{LYP1988} GGA correction. However they did not 
reoptimize the three parameters $a_0$, $a_x$, and $a_c$.  This gave the
B3LYP functional \cite{SDCF1994} which was then extensively tested and was also
found to work well.  In Mike Frisch's own words to the computational
chemistry list (CCL\index{CCL}):
\begin{quote}
\noindent
``When I first decided to make use of Becke's parametrization based on
 adiabatic connection, we had not yet coded the PW91 correlation functional
 in {\sc Gaussian}.  We had coded the earlier Perdew correlation functional 
 (P86) but found that LYP seemed to work better for molecules.  I felt that 
 if the parameters Becke had optimized represented real physical content in the
 model and not just curve fitting, then the same values should be useful with
 other functionals of the same general type (i.e., GGAs).  So I tested the
 same parameters with BLYP instead of BPW91, and found that indeed they gave a
 similar improvement in predicted energetics.  Just as importantly, the
 parameters, which Becke fit to dissociation energies of neutral molecules at
 fixed geometries, also improved predicted structures and vibrational
 frequencies as compared to pure DFT, and also worked well for ionization
 potentials and electron affinities.  So both the transferability of the
 parameters to different functionals and the fact that parameters fit to one
 property improve virtually all other properties confirmed that Becke's scheme
 improves the physical content of the model and is not just fitting a 
 particular type of property.

 The reason for including VWN was that, unlike most correlation functionals,
 LYP does not have distinct local and gradient-corrected terms.  So to adjust
 the amount of non-local correlation from LYP as required by Becke's
 parameters, I needed to use a separate local correlation functional.
 That is, instead of Becke's:
 \begin{verbatim}
   1.0 x (local correlation) 
   + 0.81 XC(gradient-correction for correlation)
  \end{verbatim}
 I did
 \begin{verbatim}
   0.81 (LYP local+gradient-correction) 
   + 0.19 (VWN3 local)
 \end{verbatim}
 Unfortunetly, I wasn't as precise as I should have been in the paper and
 didn't specify which version of VWN (VWN3) I used for the local
 correlation.  This has led to a bit of confusion, with some people using
 VWN5 in their implementations for the local part.  The difference between
 the two versions is a small variation in total energy, but the predictions
 are basically the same regardless of which local functional is used to
 provide the small non-gradient-corrected part.
 We have since coded PW91 and found that B3PW91 is not quite as good as
 B3LYP.  I think this reflects the fact that PW91 isn't as good as LYP for
 molecules, not any difference in optimal values for the 3 parameters.  PW91
 is exact for the uniform electron gas, which physicists like, while LYP is
 wrong in this limit.  However, LYP was designed to make He come out right.
 Since He is a better example of the highly non-uniform electron density
 in molecules than the uniform electron gas, it is not surprising that BLYP
 and B3LYP are (slightly) better approximations for molecules than BPW91 and
 B3PW91.'' \cite{Frisch}
\end{quote}
Usually those who cite the B3LYP functional also often cite Axel's 
[{\bf B1993}] article.  Needless to say, Axel was not entirely amused.
Filipp Furche wrote to me that:
\begin{quote}
\noindent
``I mostly remember him [Axel] as a modest man who did not 
like to talk much about himself. However, when the conversation 
came to B3LYP and the fact that Axel had proposed a rather different 
functional in the paper almost everyone is citing for it, namely, 
B3PW91, he became quite animated, and we recounted how several mistakes 
in the {\sc Gaussian} implementation led to B3LYP. At some point he exclaimed: 
`They LYPified my functional!'~''
\end{quote}
To make matters even more confusing, as already mentioned,
the B3LYP functional in some programs
is not the same as in {\sc Gaussian} because the VWN (i.e., RPA) 
parameterization of the LDA has been ``corrected'' with the VWN5 (i.e., what
everyone else calls the VWN) parameterization.  In such cases, there is usually
some option, such as {\tt B3LYP/G} that has been introduced so as to give
the same results as in {\sc Gaussian}.

% \input{./tables/functionals.tex}
% ========================================
% File: functionals.tex
% Last updated: 2 June 2026
% ========================================

\begin{table}
\begin{center}
\begin{tabular}{ccc}
\hline \hline
Name & Comments & Reference \\
\hline
DH24, SOS-DH24 & double hybrid & {\bf B2024} \\
revDH23, SOS-DH23 & double hybrid & {\bf B2023} \\
DH23 & double hybrid & {\bf BSM2023} \\
B22plus & ``super'' hybrid & {\bf B2022b} \\
B22  & ``super'' hybrid  & {\bf B2022} \\
B13  & HF + DC + NDC + SC  & {\bf B2013b}, {\bf B2013c} \\
B09  & relativistic kinetic energy mGGA & {\bf B2009} \\
DF07 & HF + DC + NDC + XDM      & {\bf BJ2007b} \\
XDM  & van der Waals & {\bf BJ2005b}, {\bf JB2005} \\
B05  & HF + DC + NDC   & {\bf B2003}, {\bf B2005} \\
B98  & 2nd order GGA hybrid         & {\bf SB1998b}, {\bf SB1998} \\
B97  & GGA hybrid                   & {\bf B1997}, {\bf B1998} \\
B96j & current density functional   & {\bf B1996b} \\
B88cBR & BR + DC & {\bf B1994} \\
B3PW91 & Hybrid xc                  & \framebox{\bf B1993} \\
BH\&H & Hybrid xc                    & {\bf B1993b} \\
ELF  & conceptual tool              & {\bf BE1990} \\
BR   & Becke-Roussel mGGAx          & {\bf BR1989} \\
B88c & mGGAc                        & {\bf B1988c} \\
B88x & GGAx                 & \framebox{\bf B1988} \\
B86x & GGAx                 & {\bf B1986}  \\
X$\alpha\beta$ & GEAx & {\bf B1983}, {\bf B1985} \\ 
\hline \hline
\end{tabular}\\
\caption{
\label{tab:functionals}
Some of Axel Becke's most famous DFAs.
His work picks up where the DFT column of Table~\ref{tab:flavors} leaves off.
The notation includes ``x'' for exchange, ``c'' for correlation, 
and ``xc'' for exchange-correlation.  See text for additional information.
}
\end{center}
\end{table}
Needless to say, Axel was becoming quite famous for his constant improvement
of DFAs.  Dennis R.\ Salahub wrote to me:
\begin{quote}
\noindent
``I don't have any specific reminiscences about Axel.  We met 
at many conferences over the years and I'd often ask him the 
same question: Got any new functionals for us, Axel?  He always 
had something.''
\end{quote}
[{\bf B1995}] provides a nice overview of Axel's work developing functionals
up till 1995.  {\bf Table~\ref{tab:functionals}} tries to summarize some
of Axel's more important functionals. 

% -------------------------------------------
\subsection{A Legendary Dinner}
% -------------------------------------------

This subsection begins with when Axel met John Pople and Walter Kohn and
ends with the Chemistry Nobel Prize awarded to Kohn and Pople.  According
to Russell Boyd, Axel's first conference presentation was entitled 
``Numerical Hartree-Fock-Slater Calculations'' and was given at the
64th Chemical Conference and Exhibition of The Chemical Institute of 
Canada in Halifax in 1981.  Russ continues, ``It was there that Axel 
first met John Pople.  Two years later he met Walter Kohn in Portugal.''
Andreas Savin tells me regarding one of these early conferences,
``Axel was very excited to
meet John Pople. He told me that if he succeeds to convince Pople about
the quality of DFT, it would be great for DFT. He met Pople in a break,
and I asked him how the meeting went. Axel was very upset: `Do you know
what Pople asked me? Why does H$_2^+$ not work? I told him to take
Hartree-Fock for H$_2^+$.' ''  Of course, H$_2^+$ is a single-electron 
system where self-interaction errors (SIEs\index{SIE}) in DFT are expected 
to be very large
with most DFAs and the H$_2^+$ potential energy curve calculated using 
these same DFAs is notoriously wrong.  However John Pople's attitude 
towards DFT would change dramatically when Axel and John Pople met at 
another conference a decade later.

The International Congress of Quantum Chemistry (ICQC\index{ICQC}) is one
of several important long established meetings in the field.  The first
one was held in Menton, France, in 1973, a city on the beautiful French
Riviera bordering Italy and famous for its lemon trees.  Subsequent ICQC
have been in a variety of locations, but it returned to Menton in 1991,
2000, and 2018.

% ----------------------
\begin{figure}
\begin{center}
\includegraphics[width=0.6\textwidth]{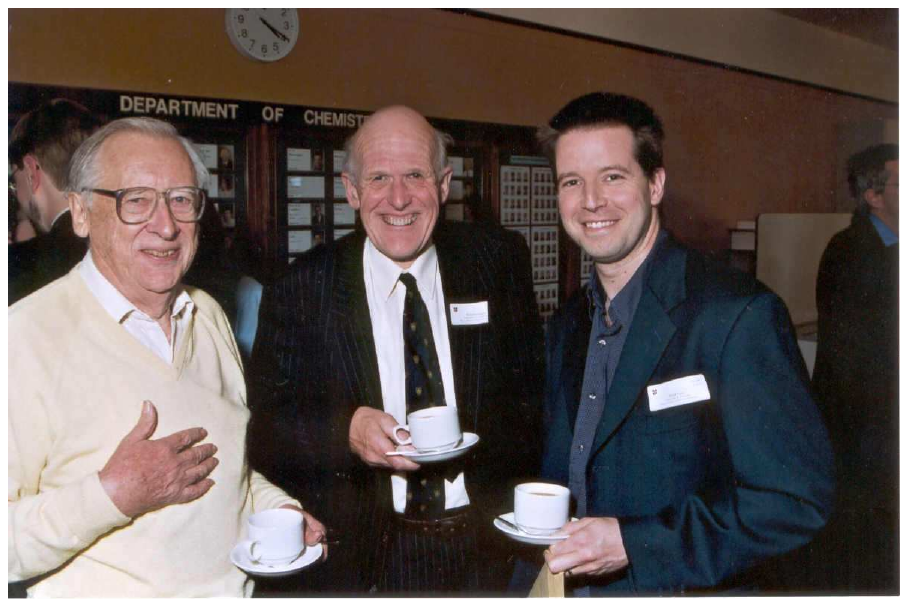}
\end{center}
\caption{
From left to right John Pople, Nicholas Handy, and David Tozer.
Used with permission from David Tozer.
\label{fig:DJT2}
}
\end{figure}
% ----------------------
According to Axel, the seventh ICQC meeting in Menton in 1991 was of particular
importance regarding the incorportation of DFT into the traditional
{\em ab initio} program {\sc Gaussian}.  
There is a {\tt YouTube} video \cite{FreasierVideo} which claims that 
it was a luncheon discussion between John Pople and Axel which ultimately
led to DFT being incorporated into the most widely used of all the {\em ab 
initio} quantum chemistry programs, namely the {\sc Gaussian} program.
This is at least partially confirmed by Axel's own written words:
\begin{quote}
\noindent
``The 1991 Menton ICQC was personally memorable for other reasons as
well.  My first real conversation with John Pople had taken place that week. 
It was undoubtedly the most important lunch of my career! [...] The
release of {\sc Gaussian92/DFT} ushured in a new era of DFT calibration and
development, and widely expanded the standard vocabularly of the quantum
chemist.'' \cite{Queens}
\end{quote}

For me, it was difficult to
say which of two {\em ab initio} quantum chemistry programs were the first
to break the taboo on DFT.  Was it Nicholas Handy of Cambridge Analytical
Derivative Package ({\sc CADPAC}) fame or was it John Pople of {\sc Gaussian}
fame who were first persuaded by Axel?  Articles reporting the implementation
of DFT in {\sc CADPAC} \cite{MLHA1992} and in {\sc Gaussian} \cite{PGJ1992}
appeared in the same issue of the journal {\em Chemical Physics Letters}
in 1992.
However only the {\sc CADPAC} article reports numbers and the {\sc Gaussian}
article thanks Nicholas Handy in the acknowledgements for useful comments.
(There is also a slightly earlier announcement of DFT in {\sc Gaussian}
\cite{GJPF1992}.)

Although initially tempted to give priority to Handy,
the situation is actually more complicated than I had first thought as
John Pople and Nicholas Handy ({\bf Fig.~\ref{fig:DJT2}}) were in close
communication at this time.  Handy wrote that,
\begin{quote}
\noindent
``We commenced our studies into Density Functional Theory (DFT) in January
1992.  This followed on from a visit by John Pople in December 1991, who
suggested that I should investigate DFT, one of the reasons being (if I 
remember correctly) that the published calculations by Becke were
very promising and that he (Pople) had the functional BLYP giving good
results.''
\cite{CCS2004}
\end{quote}
A nontrivial concern was the use of a grid in DFT calculations which might
lead to difficulties of reproducing results between different programs.
Handy wrote,
\begin{quote}
\noindent
``We found that two entirely distinct programs, one written in Carnegie-Mellon
[{\sc Gaussian}] and one in Cambridge [{\sc CADPAC}], gave exactly the same
Kohn-Sham energy (to eight decimal places).  I realised that DFT was another
quantum chemistry because all calculations should be completely reproducible
given the minimal data (molecule, geometry, basis set, functional).''
\cite{CCS2004}
\end{quote}
I think we can thank the Becke grid for this!  See also Ref.~\cite{HTL+1993}.

But let us stop here and take a moment to let what we have learned sink in:
Axel Becke, the scientist who
did much of his work alone on a workstation using his own program {\sc NUMOL},
has managed to persuade John Pople to put DFT into {\sc Gaussian} and
Nicholas Handy to put DFT into {\sc CADPAC}.  Both of these
programs are commercial {\em ab initio} programs with a healthy user base.  
The separation between the DFT community with its specialized computer 
programs and the {\em ab initio} community with its specialized computer 
programs ceased to exist almost overnight!  (John Pople passed away in 2004; 
Nicholas Handy passed away in 2012.)

That brings me to the 1998 Nobel Prize in Chemistry.  
This was one of several Nobel Prizes recognizing the importance of 
theoretical and computational chemistry (see the article of Russell Boyd
\cite{B2024a} for a review of some of the others).  On the surface, this
prize was shared between two people for two different reasons.  Half was
awarded to Walter Kohn ``for his development of the density-functional theory''.
The other half was awarded to John Pople ``for his development of computational
methods in quantum chemistry''.  Walter Kohn, a chemist-friendly physicist,
won the prize for his pioneering work in  and untiring promotion of DFT in both chemistry and physics.
John Pople created the first widely-distributed {\em ab initio} program based
upon gaussian-type orbitals (GTOs\index{GTO}) {\em and} he was convinced by
Axel to put DFT into {\sc Gaussian}.  It is very tempting to think of this
as a Nobel Prize for the coming of age of DFT in chemistry.  If so, then
Axel Becke surely played a huge role bringing the DFT and {\em ab initio}
quantum chemistry communities together.  And some of us might ask ourselves
why the prize was not also awarded to Axel?  In fact, Henry F.\ Schaefer~III
has written:
\begin{quote}
\noindent
``On Nobel prize Tuesday, October 13, 1998, I presented the annual Kenneth S.
Pitzer Lecture at the University of California at Berkeley. In addition to my
prepared lecture, I presented a few observations on the John Pople -- Walter
Kohn prize. At the end of my lecture my colleague and friend, Professor
Robert A. (`Bob') Harris, stood up. Bob stated with no hesitation `If Pople and
Kohn deserved the prize, then Axel Becke should have shared it.' Wisdom for
the ages.'' \cite{NBB+2026}
\end{quote}
I had the opportunity to meet
Bj\"orn Roos who had been on the Nobel committee that selected Kohn and Pople.
I said I had a question for him.  He replied that he thought he knew what
my question was but that members of the Nobel committee are forbidden to give
out any information about how they make their decisions.  So I must leave it
at that.
(Bj\"orn Roos passed away in 2010; Walter Kohn passed away in 2016.)

\begin{quote}
\noindent
{\sf
I would like to add one more thought here:
DFT in the form of HFS (X$\alpha$) had been rejected by the {\em ab initio}
quantum chemistry community as unreliable.  What, if anything, had changed?
It is true that Axel was one of the first to make credibly fully numerical
(basis-set free) calculations in DFT, hence finally allowing a reliable
evaluation of the quantitative limits of DFAs.  Then he (and
others) had made major improvements in the energies obtained by 
DFAs.  But DFT calculations were
still expected to fail from time to time and yet it no longer mattered!
How could this be?
Could it be that something else was also going on?  Could it be that the
time was ripe to adopt a new method with near {\em ab initio} accuracy
but which could be applied to much larger molecules of a size more interesting
for bench chemists and (presumably also) to granting agencies?
I suspect this was also part of the answer as to why DFT was poised to
enter mainstream quantum chemistry. Another reason, though I do not
think that this was as evident to the average quantum chemist, is that
it was beginning to be better understood why some functionals worked
better than others for specific applications and when failures might
be expected. --- MEC
}
\end{quote}
% =================================================
\section{Axel's Tree}
\label{sec:ladder}
% \input{ladder.tex}

% ===============================================================
% File ladder.tex .
% Last modified: 18 July 2026
% ===============================================================

Axel's introduction of hybrid functionals into DFT changed the standard
from KS DFT into GKS DFT.  This led to (and continues to cause) confusion
in how to interpret DFT calculations.  Some clarity came from John Perdew's
vision of Jacob's ladder (see below).  In a way, Axel's subsequent work would involve
climbing Perdew's ladder, but not quite.  A closer examination of Axel's
work suggests that Axel was indeed building a ladder (or perhaps better 
described as
a sequence of correction terms) but not necessarily
the same ladder as John Perdew.  Arguably Axel sometimes even branched
off here and there in different directions such as when he added a dependence
in his functionals on the current density [{\bf B2002}], so the term ``Axel's
tree'' might be more appropriate.  This section begins
with a brief description of Perdew's ladder and then goes on to examine
the various rungs of Axel's ladder.  Some of Axel's papers involve improved
parameterizations of his previous functionals.  I will not dwell on these
as they often do not present new ideas.  Instead I will stick to what I have 
identified as Axel's most important idea articles --- those which add new 
bits and pieces to his ladder.  In so far as possible, I will try to use
Axel's own vocabulary for these pieces. 

% -------------------------------------------
\subsection{Jacob's Ladder}
% -------------------------------------------

%---------------------------------------------------------------
\begin{figure}
\begin{center}
\includegraphics[width=\textwidth]{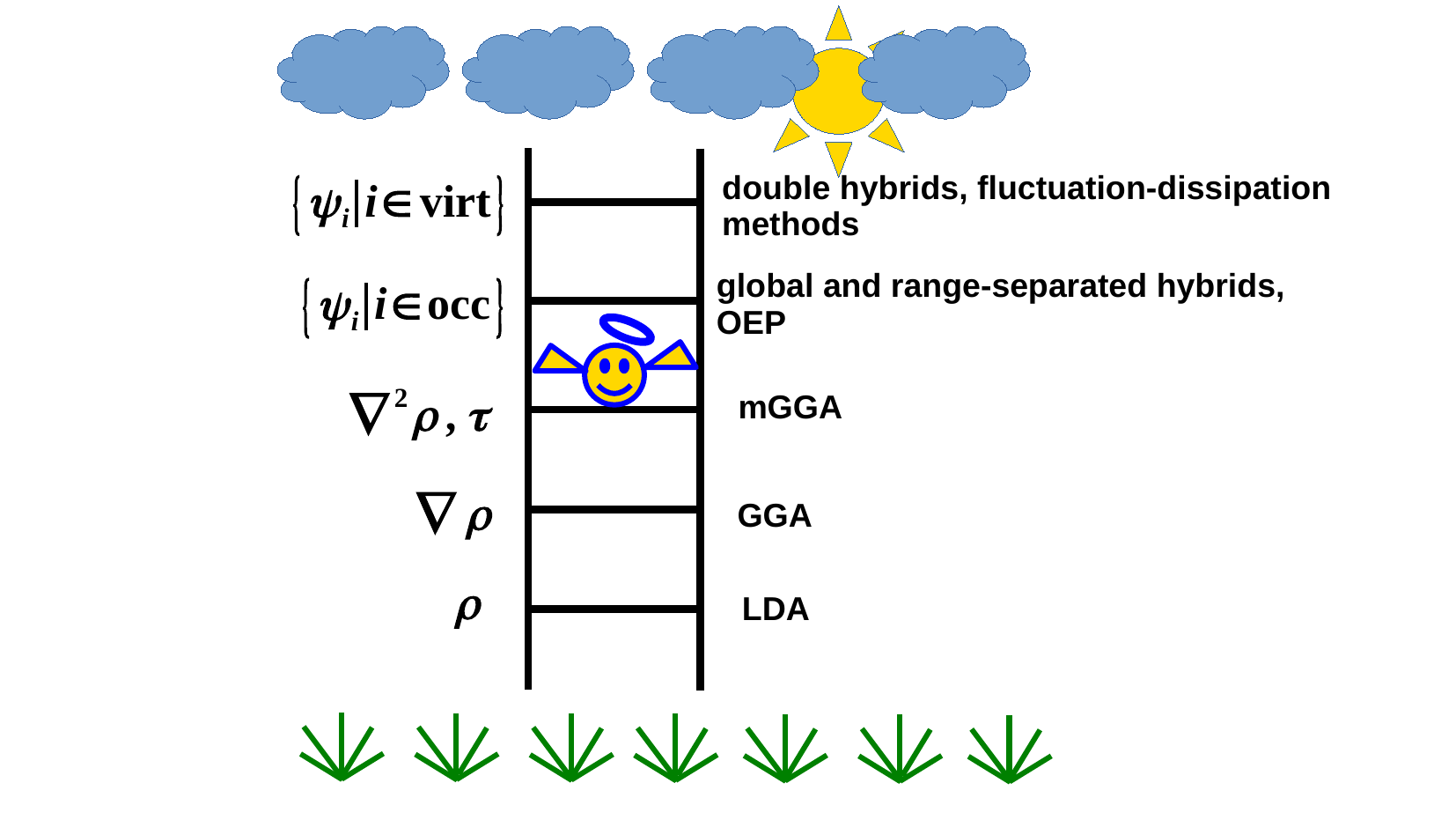}
\end{center}
\caption{
Jacob's ladder \cite{PS2001,PRC+2009}.
\label{fig:jacob}
}
\end{figure}
%---------------------------------------------------------------
Many of us who were able to attend the tenth ICQC meeting in Menton
in 2000 found it remarkable, in part, because of a particularly 
inspirational talk by John Perdew where he presented his idea of a 
Jacob's ladder of functionals.  This was a step beyond Becke's three 
generations of DFAs.  Here John Perdew envisaged a ladder similar to 
that described in the Bible:
\begin{quote}
\noindent
``Jacob left Beersheba and set out for Harran. When he reached a certain place,
he stopped for the night because the sun had set. Taking one of the stones there,
he put it under his head and lay down to sleep. He had a dream in which he saw a 
stairway resting on the earth, with its top reaching to heaven, and the angels of
God were ascending and descending on it.''
--- Genesis 28: 10-12
\end{quote}
A version of Perdew's ladder is shown in {\bf Fig.~\ref{fig:jacob}}.  Users of DFT
are represented by an angel (in the spherical approximation) ascending the ladder
rung-by-rung to obtain increasingly sophisticated versions of DFAs where each rung
represents a new type of functional dependence.  It is hoped that by adding this additional
flexibility in creating new functionals, then greater accuracy can be achieved until finally
our angel reaches DFT heaven (which is presumably the exact solution of the nonrelativistic
Schr\"odinger equation).  While this is more a hope than a guarantee, it does seem to be
the empirically-observed average trend among DFAs as we climb the ladder.  
Perdew emphasized that, since DFAs higher on
the ladder are more computational-resource intensive, that our angel also needs to be able
to descend the ladder to be able to access less compute intensive DFAs for applications to
larger systems.

% -------------------------------------------
\subsection{Becke-Roussel (BR) Functional}
% -------------------------------------------

\begin{quote}
\noindent
{\sf
During my time in Montreal, I saw a number of very talented individuals
pass through the lab of Dennis Salahub.  One of these was Emil Proynov
who chose to work on the daunting task of improving functionals.  I remember
that he was particularly impressed by the Becke-Roussel functional. 
Another of the very talented individuals that I met during my time with
Dennis Salahub was Martin Kaupp who has gone on to develop functionals
for strong correlation. --- MEC
}
\end{quote}

We have noted how GGAs grew more or less systematically out of the study of 
xc-holes.  Axel's next step was to abandon the condition that the xc-hole
be correct in the limit of the HEG.  Instead he reasoned that it is more
important for molecular calculations that the xc-hole be correct for atoms.
This led to the Becke-Roussel (BR\index{BR}) functional [{\bf BR1989}]
which will be described here in detail, more or less following Appendix A 
of [{\bf B2022b}].

The BR functional is an x-only functional which begins with an assumed
form for the x-hole $\rho_x(1,2)$ [Eq.~(\ref{eq:GGA.16})].  We calculate
\begin{equation}
   \rho_x(1,2) = -\frac{\psi(1)\psi^*(2)\psi(2)\psi^*(1)}{\psi(1)\psi^*(1)}
               = -\psi^*(2)\psi(2) = -\frac{a^3}{8\pi} e^{-a r_2} \, ,
   \label{eq:BR.1}
\end{equation}
for a hydrogenic wave function and neglecting spin.  Setting 
${\vec r} = {\vec r}_1$ and ${\vec s} = {\vec r}_2 - {\vec r}_1$, then 
spherical averaging of ${\vec s}$ gives,
\begin{equation}
  \rho_x(r,s) = \frac{a}{16\pi rs} \left[ \left( a \vert r- s \vert
  + 1 \right) e^{-a \vert r - s \vert } - \left( a \vert r + s \vert
  + 1 \right) e^{-a \vert r + s \vert } \right] \, .
  \label{eq:BR.2}
\end{equation}
which is then rewritten with $r$ renamed $b$.  This
is Taylor expanded to give the spherically averaged x-hole [$\rho_{0,0}({\vec r},\Delta r)$ in Eq.~(\ref{eq:GGA.17a})] as
\begin{equation}
  h_{x\sigma}^{\mbox{\tiny BR}} ({\vec r},s) =  -Ae^{-x} - 
  \frac{Aa^2}{6} \left(1-\frac{2}{x} \right) e^{-x} s^2 + \mbox{HOT} \, ,
  \label{eq:BR.3}
\end{equation}
where $A=a^3/(8\pi)$ and $x=ab$.  This is compared against the known
exact expansion of the x-hole,
\begin{equation}
  h_{x\sigma}^{\mbox{\tiny exact}}({\vec r},s) = -\rho_\sigma + Q_\sigma s^2 + 
  \mbox{HOT} \, .
  \label{eq:BR.4}
\end{equation}
where 
\begin{eqnarray}
   Q_\sigma & = & \frac{1}{6} \left( \nabla^2 \rho_\sigma - 2 D_\sigma \right)
   \nonumber \\
   D_\sigma & = & \tau_\sigma - \frac{1}{4} 
   \frac{ \vert {\vec \nabla} \rho_\sigma \vert^2}{\rho_\sigma} 
   \nonumber \\
   \tau_\sigma  & = & \sum_i \vert {\vec \nabla} \psi_{i,\sigma} \vert^2 \, .
    \label{eq:BR.5}
\end{eqnarray}
Here $\tau_\sigma$ is the spin-$\sigma$ kinetic energy density.  Identifying
the two coefficients in Eqs.~(\ref{eq:BR.4}) and (\ref{eq:BR.5}) gives two
contraints while normalizing the x-hole to contain exactly one hole gives
a third constraint.  The latter leads to $8\pi A = 1$.  Combining these
conditions leads to a nonlinear equation (!) which must be solved 
numerically for $x$:
\begin{equation}
  \frac{xe^{-2x/3}}{x-2} = \frac{2}{3} \pi^{2/3} \frac{\rho_\sigma^{5/3}}
  {Q_\sigma} \, .
  \label{eq:BR.6}
\end{equation}
% which must be solved numerically.  
This nonlinear equation certainly
complicates the problem of taking functional derivatives, but this is
a mere technical concern.  Once solved, $b$ is obtained by solving,
\begin{equation}
  b^3 = \frac{x^3 e^{-x}}{8\pi \rho_\sigma} \, .
  \label{eq:BR.7}
\end{equation}
Finally the spin-$\sigma$ x-energy is,
\begin{equation}
  E_{x\sigma}^{\mbox{\tiny BR}} = \frac{1}{2} \int \rho_\sigma 
  U_{x\sigma}^{\mbox{\tiny BR}} \, d{\vec r} \, 
  \label{eq:BR.8}
\end{equation}
where
\begin{equation}
  U_{x\sigma}^{\mbox{\tiny BR}} = -\frac{1}{b}\left( 1-e^{-x} - \frac{1}{2}
  x e^{-x} \right) 
  \label{eq:BR.9}
\end{equation}
is the Coulomb potential of the BR hole.  This very interesting
functional is one of the first mGGAs.  While it appeared
to be very physical and accurate, taking functional derivatives was far
from obvious, initially preventing, self-consistent calculations.
This changed in 2008 when Proynov, Gan, and Kahn found an accurate
analytical interpolation for the solution of Eq.~(\ref{eq:BR.6}) 
which allowed a self-consistent implementation of the BR functional in 
the {\sc Q-Chem} program \cite{PGK2008}.

% -------------------------------------------
\subsection{Types of Correlation}
% -------------------------------------------

Since the BR functional is x-only, we still have to add a correlation part.
This is a good place to discuss how different types of correlation are
classified in DFT and in WFT.  In each
case, the classification system used is fuzzy in the sense that borders
between the different types of correlation are not really well defined.
Nevertheless such fuzzy concepts have often been found to be useful in
methods development.

In the first instance, correlation is divided into ordinary (weak) correlation
and ``strong correlation.''  To paraphrase Lucia Reining (who I find has a 
gift for explaining things clearly), ``strong correlation is anything that 
we do not yet understand.''  In solid-state physics, this might be the 
problem of the relative energies of the $3d$ and $4s$ orbitals in first-row 
transition metal compounds, which I have often heard called strong correlation.
It is often addressed by DFT+Hubbard models.  To be fair, Lucia recognizes
that the U in DFT+U is a way to approximate GKS and thereby correct the
orbital energies, rather than as a form of correlation.  Another type of
strong correlation is the formation of Cooper pairs in superconductors.
This can be addressed by a special reformulation of DFT for superconductors.
But we will not be talking about either of these types of ``correlation'' here.

\paragraph{WFT}
Bartlett and Stanton gave a convenient WFT classification of different
types of correlation \cite{BS1994}:
Electron correlation is divided into dynamical correlation present
when a single determinant is a good first approximation to the wave function,
static (or degeneracy) correlation which is typically due to energetically
degenerate determinants resulting from symmetry (e.g., spin multiplets),
and nondynamic (or quasi-degeneracy) correlation.   Dynamic correlation is
ordinary (weak) correlation, while both static and nondynamic correlation are
strong correlation.

\paragraph{DFT}
A classification based upon the density, or even better, upon the xc-hole, 
is more useful for DFT than is one
based upon $N$-electron wave functions.  Nevertheless when DFT and WFT
use the same name for different types of correlation, then we should expect
at least a rough correspondance between the two classification schemes.
In DFT, dynamic correlation is often described as mainly on the atoms.
Left-right correlation describes orbital localization after symmetry
breaking in diatomics. Nevertheless open-shell atoms have different spin
multiplet energies and this is clearly what WFT calls static correlation.

Axel tries to be more precise in [{\bf B2013c}] where he distinguishes
four types of correlation, namely
\begin{description}
  \item[dynamical] ``is a {\em local} correlation of shorter range,
  acting over distances of atomic size or less.  It arises, for the 
  most part, from well-known interelectronic cusp conditions.''
  \item[static] ``also called `nondynamical' or `left-right' correlation
  is {\em defined} in this work as being {\em multi-center} in range.
  It distributes electrons over various centers in a molecule such that
  clustering on any given center is prevented.''
  \item[strong] is needed for ``accurate computations on 
  dissociating chemical systems without breaking space or spin symmetries 
  and without using multi-determinantal reference states.''
  \item[dispersion] ``correlations arise from instantaneous 
  multipole-multipole interactions between electrons on different
  centers. [...] the effects of dispersion are significant at long
  range where static and dynamical correlations are unimportant.''
\end{description}
Note how Axel's DFT definitions correspond roughly, but not exactly, to
the Bartlett-Stanton WFT definitions. Axel then embarked upon a program
that would build a sort of series where each type of correlation 
constitutes a new term that may be added.  In the process of building
this series, Axel will also climb Perdew's ladder.

% -------------------------------------------
\subsection{Dynamical Correlation (DC)}
% -------------------------------------------

The BR x-functional is one example of what Axel called a ``coordinate
space model'' to contrast it with momentum space work by Langreth 
and Perdew.  In [{\bf B1988c}], Becke also presented a ``coordinate space 
model'' for correlation to be used together with the BR x-functional
(i.e., as the B88cBR functional).  The derivation is similar in philosophy
to that of the BR functional but is based upon the c-hole rather than
the x-hole and in particular focuses on the electron-electron 
Fermi (parallel spin, par\index{par}) and Coulomb (opposite spin, 
opp\index{opp}) cusp conditions.  As such, it is a direct correlation
(DC\index{DC}) functional.  This gives for the Coulomb correlation,
\begin{equation}
  E_{\mbox{\tiny DC}}^{\mbox{\tiny opp}} = -0.8 \int \rho_\uparrow
  \rho_\downarrow z_{\mbox{\tiny opp}}^2 \left[ 1- \frac{\ln(1+z_{\mbox{\tiny opp}})}{z_{\mbox{\tiny opp}}} \right] \, d{\vec r}
  \label{eq:DC.1}
\end{equation}
where the $z_{\mbox{\tiny opp}}$ correlation length is derived
from the BR x-potential,
\begin{equation}
  z_{\mbox{\tiny opp}} = c_{\mbox{\tiny opp}} \left( \vert U_{x\uparrow}^{\mbox{\tiny BR}} \vert^{-1} + \vert U_{x\downarrow}^{\mbox{\tiny BR}} \vert^{-1} \right) 
  \, ,
  \label{eq:DC.2}
\end{equation}
with $c_{\mbox{\tiny opp}} = 0.63$.  For the Fermi correlation,
\begin{equation}
  E_{\mbox{\tiny DC}}^{\mbox{\tiny par}} = E_{\mbox{\tiny DC}}^{\uparrow \uparrow} + E{\mbox{\tiny DC}}^{\downarrow \downarrow} \, ,
  \label{eq:DC.3}
\end{equation}
with
\begin{equation}
  E_{\mbox{\tiny DC}}^{\sigma \sigma} = -0.01 \int \rho_\sigma
  D_\sigma z_{\sigma \sigma}^4 
  \left[ 
        1 - \frac{2}{z_{\sigma \sigma}}
  \ln \left(1+\frac{z_{\sigma \sigma}}{2} \right) 
  \right] \, d{\vec r}
  \, ,
  \label{eq:DC.4}
\end{equation}
with the $z_{\sigma \sigma}$ correlation lengths derived from the BR
x-potential,
\begin{equation}
  z_{\sigma \sigma} = 2 c_{\sigma \sigma} 
  \vert U_{x\sigma}^{\mbox{\tiny BR}} \vert^{-1} \, ,
  \label{eq:DC.5}
\end{equation}
and $c_{\sigma \sigma} = 0.88$.

% -------------------------------------------
\subsection{Nondynamical Correlation (NDC)}
% --------------------------------------------

An interesting problem with the way Axel has defined nondynamical correlation
is that it also applies to the fact that the potential energy curve 
(PEC\index{PEC}) of H$_2^+$ is wrong for GGAs and hybrid functionals.
The reason for this can be given in several different ways, with the
simplest explanation being an incomplete canceling of the coulomb term 
by the approximate xc term (i.e., the SIE).
Although the proper dissociation 
of H$_2^+$ is trivial for WFT, most DFAs lead to an improper dissociation of 
H$_2^+$ which should dissociate to two 
[H H$^+$ $\leftrightarrow$ H$^+$ H ]
following a classic Morse-like or Lennard-Jones-like potential energy
curve with a total energy of -0.5 Ha at infinite bond distance.  However
most DFAs start to dissociate correctly, cross a barrier which is too low,
and then the total energy becomes much too negative at large bond distance.
The problem is immediately fixed by using 100\% exact exchange.
This led to the formulation of the B05 functional for nondynamical
correlation (NDC\index{NDC}) [{\bf B2003}, {\bf B2005}].

The BR x-functional is no longer used because we have 100\% exchange.
However the basic argument behind the BR functional is used to obtain
a measure of the multicenter delocalization of the xc-hole.  The idea
is to set the Slater potential,
\begin{equation}
  U_{x\sigma}^{\mbox{\tiny exact}}({\vec r}_1) = -\int 
  \frac{ \gamma_\sigma({\vec r}_1,{\vec r}_2) \gamma_\sigma({\vec r}_2, 
  {\vec r}_1) }
  {\rho_\sigma({\vec r}_1) \vert {\vec r}_1 - {\vec r}_2 \vert} \, d{\vec r}_2
  \label{eq:NDC.1}
\end{equation}
equal to the BR x-hole Coulomb potential,
\begin{equation}
  U_{x\sigma}^{\mbox{\tiny BR}} = -\frac{8\pi A}{a^2 x} \left( 1- e^{-x} - \frac{1}{2} x e^{-x} \right) \, ,
  \label{eq:NDC.2}
\end{equation}
with $x = ab$.  This leads to a new nonlinear equation to solve for $x$, 
namely,
\begin{equation}
  \frac{x-2}{x^2} \left( e^x - 1 - \frac{x}{2} \right) 
  = -\frac{3}{4\pi} \frac{Q_\sigma}{\rho_\sigma^2} U_{x\sigma}^{\mbox{\tiny
  exact}} \, .
  \label{eq:NDC.3}
\end{equation}
Then 
\begin{equation}
   A = \rho_\sigma e^x \, 
   \label{eq:NDC.4}
\end{equation}
and 
\begin{equation}
  a^2 = \frac{6 Q_\sigma}{\rho_\sigma} \frac{x}{(x-2)} \, .
  \label{eq:NDC.5}
\end{equation}
Also $b=x/a$.  Finally the {\em effective} charge of the BR hole,
\begin{equation}
  N_{x\sigma}^{\mbox{\tiny eff}} = \frac{8\pi A}{a^3} 
  \label{eq:NDC.6}
\end{equation}
(which may be less than unity!) is a measure of the delocalization
of the hole onto multiple centers.  This will allow us to calculate the
opposite spin (opp) and parallel spin (par) NDC energies and sum them to get the total
NDC energy,
\begin{equation}
  E_{\mbox{\tiny NDC}} = E_{\mbox{\tiny NDC}}^{\mbox{\tiny opp}} 
  + E_{\mbox{\tiny NDC}}^{\mbox{\tiny par}} \, .
  \label{eq:NDC.7}
\end{equation}
The opposite-spin NDC is calculated as,
\begin{equation}
  E_{\mbox{\tiny NDC}} = \frac{1}{2} \int f_{\mbox{\tiny opp}} \rho_\uparrow
  U_{x\downarrow}^{\mbox{\tiny exact}} \, d{\vec r} 
  + \frac{1}{2} \int f_{\mbox{\tiny opp}} \rho_\downarrow
  U_{x\uparrow}^{\mbox{\tiny exact}} \, d{\vec r}  \, ,
  \label{eq:NDC.8}
\end{equation}
where
\begin{equation}
  f_{\mbox{\tiny opp}} = \min \left( \frac{1-N_{x\uparrow}^{\mbox{\tiny eff}}}{N_{x\downarrow}^{\mbox{\tiny eff}}} , \frac{1-N_{x\downarrow}^{\mbox{\tiny eff}}}{N_{x\uparrow}^{\mbox{\tiny eff}}} , 1 \right) \, .
  \label{eq:NDC.9}
\end{equation}
The parallel-spin NDC is calculated as,
\begin{equation}
  E_{\mbox{\tiny NDC}} = 
  \frac{1}{2} \int \rho_\uparrow U_{\mbox{\tiny NDC}}^{\uparrow \uparrow} \, {\vec r}
  + \frac{1}{2} \int \rho_\uparrow U_{\mbox{\tiny NDC}}^{\downarrow \downarrow} \, d{\vec r} \, ,
  \label{eq:NDC.10}
\end{equation}
where 
\begin{equation}
  U_{\mbox{\tiny NDC}}^{\sigma \sigma} = -f_{\sigma \sigma}
  \frac{D_\sigma}{3\rho_\sigma} M_\sigma^{(1)} 
  \, .
  \label{eq:NDC.11}
\end{equation}
This involves the rather peculiar function,
\begin{equation}
  M_\sigma^{(n)} = 4\pi \int_0^\infty s^{n+2} \vert h_{x \sigma}^{\mbox{\tiny eff}} \vert \, ds \, ,
  \label{eq:NDC.12}
\end{equation}
with $n=1$.  
The quantity $h_{xc}^{\mbox{\tiny eff}}$ is $h_{xc}^{\mbox{\tiny BR}}$
[Eq.~(\ref{eq:BR.3})] in this model.
The same function with $n=2$ is used to determine 
\begin{equation}
  f_{\sigma \sigma} = \min \left( \frac{3 \rho_\sigma \Delta N_{\mbox{\tiny NDC}}^{\sigma \sigma}}{D_\sigma M_\sigma^{(2)}} , 1 \right) \, ,
  \label{eq:NDC.13}
\end{equation}
where
\begin{equation}
  \Delta N_{\mbox{\tiny NDC}}^{\sigma \sigma} = \min \left(
  1-N_{x\sigma}^{\mbox{\tiny eff}} - f_{\mbox{\tiny opp}} N_{x\sigma}^{\mbox{\tiny eff}} , N_{x\sigma}^{\mbox{\tiny eff}} \right) 
  \label{eq:NDC.14}
\end{equation}
is the amount of parallel-spin depletion needed to normalize the local
spin $(\sigma,\sigma)$ exchange NDC hole.
Note that the NDC term is fairly complicated and the details have evolved
over time.  So the treatment here is intended primarily to give the flavor
with the interested reader being referred to Axel's articles for more details.

% -------------------------------------------
\subsection{Strong Correlation (SC)}
% -------------------------------------------

Recall that wave-function theorists consider strong correlation (SC\index{SC})
to be divided into static and nondynamic correlation.  However Axel's definition
was a little different.  For Axel, SC is what is needed for ``accurate
computations on dissociating chemical systems without
breaking space or spin symmetries and without using multideterminantal
reference states'' [{\bf B2013c}].  It is quite challenging to include this type
of SC in a DFA!  But Axel did design a few functionals specifically for this
purpose.

In [{\bf B2013c}], Axel first presents the B13 functional which includes DC
and NDC but which lacks SC and then he invents an add-on to include SC.
The B13 functional is derived from the adiabatic connection formalism and
has the form,
\begin{equation}
  E_c = a_c U_c \, ,
  \label{eq:SC.1}
\end{equation}
where $1/2 \leq a_c \leq 1$ and $U_c$ is called a ``potential energy.''  
Although
\begin{equation}
  E_c^{\mbox{\tiny B13}} = a_{\mbox{\tiny DC}}^{\mbox{\tiny opp}} 
  U_{\mbox{\tiny DC}}^{\mbox{\tiny opp}} 
  + a_{\mbox{\tiny DC}}^{\mbox{\tiny par}} U_{\mbox{\tiny DC}}^{\mbox{\tiny par}}
  + a_{\mbox{\tiny NDC}}^{\mbox{\tiny opp}} U_{\mbox{\tiny NDC}}^{\mbox{\tiny opp}} 
  + a_{\mbox{\tiny NDC}}^{\mbox{\tiny par}} U_{\mbox{\tiny NDC}}^{\mbox{\tiny par}}  \, ,
  \label{eq:SC.2}
\end{equation}
Axel notes that,
\begin{equation}
  U_c \approx 
  U_{\mbox{\tiny DC}}^{\mbox{\tiny opp}} 
  + U_{\mbox{\tiny DC}}^{\mbox{\tiny par}}
  + U_{\mbox{\tiny NDC}}^{\mbox{\tiny opp}} 
  + U_{\mbox{\tiny NDC}}^{\mbox{\tiny par}}  \, ,
  \label{eq:SC.3}
\end{equation}
in Eq.~(\ref{eq:SC.1}) with $a_c \approx 0.6$.  However this formulation 
lacks size-consistency unless $a_c$ is allowed to vary.  To do so, he 
introduces the potential energy density $u_c$ defined by
\begin{equation}
  E_c = \int \alpha_c({\vec r}) u_c({\vec r}) \, d{\vec r} \, ,
  \label{eq:SC.4}
\end{equation}
with
\begin{equation}
  u_c =  u_{\mbox{\tiny DC}}^{\mbox{\tiny opp}} 
  + u_{\mbox{\tiny DC}}^{\mbox{\tiny par}}
  + u_{\mbox{\tiny NDC}}^{\mbox{\tiny opp}} 
  + u_{\mbox{\tiny NDC}}^{\mbox{\tiny par}} \, .
  \label{eq:SC.5}
\end{equation}
Usually potential energy densities are clear from context.  For example,
in Eq.~(\ref{eq:BR.8}), $u_{x\sigma}^{\mbox{\tiny BR}} = \rho_\sigma
U_{x\sigma}^{\mbox{\tiny BR}}/2$.
The quantity,
\begin{equation}
  x = \frac{u_{\mbox{\tiny NDC}}^{\mbox{\tiny opp}} + u_{\mbox{\tiny NDC}}^{\mbox{\tiny par}}}{u_c} \, ,
  \label{eq:SC.6}
\end{equation}
is then taken as a measure of SC.  His final form for the SC term was
\begin{equation}
  E_{\mbox{\tiny SC}} = c_2 \int x^2 u_c \, d{\vec r} \, .
  \label{eq:SC.7}
\end{equation}

The particularly difficult SC case of two symmetry equivalent determinants
is discussed in [{\bf B2013}].  This includes square H$_4$, the $D_{2d}$
transition state for rotation around the double bond in ethylene, % (H$_2$C=CH$_2$), 
the $D_{4h}$ transition state for automerization of cyclobutadiene
between rectangular $D_{2h}$ structures, and the $D_{8h}$ transition state
for double-bond shifting in cycloocatetraene between $D_{4h}$ localized
double-bond structures.  Some one used to excited states will immediately
think of these as areas where the adiabatic potential energy surface arises
from configuration mixing between diabatic potential energy surfaces 
corresponding to different electronic configurations.  Axel describes
this using a fractional occupation (ensemble) formalism.  The 1-RDM takes 
the form for a two-determinant wave function,
\begin{equation}
  \gamma = 2 \sum_{i=1}^{\mbox{\tiny HOMO}-1} \psi_i \psi_i^*
  + 2(1-f) \psi_{\mbox{\tiny HOMO}} \psi_{\mbox{\tiny HOMO}}^*
  + 2f \psi_{\mbox{\tiny LUMO}} \psi_{\mbox{\tiny LUMO}}^* \, ,
  \label{eq:SC.8}
\end{equation}
where $0 \leq f \leq 1$ is the fractional occupation number, and the density
is (as usual) the diagonal of the 1-RDM.  This is to be calculated
with the best available orbitals.  In [{\bf B2015}], Axel uses averaged
density self-interaction corrected LDA orbitals.  The next term is a fairly
elaborate one which includes both DC, NDC, and SC, with SC given in the form,
\begin{equation}
  \Delta_{\mbox{\tiny SC}}^{\mbox{\tiny B13}} = 
  \int \left( c_2 x^2 + c_3 x^3 \right) x^n u_{\mbox{\tiny c}} \, d{\vec r} \, .
  \label{eq:SC.9}
\end{equation}
The fractional occupation number $f$ is varied to obtain the lowest energy 
[Eq.~(\ref{eq:SC.8})].  Although the procedure is not self-consistent,
the optimized $f$ is then used to calculate new orbitals. 
The reader interested in the exact values of the fitting parameters
is referred back to the original references.
Suprisingly good barrier heights and potential energy curves were found.

% -------------------------------------------
\subsection{Dispersion}
% -------------------------------------------

Elementary textbooks describe London or dispersion forces as being due
to instantaneous fluctuations of molecular dipoles.  This, and other types
of attraction at a distance, are known as vdW forces.
Weakly bound molecules, such as noble gas dimers, are often referred to as
vdW molecules, although binding is not just due to long-range vdW forces.
A major problem in DFT is how to find a functional of the density that
can model long-range forces between objects with nonoverlapping densities.
Clearly the usual DFAs cannot describe dispersion!

[{\bf PB1995}] showed that the half-and-half hybrid gave at least weak
binding for noble gas dimers.

But it was not until 2005 that Axel Becke and his then PhD student Erin Johnson
figured out that dispersion interactions could be included through the
dipole moment of the xc-hole [{\bf BJ2005b}],
\begin{equation}
  {\vec d}_X(1) = \int \rho_{xc}(1,2) {\vec r}_2 \, d2 \, .
  \label{eq:XDM.1}
\end{equation}
In practice, correlation was neglected and so only the exchange-hole
dipole moment (XDM\index{XDM}) was calculated.  This gave remarkably good
results for vdW coefficients between species A and B when calculated as,
\begin{equation}
  C_6(A-B) = \frac{\langle d_X^2 \rangle_A \langle d_X^2 \rangle_B
  \alpha_A \alpha_B}{\langle d_X^2 \rangle_A \alpha_A +  \langle d_X^2 \rangle_B  \alpha_B} \, ,
  \label{eq:XDM.2}
\end{equation}
where $\alpha_{A}$ ($\alpha_B$) is the dipole polarizability of species A (B). 
This has been followed up by an extensive series of papers from Axel
and his PhD students E.\ Johnson \cite{J2007}, A.A. Arabi \cite{A2012},
and F.O.\ Kanneman \cite{K2013} on dispersion interactions, with continuing 
work from Erin.  Note that the theory has been extended to include higher
multipole moments as well as 3-body dispersion interactions.
Articles involving Axel include Refs.~[{\bf JB2005}, {\bf BJ2006b}, 
{\bf JB2006b}, {\bf JB2006}, {\bf BJ2007}].

Getting $C_6$ van der Waals (vdW\index{vdW}) coefficients 
right does not mean getting the weak binding of vdW complexes right. 
This challenge had been previously examined in [{\bf PB1995}]. 
It was taken up once again in [{\bf BJ2007b}, {\bf JB2008},
{\bf KB2009}, {\bf JBSD2009}, {\bf BAK2010}, {\bf KB2010}, 
{\bf AB2012}] where dynamical, nondynamical, and dispersion effects 
were combined in a new DFA.

Further improvements were made in 2022 with the creation of the 
``super'' hybrids [{\bf B2022}, {\bf B2022b}] which mix in some 
GGA and some mGGA exchange.

% -------------------------------------------
\subsection{Double Hybrid Functionals}
% -------------------------------------------

\begin{quote}
\noindent
``I think KS-DFT is about {\em occupied orbitals only} (I hesitate to 
suggest the acronym `OOO').''  {\bf B2014}
\end{quote}
I can only guess why Axel hesitated about the acronym OOO.  However a 
quick search on the internet indicates that OOO is bureaucratic jargon
for ``out of office'' \cite{OOO}.  Did Axel know this other use of the
acronym?

In the last three papers listed in Appendix~\ref{sec:CVaxel}, Axel
finally abandons the OOO principle and included information from
virtual orbitals. The new term in the energy expression is,
\begin{equation}
  \alpha_{\mbox{\tiny PT2}}^{\mbox{\tiny opp}} 
  \left( E_{\mbox{\tiny PT2}}^{\mbox{\tiny opp}} - \frac{1}{2}E_{C_6}^{\mbox{\tiny XDM}} \right) + \alpha_{\mbox{\tiny PT2}}^{\mbox{\tiny par}} 
  \left( E_{\mbox{\tiny PT2}}^{\mbox{\tiny par}} - \frac{1}{2}E_{C_6}^{\mbox{\tiny XDM}} \right) \, ,
  \label{eq:double.1}
\end{equation}
where PT2\index{PT2} is calculated by second-order perturbation theory.
These are designed to go to zero at very large distances.

% -------------------------------------------
\subsection{Section Summary}
% -------------------------------------------

%---------------------------------------------------------------
\begin{figure}
\begin{center}
\includegraphics[width=\textwidth]{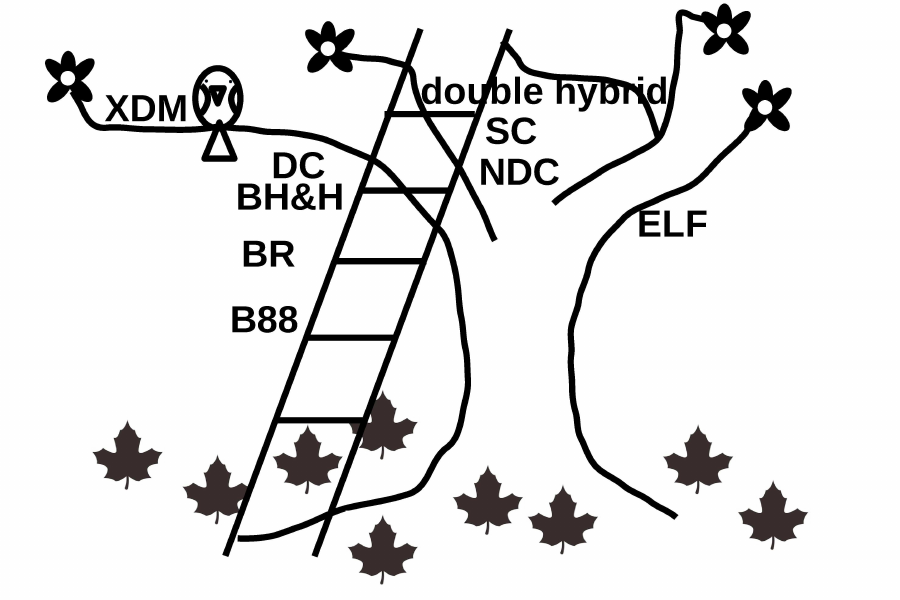}
\end{center}
\caption{
Axel's tree:  His functionals roughly climb Perdew's ladder from GGAs
to mGGAs to double hybrids, but following his physical intuition as
much as any strict mathematical principles and occasionally going off
on side branches.  I was unable to resist putting a few maple leaves 
to represent Axel's deep roots in Canada.
\label{fig:AxelTree}
}
\end{figure}
%---------------------------------------------------------------
This section shows how Axel has climbed Perdew's ladder from GGAs to mGGAs
to double hybrids.  But, together with his students, he has also added
van der Waals corrections based upon the xc-hole multipole moments in the
XDM.  This is not an obvious part of Perdew's turn of the century vision
of DFT.  {\bf Figure~\ref{fig:AxelTree}} is provides an extremely 
approximate picture of how Axel's functionals correspond to Perdew's ladder.

Axel did not restrict himself to functionals with obvious
functional derivatives, but rather included the solution of nonlinear equations
and taking the minima over several parameters in his energy functionals.
His physical intuition and his (non-self-consistent) results are impressive,
justifying his calling his work ``density-functional {\em theory}'' as 
opposed to just ``density-functional fits'' [{\bf B2022}, {\bf B2022b}]
because every term is based upon an underlying physical idea.
% Unfortunately the part of Axel's work based upon the BR functional
% (or inverting the BR procedure) is unlikely to be incorporated into 
% distributed programs until someone figures out how to calculate 
% the functional derivatives needed for self-consistency and analytical
% derivatives needed for automated geometry optimizations.  But this is a
% technical problem which is undoubtedly easier to solve than is having the
% physical intuition and creating the initial functionals.  
Axel himself had so many ideas for functionals that he seemed never to 
have the time to implement all of them self-consistently.  Indeed finding
the functional derivatives was  not {\em a priori} evident for many of
Axel's functionals. Martin Kaupp has written to me that he and
Alexei Arbuznikov were inspired by a talk that Axel on his B05 functional 
given at the World Association of Theoretical and Computational Chemists (WATOC) 
in Lugano, Switzerland, in 2002:
\begin{quote}
\noindent
``Axel gave the first presentation on what would become
B05 (the initial paper probably should be termed B03). [...]
In his usual clear and didactical way, where things seem simple initially,
until you think about some deeper complications, Axel layed out his idea of a 
coordinate-space model of static correlation. [...] Alexei and I were immediately 
intrigued and went to talk with Axel during the coffee break. We asked him about 
a self-consistent implementation of the approach, but due to the complicated and
numerically demanding form of the reverse Becke-Roussel machinery involved in the 
estimate of local exchange-hole normalizations, Axel said he shudders to even think 
about an SCF implementation.''
\end{quote}
Fortunately this technical problem has been worked on
and solved using various techniques, so that there are now self-consistent
implementations of the B88cBR and B05 functionals in 
{\sc Q-Chem} \cite{PGK2008,PLSK2012} and {\sc ReSpect} \cite{AK2006,AK2009}
and of the B13 functional 
in {\sc TurboMol} \cite{WAK2021}.  
Hence enough technical problems have been surmounted that many of Axel's more recent
ideas are becoming part of some general all-purpose quantum chemistry codes.

This is neither the place to go into even more technical details than the ones already
given nor to discuss too much of other people's work.  Nevertheless, it does seem
approriate to mention work on functionals for strong correlation that has focused on
overcoming the ``zero-sum game'' \cite{JPK+2017,J2021,KWA+2024}.  To quote the Wikipedia
\cite{zerosum}, ``Zero-sum game is a mathematical representation in game theory and 
economic theory of a situation that involves two competing entities, where the result 
is an advantage for one side and an equivalent loss for the other.''  In constructing
hybrid DFAs, there is also a zero-sum trade-off.  To quote Ref.~\cite{J2021}: 
``Model 1 `exact exchange' is exact for one-electron systems and accurate for radicals,
but underbinds ccovalent bonds.  Model 2 `local exchange-correlation' fixes this
underbinding, but over-delocalizes electrons and breaks exact constraints.  Their 
combination gives a pervasive and resilient zero-sum tradeoff between performance for
different properties.'' Axel Becke's ideas are proving critical for getting around the 
zero-sum game.

% =================================================
\section{Some Miscellaneous Topics}
\label{sec:misc}
% \input{misc.tex}

% ===============================================================
% File misc.tex .
% Last modified: 14 July 2026
% ===============================================================

There are some topics studied by Axel which do not fit into the
larger story told here, but which are nevertheless important contributions
in their own way.  Let us look at a few of them.

% -------------------------------------------
\subsection{Conceptual DFT}
% -------------------------------------------

\begin{quote}
\noindent
{\sf
Visualisation of bonds is a chemist's dream.  Axel's ELF revealed the
expected bonds and lone pairs from only the density and the kinetic
energy density, but it could also be applied to molecules such as
alkali metal clusters where the nature of the bonding was less clear.
--- MEC
}
\end{quote}

Conceptual DFT focuses on finding quantities which may be used to 
extract ``chemical information'', such as the location of bonds, from
DFT.  
In [{\bf BE1990}], together with his postdoc Edgecombe, presented
an interesting by-product of their work on DFAs.  This was the
electron localization function (ELF\index{ELF}).  In particular,
\begin{eqnarray}
  \mbox{ELF} & = & \frac{1}{1+\chi_\sigma^2} \nonumber \\
  \chi_\sigma & = & \frac{D_\sigma}{D^0_\sigma} \nonumber \\
  D_\sigma & = & \tau_\sigma - \frac{1}{4} \frac{\vert {\vec \nabla} \rho_\sigma
  \vert^2}{\rho_\sigma} \nonumber \\
  D_\sigma^0 & = & \frac{3}{5} \left( 6 \pi^2 \right)^{2/3} \rho_\sigma^{5/3}
  \, .
  \label{eq:ELF.1}
\end{eqnarray}
Note that $D_\sigma^0$ is just $D_\sigma$ for a HEG of density $\rho_\sigma$.
The ELF function does something fascinating: Starting only from the 
charge density and from the local kinetic energy, the ELF shows where
in space electrons tend to pair up to form bonds and lone pairs.  This
was further explored in [{\bf SBF+1991}].  Sam Trickey has pointed out
to me that $\chi_\sigma$ is basically the iso-orbital indicator \cite{FS2019},
used in ``essentially all modern meta-GGA [xc] DFAs as well ... as the 
enhancement factor in orbital free [kinetic energy] KE densities.''

Two other interesting functions
of the density matrix were later presented by Schmider and Becke
[{\bf SB2000}] and [{\bf SB2002}].  These are the localized-orbital locator 
(LOL\index{LOL}) and the parity function (P\index{P}).
Obtaining information about spin-contamination from the density is also an 
interesting problem to which Axel has made an important contribution
[{\bf WBS1995}].
See also [{\bf KBP1996}].

% -------------------------------------------
\subsection{Current Density Functionals}
% -------------------------------------------

The current density,
\begin{equation}
  {\vec J}_\sigma = -\frac{i}{2} \sum_k n_k \left( 
  \psi_{k\sigma}^* {\vec \nabla} \psi_{k\sigma} - \psi_{k\sigma}
  {\vec \nabla} \psi_{k\sigma}^* \right) \, ,
  \label{eq:j.1}
\end{equation}
arises naturally in the context of TD-DFT \cite{RG1984} 
(Ref.~\cite{U2013} provides a very complete review 
of TD-DFT).  In particular, Vignale has done extensive work developing 
current DFT \cite{VR1987,VR1988}. Very interestingly, the current
density has a place in DFAs for ordinary (time-independent) ground-state DFT.

An old problem for DFAs applied to open-shell atoms has been 
that degenerate states may have different densities.  For example,
it is well known that DFAs applied to the $2p_x$, $2p_y$, and $2p_z$ densities
will give the same energies but will give different energies when applied
to the $2p_{-1}$, $2p_0$, and $2p_{+1}$ densities.  Axel examined the
effects of including a current density, in DFAs in [{\bf B1996}].  
In [{\bf B2002}], Axel shows that including the current density in DFAs can 
help reduce energy differences between degenerate states with different 
densities.  Notice how the inclusion of the current density seems to add a 
new rung (or perhaps a new branch?) to Perdew's ladder!

The B3LYP functional is known to be less accurate for transition metals
than for purely organic compounds.  This may be partly because of the
greater availability of organic compounds in the training set or because
of the greater complexity  of transition metal chemistry due to the different
ways to fill the $d$ orbitals.  [{\bf JDB2007}] shows that the current
density can reduce the problem of different energies from different
fillings of the $d$ orbitals. 
% Thereafter more care was made to include
% transition metals in training sets when developing new functionals 
% {\bf JB2009}, {\bf B2013b}, {\bf JB2017}.

% -------------------------------------------
\subsection{Excited States}
% -------------------------------------------

Since I have a long-standing interest in methods for treating electronic
excited states and particularly from time-dependent (TD\index{TD}) DFT via
response theory, I was interested to find that Axel was also making 
contributions in this area without TD-DFT.  
(Refs.~\cite{C1995b,C2009,CH2012,AJ2013,U2013} provide reviews of TD-DFT).
Axel's contributions came in different ways and were often natural outgrowths
of other ideas on which he had been working.

\paragraph{Strong Correlation and Fractional Occupation Numbers}
Axel usually focused only on the ground state.  However the potential 
energy surfaces of ground and excited states are often intricately 
intertwined.  This is notably true when using response theory to obtain
information about excited states \cite{C1995b} because the quality of 
excitation energies depends upon the quality of the description of the
ground state.  From another point of view, many important features on the
ground state potential energy surface may be seen as coming from interactions
with excited state potential energy surfaces.  This includes avoided crossings
in the potential energy curves of diatomic molecules and conical intersections
in polyatomic molecules.  These tend to be areas where diabatic states (which
typically have well defined electronic configurations) undergo configuration
mixing to form photochemical funnels in the manifold of potential energy
surfaces.  This means that they are also often regions where bonds are
formed and/or broken, which is typical signature of strong correlation.
In fact, it is difficult for me to imagine how transition states on the
ground potential energy surface can arise without some type of interaction
with a low-lying excited state.

Hence it follows quite naturally that Axel's work on strong correlation
and his study of transition states would lead to work on excited states 
as well.  Our own work, as well as previous work by others, had led us
to wonder about the use of fractional occupation numbers for describing
strong correlation.  We summarized our arguments regarding fractional
occupation numbers in formal DFT on pages 297-298 of Ref.~\cite{CH2012}
when we discussed the problem of non-interacting $v$-representability
(NVR\index{NVR}). In brief, not every interacting system has a density
which corresponds to the ground state of a non-interacting system (i.e., 
is NVR).  But, if a density is NVR, formal Kohn-Sham (KS\index{KS}) DFT 
then requires that the highest partially occupied orbitals all have the 
same fractional occupation or else the xc-potential will become 
orbital-dependent, which is a violation of the original formulation of 
KS DFT.  However Axel's most recent work is typically based upon generalized
KS DFT where there is some orbital dependence and hence fractional 
occupation numbers.  Articles [{\bf B2013},{\bf B2015b}] show that
there is much to be gained with the B13 functional by adjusting the
fractional occupation numbers so as to minimize the energy.  In particular,
the description of transition states and charge transfer is markedly improved
for the ground state.  Axel then goes on to combine this approach with 
a small configuration interaction WFT calculation to obtain important
excited states in ethylene [{\bf B2015}].  I find this very interesting, 
though it was never developed into a systematic approach for calculating
excited states.

\paragraph{Revisiting the Ziegler-Rauk-Baerends Multiplet Sum Method}
Another path that led Axel to excited states was the possibility of using
the adiabatic connection formalism to justify and improve upon the 
Ziegler-Rauk-Baerends multiplet sum method for describing the open-shell singlet/triplet
splitting energy in multiplets [{\bf B2016}].  The same problem is approached
using the virial theorem in [{\bf B2018a}] and re-examined again in 
[{\bf B2018b}] and in [{\bf B2019}].  
[{\bf FBJ2018}] shows that CT
excitations are better described with Axel's virial theorem approach
than with TD-DFT when the B3LYP functional is used.
This virial exciton model has been applied
both to piezoelectricity [{\bf FBJ2020}] and to photoluminescence 
[{\bf FBJ2021}].
Of course, the atomic multiplet problem which Axel attacked using the current 
density is also an example of work involving excited states.

% =================================================
\section{Conclusion}
\label{sec:conclude}
% \input{conclude.tex}
% ===============================================================
% File conclude.tex .
% Last modified: 7 July 2026
% ===============================================================

\begin{quote}
\noindent
{\sf
As I think back, one image of Axel sticks with me --- a photo of Axel
at Queen's in front of his workstation.  Axel was an ``inventor of new
ideas''
and he tested his ideas using his very own experimental apparatus ---
initially his program for diatomics, then {\sc NUMOL} for polyatomic
molecules, and lastly sometimes {\sc NUMOL} fed with data generated
by other programs such as {\sc Gaussian}.  He did not need a big 
computer system for most of his work.  He concentrated on ideas
for new density-functional approximations that captured the physics
he had in his head.  
Modest computational resources sufficed for testing
most of his ideas.  
Often the ideas came so fast, that Axel rarely
had time to implement them self-consistently before testing.
He co-authored very few application papers and
had a very large percentage of single-author papers.  Not a computational
chemist, he was a true theoretician.
--- MEC
}
\end{quote}

The title of this manuscript was inspired by many things, but there is
one in particular that I would like to describe here:  I was at a conference 
and a speaker began his talk by telling a little story about having been 
in the library where he got into a conversation with a chemist.  Somehow 
the conversation turned to DFT about which the chemist claimed to know 
little or nothing.  However the very same chemist explained that he did do 
calculations using a marvelously successful new method called B3LYP!
% how he had been in the 
% library and someone had asked him about the conference or about what he did
% (I forget which).
% So Mel started to explain about DFT but he did not get very far before the
% person who asked the question interrupted with the comment that he ``did not
% know anything about DFT but he did use B3LYP.''  
It is somehow ironic to me
that Axel is known primarily for the articles that I boxed: \framebox{\bf B1988}
and \framebox{\bf B1993}.  Sure, these are not only his most cited articles, but
also some of the most cited articles in both the chemistry and physics literatures.
However focusing too much on these two articles ignores the fact that year after
year, paper after paper, Axel was making incremental contributions to improving
DFAs.  Taken together, we need to multiply his impact on the field by some
large factor whose size I cannot even guess.  Axel was a creator who learned
from others and discovered much on his own from which others learned.  

I hope that this article has helped to convince you that Axel was indeed
the right person in the right place at the right time.  According to Axel,
\begin{quote}
\noindent
``I attended my first DFT conference in 1983, a two week NATO workshop in Alcabideche, 
Portugal.  The intensity of the debates and the level of scientific excitement
at the meeting were impressive.  I was hooked.'' --- [{\bf B2014}]
\end{quote}
We know that cannot be quite the whole story as we know that Axel 
``was determined
to be an engineer some day'' \cite{Queens}
% saw himself intially as an Engineering Physicist 
and perhaps also an inventor.
\begin{quote}
\noindent
``To invent, you need a good imagination and a pile of junk.'' --- Thomas A.\ Edison
(as quoted in Ref.~\cite{A1998})
\end{quote}
He defended his Ph.D.\ thesis at McMaster on ``Numerical Hartree-Fock-Slater 
Calculations on Diatomic Molecules'' in July 1981.  
Concretely this meant that Axel had already found his ``pile of junk'' and
had built a fundamental apparatus for investigating DFT.  
Now the NATO workshop was inspiring him to go further in his inventing. 
Somehow he ended up in a chemistry department, rather than a physics 
department, so I also feel free to quote a very well-known chemist:
\begin{quote}
\noindent
``A detective with his murder mystery, a chemist seeking the structure of a new
compound, use little of the formal and logical modes of reasoning.  Through a 
series of intuitions, surmises, fancies, they stumble upon the right explanation,
and have a knack of seizing it when it once comes within reach.'' --- G.N.\ Lewis
(Chapter 1, page 6 of Ref.~\cite{L1926})
\end{quote}
Axel had a mystery to solve, making better and better functionals {\em and} he had
a program that he could use to study them.  Later, with Ross Dickson, they would create
the basis-set free numerical program {\sc NUMOL} for studying improved DFAs in polyatomic 
molecules.  This allowed Axel to convince others that he had reached the basis set limit
for the DFAs and so could provide reliable tests.  But Axel also did not lack any of that 
imagination of which Edison spoke nor that series of ``intuitions, surmises, fancies'' of 
which Lewis spoke.  And this allowed Axel's functionals to overcome much of the overbinding
problem with the LDA.  The new GGAs of Axel (and others!) convinced 
 John Pople to put DFT in {\sc Gaussian} and Nicholas Handy to 
put DFT in {\sc CADPAC}, both in 1992.  Probably
this was also the right moment for DFT as well because {\em ab initio} quantum chemistry,
while working wonderfully well for molecules in the gas phase, was being increasingly asked
to look at larger molecules of chemical interest.  DFT was ready and able to respond to this
call --- enough so, that DFT has largely replaced Hartree-Fock calculations as the basic workhorse of
quantum chemistry.  DFT still sometimes gave incorrect answers, but it was no longer 
condemned the way X$\alpha$ was once condemned.  Instead it was understood 
that the same DFAs that allow us to extend the accuracy of {\em ab initio}
calculations to larger molecules do sometimes fail and somehow that is OK,
perhaps because these failures are increasingly well understood, allowing
us to fix or at least avoid the most problematic cases.

As emphasized by a remarkable recent article co-written by 70 workers in 
the field \cite{THS+2022}, much has been accomplished
in DFT, but the area remains very active because many questions remain and
many problems still remain to be solved.
Axel mentioned regrets in Ref.~[{\bf B2014}].  His initial work was
based upon atoms and only used atoms to obtain the parameters needed in his functionals.
Axel regrets ever having started parameterizing using molecules as there are then so many more
molecules and molecular properties which can be fit.  He also preferred an OOO
approach [{\bf B2014}] which may be going by the way side as the 
highest rung of Jacob's ladder is being explored and, ``The floodgates are open now.  The 
virtual orbitals are pouring in'' [{\bf B2014}].  But also Axel mentioned many victories in 
Ref.~[{\bf B2014}] over dispersion and strong correlation (these have been discussed above).
John Perdew wrote to me that he had observed a shift in Axel's functionals:
\begin{quote}
\noindent
``[Axel's] early density functionals had few parameters and were shaped
more by physical insight.  At mid-career, he developed functionals with many
empirical parameters fitted to molecular energy differences.  But near the end
of his career he returned to few-parameter functionals, in which each parameter
could be guessed and was only fitted as a sort of final refinement (e.g., B2022).''
\end{quote}
Yes! the floodgates had opened, but as time went on, Axel began to close the
floodgates.

People in electronic structure are now battling with what more and more feel to be the
next frontiers: quantum computing (QC\index{QC}) and artificial intelligence (AI\index{AI}).  
AI is already here.  Deep neural networks have been used to build new DFAs such as
DM21 (deep mind 2021) \cite{KMT+2021} and Skala \cite{KSKW2026} to mention only two
of several cases.  Such DFAs typically involve both an analytic part and an AI part where
the machine-learning part takes key DFT quantities, such as those introduced by Axel
and others as measures of different types of correlation, as input from multiple places
in molecules, thus establishing global or semi-global relations between local or semi-local
DFT quantities as additional information for calculating the xc-energy.  
% The problem of 
% ``analytic derivatives'' of AI functionals also has to be solved to make these new
% functionals practical for routine calculations.  Mostly AI is not so much improving
% DFT so much as 
AI is also increasing the use of DFT by using DFT to train neural networks to obtain increasingly 
realistic force fields or guiding DFT calculations as they search critical parts of 
potential energy surfaces.  QC is not yet something to which many have access, but it 
has already set more than one researcher dreaming of how quantum computing calculations 
could improve quantum electronic structure calculations!  
% Could some future DFAs be composed of qubits?  
In 2024, IBM corporation was granted the patent US11941192B2
for ``Density-functional theory determinations using a quantum computing system''
\cite{IBM}.

It seems appropriate to end with some words from Axel about how he saw the future of 
research and research funding:
\begin{quote}
\noindent
``Small scale, independent research can survive only under the protective umbrella of 
broad and democratic funding such as NSERC [Natural Science and Engineering 
Research Council of Canada] has provided.  It is a mistake to believe, as many do, 
that scientific progress can best be made by strategic funding of target-driven 
cooperative networks.  Discoveries and innovations are serendipitous and unforeseen.  
I sincerely hope that NSERC will continue to nurture the explorations of our young and 
our independent Canadian university researchers.'' 
\cite{Queens}
\end{quote} 
I hear the echo of a (justifiably) proud Axel Becke in these words but also a cry for
the funding of small individual research which I would like to believe
is not specific to Canada.
% =================================================
\section*{Acknowledgements}
\label{sec:thanks}
% \input{thanks.tex}

% ===============================================================
% File thanks.tex .
% Last modified: 9 July 2026
% ===============================================================

In trying to describe Axel's career and provide anecdotes showing 
what Axel has meant to many of us, I have certainly written an article
with many aspects of a historical article.  I would therefore like to
begin by acknowledging Aaran J.\ Ihde whose wonderful book on the history
of modern chemistry \cite{I1964} perked my interest in and made
me feel the importance of the history of science when I first picked
up a copy in a used book store during my undergraduate years.
I have been a long-time member of the Division of the History of Chemistry
of the American Chemical Society and had the pleasure of presenting
the Divisions' Medal for a Citation for a Chemical Breakthrough to the
French Academy of Sciences \cite{Lavoisier}.  In writing this article,
I may have fallen short of what is expected of history students \cite{MP2015,S2021}
or fallen into any one of the many subtle traps in which historians can
easily find themselves \cite{F1970}.  Writing history well begins with getting 
good information, requires a lot of thought, and also requires others who 
are willing to help through fact checking and critical feedback.  This 
necessarily involves a great number of people and this work is no exception to that rule.
I trust that there is much good in this article but I am sure that there
are also shortcomings for which I take full responsibility.

I am grateful to a number of people for their permissions and comments:  
Neepa Maitra and Sam Trickey are thanked for comments that helped me
to improve the Focus volume description appearing at Ref.~\cite{ElecStruc}.
Sam Trickey is the only person that I have met who knew John Slater personally
\cite{T2010} and I have greatly profited from Sam's recollections.  
Sam is also to be credited for reminding me of somethng that I already 
knew, namely 
that LCAO X$\alpha$ programs pre-existed Axel's
work.  Richard Florizone, then president of Dalhousie University, wrote a 
wonderful letter nominating Axel Becke for the Herzberg medal and summarizing 
the Axel's career.  His permission to let me use information from his nomination
letter saved me from having to do a certain amount of bibliographical work.  
Erin Johnson was very helpful in letting me know about other activities
planned to honor Axel.  Mary Anne White did a marvelous job in providing
photos of Axel and in pointing out to me key references, notably 
Ref.~\cite{Queens}.  Ross Dickson helped me to fill in some missing information.
Filipp Furche is thanked for providing a copy of an unpublished manuscript
written by Axel.  Evert Jan Baerends is thanked for providing me with a preprint
of his manuscript of memories of Axel Becke (part of Ref.~\cite{NBB+2026}).
Russell Boyd is thanked for sharing a preprint of the entire 
Ref.~\cite{NBB+2026} and Sture Nordholm is thanked for his explanation of
the origin of this reference.
David Tozer is thanked for 
clarifying the relation between Nicholas Handy and John Pople, supplying 
copies of 
Refs.~\cite{CCS2004,HTL+1993}, and the photo of Pople, Handy, and himself. 
Andreas Savin and Emil Proynov are thanked for 
numerous discussions.
Russell Boyd is thanked for clarifying the authorship of Axel's
Dalhousie University obituary.  Arka Prava Sarkar is thanked for helping
me obtain copies of Refs.~[{\bf SB2000},{\bf JB2009},{\bf BAK2010}];
Pratyush Bhattacharjya is thanked for helping me obtain a copy of 
[{\bf B1994}]  Walid Taouali is thanked for helping me obtain copies
of Refs.~[{\bf B1994},{\bf B1996},{\bf B2015b}].
Bruce Milne is thanked for helping me obtain an electronic copy of 
Ref.~\cite{C1992}.
Adnane Aouidate and Andr\'e Farias de Moura are also thanked for helping
me to obtain electronic copies of certain articles.
% Mel Levy is thanked for a useful discussion.  
Martin Kaupp is thanked for reading the manuscript with particular attention to
Sec.~\ref{sec:ladder} and for providing me both with an update on his recent
work and on how Axel helped to inspire that work.
Erin Johnson is thanked informing me of students and postdocs of Axel since
2015.
I am grateful 
to Russell Boyd,
to Henry Chermette,
to Erin Johnson,
to Melvyn P.\ Levy,
to Bharathi Natarajan,
to John Perdew,
to Emil Proynov, 
to Lucia Reining,
to Andreas Savin, and
to Sam Trickey for taking the time to read and comment 
on the entirety of various versions of this manuscript.  
Sam Trickey is also thanks for preparing a 3 page document n
answer to some of my questions, entitled
``Some Personal Remarks on DFT History'' (and which he calls ``a considerable
revision of remarks shared with Bob Jones when he was writing'' 
Ref.~\cite{J2015}).
To some extent, many of these people could be considered as partial
co-authors.  Certainly this is a better manuscript because of them!

%%%%%%%
% EOF %
%%%%%%%
% ---------------------------------------
\appendix
% ----------------------------------------
\section{Axel Becke's Publication List and a Few Key Life Events}
\label{sec:CVaxel}
% \input{CVaxel.tex}

% ===============================================================
% File CV.tex .
% Last modified: 6 July 2026 
% ===============================================================

\small

This appendix contains a list of Axel Becke's publications
punctuated by key events, including degrees, students and postdocs, 
and some of the many awards that he received.

\begin{center}
\rule{\textwidth}{1pt}\\
{\bf 23 October 2025}  Axel passed away in Halifax, Nova Scotia, Canada.\\
\rule{\textwidth}{1pt}
\end{center}

\begin{description}
% 1
\item[B2024] A.D.\ Becke, ``A remarkably simple dispersion damping scheme 
and the DH24 double hybrid density functional'', {\em J.\ Chem.\ Phys.}\
{\bf 160}, 204118 (2024). % {\color{red} 8 citations} 
% {\color{green} 45/91} % \cite{b2024}
% 2
\item[B2023] A.D.\ Becke, ``Doubling down on density-functional theory'',
{\em J.\ Chem.\ Phys.}\ {\bf 159}, 241101 (2023). % {\color{red} 8 citations}
% {\color{green} 44/90} % \cite{b2023}
% 3
\item[BSM2023] A.D.\ Becke, G.\ Santra, and J.M.L.\ Martin, ``A double-hybrid 
density functional based on good local physics with outstanding performance 
on the GMTKN55 database'', {\em J.\ Chem.\ Phys.}\ {\bf 158}, 151103 (2023).
% {\color{red} 18 citations} {\color{green} 43/89} % \cite{bSM2023}
% 4
\item[B2022b] A.D.\ Becke, ``Density-functional theory vs density-functional 
fits: The best of both'', {\em J.\ Chem.\ Phys.}\ {\bf 157}, 234102 (2022).
% {\color{red} 14 citations} {\color{green} 43/88} % \cite{b2022b}
% 5
\item[B2022] A.D.\ Becke, ``Density-functional {\em theory} vs 
density-functional fits'', {\em J.\ Chem.\ Phys.}\ {\bf 156}, 214101 (2022).
% {\color{red} 46 citations} {\color{green} 42/87} % \cite{b2022}
% 6
\item[FBJ2021] X.\ Feng, A.D.\ Becke, E.R.\ Johnson, ``Theoretical 
investigation of polymorph- and coformer-dependent photoluminescence in 
molecular crystals'', {\em CrystEngComm} {\bf 23}, 4264 (2021).
% {\color{red} 13 citations} 
% {\color{green} 41/86} % \cite{FbJ2021}
% 7
\item[FBJ2020] X.\ Feng, A.D.\ Becke, and E.R.\ Johnson, ``Computational 
modeling of piezochromism in molecular crystals'', {\em J.\ Chem.\ Phys.}\
{\bf 152}, 234106 (2020).  % {\color{red} 4 citations} 
% {\color{green} 41/85} % \cite{FbJ2020}
% 8
\item[AID+2020] E.\ Awoonor-Williams, W.C.\ Isley, S.G.\ Daleo, 
E.R.\ Johnson, H.B.\ Yu, A.D.\ Becke, B.\ Roux, and C.N.\  Rowley,   
``Quantum Chemical Methods for Modeling Covalent Modification of 
Biological Thiols'', {\em J.\ Comp.\ Chem.}\ {\bf 41}, 427 (2020).
% {\color{red} 41 citations} {\color{green} 41/85} % \cite{AID+b2020}
\end{description}

\begin{center}

\rule{\textwidth}{1pt}\\
{\bf 2021} X.\ Feng obtains his Ph.D.\ thesis, co-supervised by E.R.\ Johnson
and A.D.\ Becke: ``A Novel Theoretical Approach to Model Electronic 
Excitations in Molecular Crystals''. \cite{F2021} \\
\rule{\textwidth}{1pt}\\
{\bf 2019} S.G.\ Dale finishes working with Axel as a postdoctoral
fellow/research associate.
\rule{\textwidth}{1pt}

\end{center}

\begin{description}
% 9
\item[B2019] A.D.\ Becke, ``Dependence of the virial exciton model on basis 
set and exact-exchange fraction'', {\em J.\ Chem.\ Phys.}\ {\bf 150},
241101 (2019).  % {\color{red} 4 citations} {\color{green} 41/84} % \cite{b2019}
% 10
\item[FBJ2018] X.\ Feng, A.D.\ Becke, E.R.\ Johnson, ``Communication: Becke’s 
virial exciton model gives accurate charge-transfer excitation energies'',
{\em J.\ Chem.\ Phys.}\ {\bf 149}, 231101 (2018).
% {\color{red} 10 citations} {\color{green} 40/83} % \cite{FbJ2018}
% 11
\item[DBJ2018] S.G.\ Dale, A.D.\ Becke, and E.R.\ Johnson, ``Density-functional 
description of alkalides: introducing the alkalide state'', {\em Phys.\ Chem.\ 
Chem.\ Phys.}\ {\bf 20}, 26710 (2018).
% {\color{red} 10 citations} {\color{green} 40/82} % \cite{DbJ2018}
% 12
\item[LDT+2018] L.M.\ LeBlanc, S.G.\ Dale, C.R.\ Taylor, A.D.\ Becke,
G.M.\ Day, and E.R.\ Johnson, ``Pervasive Delocalisation Error Causes 
Spurious Proton Transfer in Organic Acid–Base Co-Crystals'',
{\em Ang.\ Chemie, Int.\ Ed.}\ {\bf 57}, 14906 (2018).
% {\color{red} 72 citations} {\color{green} 39/81} % \cite{LDT+2018}
% 13
\item[B2018b] A.D.\ Becke, ``Communication: Optical gap in polyacetylene 
from a simple quantum chemistry exciton model'', {\em J.\ Chem.\ Phys.}\
{\bf 149}, 081102 (2018).
% {\color{red} 13 citations} {\color{green} 39/81} % \cite{b2018b}
% 14 conference abstract
% 15
\item[BDJ2018] A.D.\ Becke, S.G.\ Dale, and E.R.\ Johnson, ``Communication: 
Correct charge transfer in CT complexes from the Becke'05 density 
functional'', {\em J.\ Chem.\ Phys.}\ {\bf 148}, 211101 (2018).
% {\color{red} 22 citations} {\color{green} 39/82} % \cite{bDJ2018}
% 16 conference abstract
% 17 conference abstract
% 18
\item[B2018a] A.D.\ Becke, ``Singlet-triplet splittings from the virial 
theorem and single-particle excitation energies'', {\em J.\ Chem.\ Phys.}\
{\bf 148}, 044112 (2018).
% {\color{red} 24 citations} {\color{green} 39/81} % \cite{b2018a}
% 19
\end{description}

\begin{center}

\rule{\textwidth}{1pt}\\
{\bf 2017} S.G.\ Dale begins work with Axel as a postdoctoral
fellow/research associate.\\
\rule{\textwidth}{1pt}

\end{center}

\begin{description}
\item[DJB2017] S.G.\ Dale, E.R.\ Johnson, and A.D.\ Becke, ``Interrogating 
the Becke'05 density functional for non-locality information'',
{\em J.\ Chem.\ Phys.}\ {\bf 147}, 154103 (2017).
% {\color{red} 15 citations} {\color{green} 38/80} % \cite{DJb2017}
% 20
\item[JB2017] E.R.\ Johnson and A.D.\ Becke, ``Communication: DFT treatment 
of strong correlation in $3d$ transition-metal diatomics'', {\em J.\ Chem.\ 
Phys.}\ {\bf 146}, 211105 (2017).
% {\color{red} 44 citations} {\color{green} 38/79} % \cite{Jb2017}
% 21
\item[B2017] A.D.\ Becke, ``Forward'', in {\em Non-Covalent Interactions in 
Quantum Chemistry and Physics}, ed.\ A.\ Otero-de-la-Roza and G.A.\ DiLabio
(Elsevier: Amsterdam, Netherlands, 2017).
% {\color{red} 0 citations} {\color{green} 38/78} % \cite{b2017}
% 22
\item[B2016] A.D.\ Becke, ``Vertical excitation energies from the 
adiabatic connection'', {\em J.\ Chem.\ Phys.}\ {\em 145}, 194107 (2016).
% {\color{red} 22 citations} {\color{green} 37/77} % \cite{b2016}
% 23 abstract
% 24
\end{description}

\vspace{0.25cm}

\begin{center}

\rule{\textwidth}{1pt}\\
{\bf 2016} Axel awarded the Canada Council Killam Prize in Natural Sciences.\\
\rule{\textwidth}{1pt}\\
{\bf 2016} X.\ Feng begins doctoral studies with Erin and Axel\\
{\bf 2015} Axel retires as Professor and Killam Chair in Computational Science,
Harry Shirreff Professor of Chemical Research, Department of
Chemistry, Dalhousie University, Halifax, Nova Scotia, Canada and is granted
Emeritus status.\\
\rule{\textwidth}{1pt}\\
{\bf 2015} Axel is Awarded the Gerhard Herzberg Canada Gold Medal for Science and
Engineering, Canada's highest award in science, part of the 1 million
Canadian dollar award was used to establish the Herzberg-Becke Chair
in Theoretical Chemistry at Dalhousie University which allowed Prof.\
Erin Johnson to return to Canada after her ``exile'' as a faculty member 
at the University of Califonria in Merced, U.S.A.\\
\rule{\textwidth}{1pt}\\
{\bf 2015}  Axel is awarded the Chemical Institute of Canada Medal.\\
\rule{\textwidth}{1pt}

\end{center}

\begin{description}

\item[B2015] A.D.\ Becke, ``Excited-state surfaces of ethylene from the B13 
strong-correlation density functional'', {\em Mol.\ Phys.}\ {\bf 113},
1884 (2015).  % {\color{red} 6 citations} {\color{green} 36/76} % \cite{b2015}
% 25
\item[B2015b] A.D.\ Becke, ``Fractional Kohn-Sham occupancies from a 
strong-correlation density functional'', in {\em Density Functionals:
Thermochemistry}, edited by E.R.\ Johnson, page 175, volume {\bf 365}
of the {\em Topics in Current Chemistry Series} (Springer: 2015).
% {\color{red} 7 citations} {\color{green} 35/75} % \cite{b2015b}

\end{description}

\begin{center}

\rule{\textwidth}{1pt}\\
{\bf 2014} Axel is awarded the Theoretical Chemistry Award of the American
Chemical Society.\\
\rule{\textwidth}{1pt}

\end{center}

\begin{description}

% 26
\item[B2014] A.D.\ Becke, ``Perspective: Fifty years of density-functional
theory in quantum chemistry'', {\em J.\ Chem.\ Phys.}\ {\bf 140}, 18A301
(2014).  % {\color{red} 1193 citations} {\color{green} 34/74} % \cite{b2014}
% 27 ACS Award in Theoretical Chemistry
% 28
\item[B2013] A.D.\ Becke, ``Communication: Two-determinant mixing with
a strong-correlation density functional'', {\em J.\ Chem.\ Phys.}\
{\bf 139}, 021104 (2013).  % {\color{red} 21 citations} 
% {\color{green} 33/73} % \cite{b2013}
% 29
\item[B2013b] A.D.\ Becke, ``Communication: Calibration of a strong-correlation
density functional on transition metal atoms'', {\em J.\ Chem.\ Phys.}\
{\bf 138}, 161101 (2013).  % {\color{red} 24 citations} 
% {\color{green} 32/72} % \cite{b2013b}
% 30
\item[B2013c] A.D.\ Becke, ``Density functionals for static, dynamical,
and strong correlation'', {\em J.\ Chem.\ Phys.}\ {\bf 138}, 074109 (2013).
% {\color{red} 159 citations} {\color{green} 31/71} % \cite{b2013c}
\end{description}

\begin{center}

\rule{\textwidth}{1pt}
{\bf 2013} F.O.\ Kannemann obtains his 
Ph.D.\ Thesis with Axel: Development and benchmarking of a semilocal 
density-functional approiximation including dispersion \cite{K2013}.
\rule{\textwidth}{1pt}\\
{\bf 2012} Alya A.\ Arabi obtains her 
Ph.D.\ Thesis: Density functional theory: Dispersion interactions and
biological interactions (co-supervised by Axel Becke and Cherif Matta)
\cite{A2012}. \\
\rule{\textwidth}{1pt}\\

\end{center}

\begin{description}
% 31 
\item[AB2012] A.A.\ Arabi and A.D.\ Becke, ``Assessment of the PW86+PBE+XDM
density functional on van der Waals complexes at non-equilibrium geometries'',
{\em J.\ Chem.\ Phys.}\ {\bf 137}, 014104 (2012). %  {\color{red} 19 citations} 
% {\color{green} 30/70} % \cite{Ab2012}
% 32 Abstract
% 33

\item[KB2012] F.O.\ Kannemann and A.D.\ Becke, ``Atomic volumes and 
polarizabilities in density-functional theory'', {\em J.\ Chem.\ Phys.}\
{\bf 136}, 034109 (2012).  % {\color{red} 33 citations} 
% {\color{green} 30/69} % \cite{Kb2012}
% 34 Abstract
% 35
\item[BAK2010] A.D.\ Becke, A.A.\ Arabi, and F.O. Kannemann, 
``Nonempirical density-functional theory for van der Waals interactions'',
{\em Can.\ J.\ Chem.}\ {\bf 88}, 1057 (2010).
% {\color{red} 28 citations} {\color{green} 30/68} % \cite{bAK2010}
% 36 Abstract
% 37
\item[KB2010] F.O.\ Kannemann and A.D.\ Becke, ``Van der Waals interactions
in density-functional theory: Intermolecular complexes'', {\em J.\ Chem.\
Theory Comput.}\ {\bf 6}, 1081 (2010).  % {\color{red} 154 citations} 
% {\color{green} 30/67} % \cite{Kb2010}
% 38
\item[B2009] A.D.\ Becke, ``A density-functional approximation for
relativistic kinetic energy'', {\em J.\ Chem.\ Phys.}\ {\bf 131},
244118 (2009).  % {\color{red} 5 citations} 
% {\color{green} 30/66} % \cite{b2009}

\end{description}

\begin{center}

\rule{\textwidth}{1pt}\\
{\bf 2008} A.A.\ Arabi begins doctoral studies with Axel and with Cherif Matta.\\
\rule{\textwidth}{1pt}
{\bf 2007} E.R.\ Johnson obtains her
Ph.D.\ thesis with Axel: ``A density-functional theory including 
dispersion interactions''. \cite{J2007}
\rule{\textwidth}{1pt}
{\bf 2007} F.O.\ Kannemann begins doctoral studies with Axel.\\
\rule{\textwidth}{1pt}

\end{center}

\begin{description}

% 39 
\item[JB2009] E.R.\ Johnson and A.D.\ Becke, ``Tests of an 
exact-exchange-based density-functional theory on transition-metal
complexes'', {\em Can.\ J.\ Chem.}\ {\bf 87}, 1369 (2009).
% {\color{red} 26 citations} {\color{green} 29/65} % \cite{Jb2009}
% 40
\item[JBSD2009] E.R.\ Johnson, A.D.\ Becke, C.D.\ Sherrill, and G.A.\ DiLabio,
``Oscillations in meta-generalized-gradient approximation potential
energy surfaces for dispersion-bound complexes'', {\em J.\ Chem.\ Phys.}\
{\bf 131}, 034111 (2009).  % {\color{red} 162 citations} 
% {\color{green} 29/64} % \cite{JbD2009}
% 41 
\item[KB2009] F.O.\ Kannemann and A.D.\ Becke, ``Van der Waals interactions in 
density-functional theory: Rare-gas diatomics'', {\em J.\ Chem.\
Theory Comput.}\ {\bf 5}, 719 (2009).
% {\color{red} 142 citations} 
% {\color{green} 29/63} % \cite{Kb2009}
% 42 
\item[JB2008] E.R.\ Johnson and A.D.\ Becke, ``A unified density-functional 
treatment of dynamical, nondynamical, and dispersion correlations. II. 
Thermochemical and kinetic benchmarks'', {\em J.\ Chem.\ Phys.}\
{\bf 128}, 124105 (2008).  % {\color{red} 39 citations} 
% {\color{green} 29/62} % \cite{Jb2008}
% 43
\item[BJ2007] A.D.\ Becke and E.R.\ Johnson, ``Exchange-hole dipole
moment and the dispersion interaction revisited'', {\em J.\ Chem.\ Phys.}\
{\bf 127}, 154108 (2007).  % {\color{red} 290 citations} 
% {\color{green} 29/61} % \cite{bJ2007}
% 44
\item[BJ2007b] A.D.\ Becke and E.R.\ Johnson, ``A unified density-functional
treatment of dynamical, nondynamical, and dispersion correlations'',
{\em J.\ Chem.\ Phys.}\ {\bf 127}, 124108 (2007).
% {\color{red} 202 citations} {\color{green} 28/60} % \cite{bJ2007b}
% 45
\item[JDB2007] E.R.\ Johnson, R.M.\ Dickson, and A.D.\ Becke, ``Density
functionals and transition-metal  atoms'', {\em J.\ Chem.\ Phys.}\
{\bf 126}, 184104 (2007).  % {\color{red} 35 citations} 
% {\color{green} 28/59} % \cite{JDb2007}

\end{description}

\begin{center}

\rule{\textwidth}{1pt}\\
{\bf 2006} Axel is elected Fellow of the Royal Society (FRS) of London.\\
\rule{\textwidth}{1pt}\\
{\bf 2006} Axel leaves Queen's University, Kingston,
Ontario, Canada to become Professor and Killam Chair in Computational Science,
Harry Shirreff Professor of Chemical Research, Department of
Chemistry, Dalhousie University, Halifax, Nova Scotia, Canada.
\rule{\textwidth}{1pt}\\
{\bf 2006} R.M.\ Dickson finishes his position with Axel as postdoctoral fellow/research associate\\
\rule{\textwidth}{1pt}

\end{center}

\begin{description}
% 46 
\item[JB2006] E.R.\ Johnson and A.D.\ Becke, ``Van der Waals interactions
from the exchange hole dipole moment: Application to bioorganic benchmark
systems'', {\em Chem.\ Phys.\ Lett.}\ {\bf 432}, 600 (2006).
% {\color{red} 59 citations} {\color{green} 28/58} % \cite{Jb2006}
% 47
\item[BJ2006] A.D.\ Becke and E.R.\ Johnson, ``A simple effective potential
for exchange'', {\em J.\ Chem.\ Phys.}\ {\bf 124}, 221101 (2006).
% {\color{red} 1397 citations} 
% {\color{green} 28/57} % \cite{bJ2006}
% 48 Abstract

% \item[59] A.D. Becke and E.R. Johnson.
% A Simple Effective Potential for Exchange.
% Journal of Chemical Physics 124, 221101 (2006).

\item[JB2006b] E.R.\ Johnson and A.D.\ Becke, ``A post-Hartree-Fock model of 
intermolecular interactions: Inclusion of higher-order corrections'',
{\em J.\ Chem.\ Phys.}\ {\bf 124}, 174104 (2006). 
% {\color{green} 28/56} % \cite{Jb2006b}

\item[BJ2006b] A.D.\ Becke and E.R.\ Johnson, ``Exchange-Hole Dipole Moment 
and the Dispersion Interaction: High Order Dispersion Coefficients'',
{\em J.\ Chem.\ Phys.}\ {\bf 124}, 014104 (2006). 
% {\color{green} 28/55} % \cite{bJ2006b}

\item[BJ2005] A.D.\ Becke and E.R.\ Johnson, ``A Density-Functional Model 
of the Dispersion Interaction'', {\em J.\ Chem.\ Phys.}\ {\bf 123},
154101 (2005). % {\color{green} 28/54} % \cite{bJ2005}

\item[DB2005] R.M.\ Dickson and A.D.\ Becke,  ``Reaction Barrier Heights 
from an Exact-Exchange-Based Density-Functional Correlation Model'',
{\em J.\ Chem.\ Phys.}\ {\bf 123}, 111101 (2005). 
% {\color{green} 28/53} % \cite{Db2005}

\item[JB2005] E.R.\ Johnson and A.D.\ Becke,  ``A Post-Hartree-Fock Model 
of Intermolecular Interactions'', {\em J.\ Chem.\ Phys.}\ {\bf 123},
024101 (2005). % {\color{green} 28/52} % \cite{Jb2005}

\item[BJ2005b] A.D.\ Becke and E.R.\ Johnson, ``Exchange-Hole Dipole Moment 
and the Dispersion Interaction'', {\em J.\ Chem.\ Phys.}\ {\bf 122},
154104 (2005).  % {\color{green} 28/51} % \cite{bJ2005b}

\item[B2005] A.D.\ Becke, ``Real-Space Post-Hartree-Fock Correlation Models'',
{\em J.\ Chem.\ Phys.}\ {\bf 122}, 064101 (2005). 
% {\color{green} 28/50} % \cite{b2005}

\end{description}

\begin{center}

\rule{\textwidth}{1pt}\\
{\bf 2004} E.R.\ Johnson begins her doctoral studies with Axel.\\
\rule{\textwidth}{1pt}\\
{\bf 2003-2004} Y.\ Huh does undergraduate research with Axel.\\
\rule{\textwidth}{1pt}\\
{\bf 2003} R.M.\ Dickson begins work with Axel as postdoctoral fellow/research associate\\
\rule{\textwidth}{1pt}

\end{center}

\begin{description}

\item[B2003] A.D.\ Becke, ``A Real-Space Model of Nondynamical Correlation'',
{\em J.\ Chem.\ Phys.}\ {\bf 119}, 2972 (2003). 
% {\color{green} 27/49} % \cite{b2003}

\item[B2002] A.D.\ Becke, ``Current Density in Exchange-Correlation 
Functionals: Application to Atomic States'', {\em J.\ Chem.\ Phys.}\
{\bf 117}, 6935 (2002). % {\color{green} 26/48} % \cite{b2002}

\item[SB2002] H.L.\ Schmider and A.D.\ Becke, ``Two Functions of the Density 
Matrix and Their Relation to the Chemical Bond'', {\em J.\ Chem.\ Phys.}\
{\bf 116}, 3184 (2002). % {\color{green} 25/47} % \cite{Sb2002}

\end{description}

\begin{center}

\rule{\textwidth}{1pt}\\
{\bf 2000} awarded the Schr\"odinger medal from the World Association of
Theoretically Oriented Chemists (WATOC).\\
\rule{\textwidth}{1pt}\\
{\bf 2000} Axel is elected Fellow of the Royal Society of Canada.\\
\rule{\textwidth}{1pt}\\
{\bf 2000} H.L.\ Schmider finishes with Axel as postdoctoral fellow/research associate.\\
\rule{\textwidth}{1pt}

\end{center}

\begin{description}

\item[SB2000] H.L.\ Schmider and A.D.\ Becke, ``Chemical content of the 
kinetic energy density'', {\em J.\ Molec.\ Struct.: THEOCHEM} {\bf 527},
51 (2000). % {\color{green} 25/46} % \cite{Sb2000}

\item[B2000] A.D.\ Becke, ``Simulation of Delocalized Exchange by Local 
Density Functionals'', {\em J.\ Chem.\ Phys.}\ {\bf 112}, 4020 (2000).
% {\color{green} 25/45} % \cite{b2000}

\item[B1999] A.D.\ Becke,  ``Exploring the Limits of Gradient Corrections 
in Density-Functional Theory'', {\em J.\ Comput.\ Chem.}\ {\bf 20}, 63 
(1999).  % {\color{green} 24/44} % \cite{b1999}

\item[SB1998] H.L.\ Schmider and A.D.\ Becke, ``Density Functionals from 
the Extended G2 Test Set: Second-Order Gradient Corrections'',
{\em J.\ Chem.\ Phys.}\ {\bf 109}, 8188 (1998). 
% {\color{green} 23/43} % \cite{Sb1998}

\item[B1998] A.D.\ Becke, ``A New Inhomogeneity Parameter in 
Density-Functional Theory'', {\em J.\ Chem.\ Phys.}\ {\bf 109},
2092 (1998). % {\color{green} 23/42} % \cite{b1998}

\item[SB1998b] H.L.\ Schmider and A.D.\ Becke, ``Optimized Density 
Functionals from the Extended G2 Test Set'' {\em J.\ Chem.\ Phys.}\
{\bf 108}, 9624 (1998). % {\color{green} 22/41} %  \cite{Sb1998b}

\end{description}

\begin{center}

\rule{\textwidth}{1pt}\\
{\bf 1998} Nobel Prize in Chemistry awarded to Walter Kohn ``for his 
development of the density-functional theory'' and to John Pople
``for his development of computational methods in quantum chemistry''.
\rule{\textwidth}{1pt}

\end{center}

\begin{description}

\item[B1997] A.D.\ Becke, ``Density-Functional Thermochemistry V: 
Systematic Optimization of Exchange-Correlation Functionals''
{\em J.\ Chem.\ Phys.}\ {\bf 107}, 8554 (1997). 
% {\color{green} 22/40} % \cite{b1997}

\item[KBP1996] W.\ Kohn, A.D.\ Becke, and R.G.\ Parr, ``Density-Functional 
Theory of Electronic Structure'', {\em J.\ Phys.\ Chem.}\ {\bf 100},
12974 (1996). 
% {\color{green} 21/39} % \cite{KbP1996}

\item[DB1996] R.M.\ Dickson and A.D.\ Becke, ``Local Density-Functional 
Polarizabilities and Hyperpolarizabilities at the Basis-Set Limit'',
{\em J.\ Phys.\ Chem.}\ {\bf 100}, 16105 (1996). 
% {\color{green} 21/38} % \cite{Db1996}

\item[B1996] A.D.\ Becke, ``Current-Density Dependent Exchange-Correlation 
Functionals'', {\em Can.\ J.\ Chem.}\ {\bf 74}, 995 (1996).  
% {\color{green} 21/37} % \cite{b1996}

\item[B1996b] A.D.\ Becke, ``Density-Functional Thermochemistry IV: A New 
Dynamical Correlation Functional and Implications for Exact-Exchange Mixing'',
{\em J.\ Chem.\ Phys.}\ {\bf 104}, 1040 (1996). 
% {\color{green} 20/36} % \cite{b1996b}

\end{description}

\begin{center}

\rule{\textwidth}{1pt}\\
{\bf 1996} H.L.\ Schmider begins with Axel as postdoctoral fellow/research associate.\\
\rule{\textwidth}{1pt}

\end{center}

\begin{description}

\item[KB1995] K.E.\ Edgecombe and A.D.\ Becke, ``Cr$_2$ in Density-Functional 
Theory: Approximate Spin Projection'', {\em Chem.\ Phys.\ Lett.}\ {\bf 244},
427 (1995).  
% {\color{green} 19/35} % \cite{Kb1995}

\item[B1995] A.D.\ Becke, ``Exchange-Correlation Approximations in 
Density-Functional Theory'', in {\em Modern Electronic Structure Theory}, 
edited by D.R. Yarkony (World Scientific,1995). 
% {\color{green} 19/34} % \cite{b1995}

\item[BSS1995] A.D.\ Becke, A.\ Savin, and H.\ Stoll, ``Extension of the 
Local-Spin-Density Exchange-Correlation Approximation to Multiplet States'',
{\em Theor.\ Chim.\ Acta} {\bf 91}, 147 (1995). 
% {\color{green} 18/33} % \cite{bSS1995}

\item[WBS1995] J.\ Wang, A.D.\ Becke, and V.H.\ Smith, Jr.,
``Evaluation of $\langle \hat{S}^2 \rangle$ in Restricted and 
Unrestricted Hartree-Fock and Density-Functional Theories'',
{\em J.\ Chem.\ Phys.}\ {\bf 102}, 3477 (1995). 
% {\color{green} 18/32} % \cite{WbS1995}

\item[PB1995] J.M.\ Perez-Jorda and A.D.\ Becke, ``A Density-Functional 
Study of Van der Waals Forces: Rare Gas Diatomics'', {\em Chem.\ Phys.\
Lett.}\ {\bf 233}, 134 (1995). % {\color{green} 18/31} % \cite{Pb1995}

\end{description}

\begin{center}

\rule{\textwidth}{1pt}\\
{\bf 1995} J.M.\ Perez-Jorda, finishes working with Axel as postdoctoral fellow/research associate\\
\rule{\textwidth}{1pt}\\
{\bf 1994-1995} K.E.\ Edgecombe, postdoctoral fellow/research associate.\\
\rule{\textwidth}{1pt}

\end{center}

\begin{description}

\item[B1994] A.D.\ Becke, ``Thermochemical Tests of a Kinetic-Energy 
Dependent Exchange-Correlation Approximation'', {\em Int.\ J.\ Quant.\
Chem.\ Symp.}\ {\bf 28}, 625 (1994).  
% {\color{green} 17/30} % \cite{b1994}.

\item[PBS1994] J.M.\ Perez-Jorda, A.D.\ Becke, and E.\ San-Fabian,
``Automatic Numerical Integration Techniques for Polyatomic Molecules'',
{\em J.\ Chem.\ Phys.}\ {\bf 100}, 6520 (1994). 
% {\color{green} 16/29} % \cite{PbS1994}

\end{description}

\begin{center}

\rule{\textwidth}{1pt}\\
{\bf 1993} J.M.\ Perez-Jorda begins work with Axel as postdoctoral fellow/research associate\\
\rule{\textwidth}{1pt}\\
{\bf 1993} R.M.\ Dickson, postdoctoral fellow/research associate\\
\rule{\textwidth}{1pt}

\end{center}

\begin{description}

\item[DB1993] R.M.\ Dickson and A.D.\ Becke, ``Basis-Set-Free Local 
Density-Functional Calculations of Geometries of Polyatomic
Molecules'', {\em J.\ Chem.\ Phys.}\ {\bf 99}, 3898 (1993). 
% {\color{green} 16/28} % \cite{Db1993}

\item[B1993] A.D.\ Becke, ``Density-Functional Thermochemistry III: the 
Role of Exact Exchange'', {\em J.\ Chem.\ Phys.}\ {\bf 98}, 5648 (1993).
% {\color{red} (cited 43,179 times)}  {\color{green} 16/27} % \cite{b1993}

\item[B1993b] A.D.\ Becke,  ``A New Mixing of Hartree-Fock and Local 
Density-Functional Theories'', {\em J.\ Chem.\ Phys.}\ {\bf 98},
1372 (1993). % {\color{green} 15/26} % \cite{b1993b}

\item[B1992] A.D.\ Becke, ``Density-Functional Thermochemistry II: the 
Effect of the Perdew-Wang Generalized-Gradient Correlation Correction'',
{\em J.\ Chem.\ Phys.}\ {\bf 97}, 9173 (1992). 
% {\color{green} 14/25} %  \cite{b1992}

\item[B1992b] A.D.\ Becke, ``Density-Functional Thermochemistry I: 
the Effect of the Exchange-Only Gradient Correction'', {\em J.\ Chem.\
Phys.}\ {\bf 96}, 2155 (1992). % {\color{green} 13/24} % \cite{b1992b}

\end{description}

\begin{center}

\rule{\textwidth}{1pt}\\
{\bf 1992} R.M.\ Dickson obtains his 
PhD Thesis with Axel at Queen's: ``Tests of a Basis-Set-Free Approach to 
Local Density Functional Calculations of the Structures of Polyatomic 
Molecules'' \cite{D1992}\\
\rule{\textwidth}{1pt}\\
{\bf 1992} {\sc CADPAC5} \cite{MLHA1992} and {\sc Gaussian92/DFT} 
\cite{PGJ1992} become 
the first {\em ab initio} quantum chemistry codes to incorporate DFT.\\
\rule{\textwidth}{1pt}\\
{\bf 1991} Axel wins the Medal of the International Academy of Quantum Molecular Science.\\
\rule{\textwidth}{1pt}

\end{center}

\begin{description}

\item[SBF+1991] A.\ Savin, A.D.\ Becke, J.\ Flad, R.\ Nesper, H.\ Preuss, and 
H.G.\ von Schnering, ``A New Look at Electron Localization'',
{\em Angew.\ Chem., Int.\ Ed.}\ {\bf 30}, 409 (1991); ``Ein neuer Blick
auf die Electronenlokalisierung'', {\em Angew.\ Chem.}\ {\bf 103}, 421 (1991).
% {\color{green} 12/23} % \cite{SbF+1991,SbF+1991b}
\item[BE1990] A.D.\ Becke and K.E.\ Edgecombe, ``A Simple Measure of 
Electron Localization in Atomic and Molecular Systems'', {\em J.\ Chem.\
Phys.}\ {\bf 92}, 5397 (1990).  % {\color{green} 12/22} % \cite{bE1990}

\end{description}

\begin{center}

\rule{\textwidth}{1pt}\\
{\bf 1990} K.E..\ Edgecombe finishes his postdoctoral fellow/research 
associate with Axel.\\
\rule{\textwidth}{1pt}

\end{center}

\begin{description}

\item[BD1990] A.D.\ Becke and R.M.\ Dickson, ``Numerical Solution of 
Schroedinger's Equation in Polyatomic Molecules'' {\em J.\  Chem.\ Phys.}\
{\bf 92}, 3610 (1990). % {\color{green} 12/21} % \cite{bD1990}

\item[B1989] A.D.\ Becke, ``Basis-Set-Free Density-Functional Quantum 
Chemistry'', {\em Int.\ J.\ Quant.\ Chem., Symp.}\ {\bf 23}, 599 (1989).
% {\color{green} 12/20} % \cite{b1989}

\item[ZTFB1989] T.\ Ziegler, V.\ Tschinke, L.\ Fan, and A.\ Becke,
``Theoretical study on the electronic and molecular structures of 
(C$_5$H$_5$)M(L) (M = rhodium, iridium; L = carbonyl, phosphine) and 
M(CO)$_4$ (M = ruthenium, osmium) and their ability to activate the 
carbon-hydrogen bond in methane'',
{\em J.\ Am.\ Chem.\ Soc.}\ {\bf 111}, 9177 (1989). 
% {\color{green} 11/19} % \cite{ZTFb1989}

\item[B1989b] A.D.\ Becke, ``Density-Functional Theories in Quantum Chemistry: 
Beyond the Local Density Approximation'', in {\em The Challenge of $d$ and $f$ 
Electrons: Theory and Computation}, edited by D.R.\ Salahub and M.C.\ Zerner, 
American Chemical Society Symposium Series {\bf 394}, 165 (1989).
% {\color{green} 11/18} % \cite{b1989b}

\item[BR1989] A.D.\ Becke and M.R.\ Roussel, ``Exchange Holes in Inhomogeneous 
Systems: A Coordinate-Space Model'', {\em Phys.\ Rev.\ A} {\bf 39}, 3761 
(1989). % {\color{green} 10/17} % \cite{bR1989}

\item[B1988] A.D.\ Becke, ``Density-Functional Exchange Energy Approximation 
with Correct Asymptotic Behaviour'', {\em Phys.\ Rev.\ A} {\bf 38}, 3098
(1988).  % {\color{red} (cited 25,058 times)} {\color{green} 10/16} %  \cite{b1988}

\item[BD1988] A.D.\ Becke and R.M.\ Dickson, ``Numerical Solution of 
Poisson's Equation in Polyatomic Molecules'', {\em J.\ Chem.\ Phys.}\
{\bf 89}, 2993 (1988). % {\color{green} 9/15} %  \cite{bD1988}

\item[B1988b] A.D.\ Becke, ``A Multicentre Numerical Integration Scheme 
for Polyatomic Molecules'', {\em J.\ Chem.\ Phys.}\ {\bf 88}, 2547 (1988).
% {\color{green} 9/14} %  \cite{b1988b}

\item[B1988c] A.D.\ Becke, ``Correlation Energy of an Inhomogeneous 
Electron Gas: A Coordinate-Space Model'', {\em J.\ Chem.\ Phys.}\ {\bf 88},
1053 (1988). % {\color{green} 8/13} % \cite{b1988c}

\end{description}

\begin{center}

\rule{\textwidth}{1pt}\\
{\bf 1988} K.E..\ Edgecombe becomes a postdoctoral fellow/research associate
with Axel.\\
\rule{\textwidth}{1pt}
{\bf 1987-1988} M.\ Roussel does undergraduate researach with Axel.\\
\rule{\textwidth}{1pt}

\end{center}

\begin{description}

\item[ZTB1987] T.\ Ziegler, V.\ Tschinke, and A.\ Becke,
``A Theoretical Study on the Relative Strengths of the Metal-Hydrogen and
Metal-Methyl Bonds in Complexes of Middle to Late Transition Metals'',
{\em J.\ Am.\ Chem.\ Soc.}\ {\bf 109}, 1351 (1987). 
% {\color{green} 7/12} % \cite{ZTb1987}

\item[TTB1987] T.\ Ziegler, V.\ Tschinke, and A.\ Becke, ``A Theoretical 
Study on the Strength of Multiple Metal-Metal Bonds in Binuclear
Complexes and Transition-Metal Dimers by a Non-Local Density-Functional
Method'' {\em Polyhedron} {\bf 6}, 685 (1987). 
% {\color{green} 7/11} % \cite{TTb1987}

\item[B1987] A.D.\ Becke, ``Density-Functional Calculations of Molecular 
Bond Energies'', in {\em Density Matrices and Density Functionals}, 
edited by R.M.\ Erdahl and V.H.\ Smith, Jr. (Reidel, Dordrecht, 1987).
% {\color{green} 7/10} % \cite{b1987}

\end{description}

\begin{center}

\rule{\textwidth}{1pt}\\
{\bf 1986} R.M.\ Dickson begins his doctoral studies with Axel.\\
\rule{\textwidth}{1pt}

\end{center}

\begin{description}

\item[B1986] A.D.\ Becke, ``On the Large-Gradient Behaviour of the 
Density-Functional Exchange Energy'', {\em J.\ Chem.\ Phys.}\ {\bf 85},
7184 (1986). % {\color{green} 6/9} % \cite{b1986}

\item[B1986b] A.D.\ Becke, ``Density-Functional Calculations of Molecular 
Bond Energies'', {\em J.\ Chem.\ Phys.}\ {\bf 84}, 4524 (1986). 
% {\color{green} 5/8} % \cite{b1986b}

\item[B1986c] A.D.\ Becke, ``Completely Numerical Calculations on Diatomic 
Molecules in the Local Density Approximation'', {\em Phys.\ Rev.\ A} {\bf 33},
2786 (1986).  % {\color{green} 4/7} % \cite{b1986c}

\end{description}

\begin{center}

\rule{\textwidth}{1pt}\\
{\bf 1985-1986} C.\ Aiken does undergraduate research with Axel.\\
\rule{\textwidth}{1pt}

\end{center}

\begin{description}

\item[B1985] A.D.\ Becke, ``Local Exchange-Correlation Approximations and 
First-Row Molecular Dissociation Energies'', {\em Int.\ J.\ Quant.\ Chem.}\
{\bf 27}, 585 (1985).  
% {\color{green} 3/6} % \cite{b1985}

\end{description}

\begin{center}

\rule{\textwidth}{1pt}\\
{\bf 1984} Axel becomes a Professor of Chemistry at Queen's University, Kingston,
Ontario, Canada.\\
\rule{\textwidth}{1pt}\\
{\bf 1983-1984} Axel is E.B.\ Eastburn Postdocotral Fellow, Department
of Chemistry, Dalhouse University, Halifax, Nova Scotia,
Canada.\\
\rule{\textwidth}{1pt}\\
{\bf 1983} Axel finishes his NSERC Postdoctoral Fellowship.

\rule{\textwidth}{1pt}

\end{center}

\begin{description}

\item[B1983] A.D.\ Becke, ``Hartree-Fock Exchange Energy of an Inhomogeneous 
Electron Gas'', {\em Int.\ J.\ Quant.\ Chem.}\ {\bf 23}, 1915 (1983). 
% {\color{green} 2/5} % \cite{b1983}

% \item[B1983b] A.D.\ Becke, ``Numerical Hartree-Fock-Slater Calculations on 
% Diatomic Molecules: Addendum'', {\em J.\ Chem.\ Phys.}\ {\bf 78}, 4787 (1983).
% \cite{b1983b}

\item[B1982] A.D.\ Becke,  ``Numerical Hartree-Fock-Slater Calculations on 
Diatomic Molecules'', {\em J.\ Chem.\ Phys.}\ {\bf 76}, 6037 (1982); 
A.D.\ Becke, ``Numerical Hartree-Fock-Slater Calculations on
Diatomic Molecules: Addendum'', {\em J.\ Chem.\ Phys.}\ {\bf 78}, 4787 (1983).
% {\color{green} 1/4} % \cite{b1982}

\end{description}

\begin{center}

\rule{\textwidth}{1pt}\\
{\bf 1981} Axel becomes Natural Science and Engineering Research Council 
(NSERC) of Canada and Killam Postdoctoral Fellow with Russell Boyd, 
Department of Chemistry, Dalhousie University, Halifax, Nova Scotia, Canada.\\
\rule{\textwidth}{1pt}\\
{\bf 1981} Axel receives his Ph.D.\ Theoretical Physics, McMaster University, Hamilton, Ontario,
Canada\\
Doctoral Thesis, ``Numerical Hartree-Fock-Slater Calculations 
on Diatomic Molecules'', supervised by Prof.\ D.W.L.\ Sprung, 
Master University, July 1981.\\
\rule{\textwidth}{1pt}\\

\end{center}

\begin{description}

\item[KBBS1979] C.M.\ Ko, J.R.\ Borysowicz, A.D.\ Becke, and D.W.L.\ Sprung,
``A Note on Least Squares Fitting with Normalization Parameters'',
{\em Nucl.\ Phys.\ A}, {\bf 319}, 175 (1979).  % {\color{green} 0/3} 
% \cite{KBbS1979}

\item[BS1977] A.D.\ Becke and D.W.L.\ Sprung, ``Least Squares Fits with 
Normalization Parameters and Linear Constraints'', {\em Nucl.\ Phys.\ A}
{\bf 284}, 425 (1977).  % {\color{green} 0/2} % \cite{bS1977}

\end{description}

\begin{center}

\rule{\textwidth}{1pt}\\
{\bf 1977} Axel receives his M.Sc.\ Theoretical Physics, McMaster University\\
Masters Thesis, ``Eikonal Distorted Wave Approximation for High Energy
Electron Scattering from Spherical Nuclei'', supervised by 
Prof.\ D.W.L.\ Sprung, McMaster University, Hamilton, Ontario, Canada\\
\rule{\textwidth}{1pt}\\
{\bf 1975} Axel receives his B.Sc.\ Engineering Physics, Queen's University\\
\rule{\textwidth}{1pt}

\end{center}

\begin{description}

\item[SKB1975] A.J.\ Springthorpe, F.D.\ King, and A.D.\ Becke,
``Te and Ge Doping Studies in Ga$_{1-x}$Al$_x$As'', {\em J.\ Electronic Mater.}\
{\bf 4}, 101 (1975). % {\color{green} 0/1} % \cite{SKb1975}

\end{description}

\begin{center}
\rule{\textwidth}{1pt}\\
   {\bf 10 June 1953} Axel is born in Esslingen, Germany.\\
\rule{\textwidth}{1pt}
\end{center}

\normalsize

%%%%%%%
% EOF %
%%%%%%%
% ----------------------------------------
\section{Brief Summary of My Own Career}
\label{sec:CVme}
% \input{CVme.tex}
% ===============================================================
% File CVme.tex .
% Last modified: 7 June 2026 
% ===============================================================

\small

\begin{description}

\item[2001-present] Chemistry Professor at the {\em Universit\'e Grenoble Alpes}
in Grenoble, France, working on the development of time-dependent (TD\index{TD})
density-functional theory (DFT\index{DFT}), DFT treatment of spin-crossover 
Fe(II) octahedral complexes, luminescence in Ru(II) octahedral complexes, 
and organic solar cells.  (The name of the University was the {\em Universit\'e
Joseph Fourier} when I first joined the staff and later became the 
{\em Universit\'e de Grenoble} before finally acquiring its present name.)

\item[1999-2000] {\em Chercheur associ\'e} (Research Associate) of the
{\em Conseil National de la Recherche Scientifique (CNRS)} (French
National Council of Scientific Research) at the {\em Universit\'e 
Paul Sabatier} in Toulouse, France, working on time-dependent (TD\index{TD})
DFT.

\item[1991-2001] Senior scientist and computing professional at the 
{\em Universit\'e de Montr\'eal} (UdM), Montreal, Quebec, Canada, working in 
the group of Prof.\ Dennis R.\ Salahub.  Computer support for several
groups and (for a while) consultant for Canada's High Perfomance Computing
network.  Maintenance and contributions to the development of the {\sc deMon}
({\em densit\'e de Montr\'eal}) DFT programs.  Expansion of research areas
into the calculation of the polarizabilities of sodium clusters and the 
development of practical calculation of molecular absorption spectra using
time-dependent (TD\index{TD}) density-functional theory (DFT\index{DFT}).

\item[1986-1991] Postdoctoral fellow and then research associate with Prof.\
Delano P.\ Chong at the University of British Columbia (UBC), Vancouver, 
British Columbia, Canada, developing Green's function theory and DFT
for the treatment of electron momentum spectroscopy (EMS\index{EMS}) 
and testing of DFT for calculating polarizabilities and 
hyperpolarizabilities.

\item[1980-1986] PhD, Theoretical Chemistry, University of Wisconsin,
with Prof.\ John E.\ Harriman, working on reduced density matrix 
theory and (during a short postdoctoral position) on Husimi functions.

\item[1979-1980] Research assistant in analytical chemistry at the
{\em Arrhenius Laboratoriet} at {\em Stockholms Universit\"atet}, 
Stockholm, Sweden.  

\item[1975-1979] BS Chemistry, University of California at Berkeley,
Berkeley, California, U.S.A.  Undergraduate research with Prof.\
Henry F.\ Schaefer III doing Hartree-Fock and configuration interaction
calculations on small exotic molecules.

\item[1957-1975] Born in Madison, Wisconsin, U.S.A., but grew up mainly
in Berkeley, California, U.S.A.

\end{description}

\normalsize
% ----------------------------------------
% \bibliographystyle{myaip}
% \bibliography{becke,refs}

% \printindex
% ----------------------------------------------------------
\end{document}